\documentclass{aa}
\usepackage{comment} 
\usepackage{graphicx}
\usepackage{xcolor}
\usepackage{amsmath}
\usepackage{natbib}
\bibpunct{(}{)}{;}{a}{}{,}
\usepackage{txfonts}
\usepackage{hyperref}
\hypersetup{
    colorlinks,
    linkcolor={blue},
    citecolor={blue},
    urlcolor={blue}
}

\usepackage[noabbrev,capitalize,nameinlink]{cleveref}
\begin{document} 
   \title{Rethinking mass transfer: a unified semi-analytical framework for circular and eccentric binaries}
   
   \subtitle{I. Orbital evolution due to conservative mass transfer}

   \author{A. Parkosidis\inst{1} \and 
           S. Toonen \inst{1} \and 
           F. Dosopoulou \inst{2} \and 
           E. Laplace \inst{3,4,1}
          }

   \institute{Anton Pannekoek Institute for Astronomy, University of Amsterdam, Amsterdam 1098 XH, The Netherlands\\
              \email{a.parkosidis@uva.nl} 
              \and
             School of Physics and Astronomy, Cardiff University, Cardiff, CF24 3AA, United Kingdom \and
             Institute of Astronomy, KU Leuven, Celestijnenlaan 200D, B-3001 Leuven, Belgium
              \and 
              Leuven Gravity Institute, KU Leuven, Celestijnenlaan 200D, box 2415, 3001 Leuven, Belgium
              }

   \date{}
  
  \abstract{Mass transfer (MT) is a fundamental process in stellar evolution. While MT in circular orbits is well studied, observations indicate that it also occurs in eccentric ones, where theoretical models are limited. We present a new semi-analytic framework for the secular orbital evolution of mass-transferring binaries, treating stars either as point-masses or as extended bodies. For the first time, a MT model is applicable to both circular and eccentric orbits and accommodates conservative and non-conservative MT across a broad range of mass ratios and stellar spins. We derive secular, orbit-averaged equations describing the orbital evolution by treating MT, mass loss, and angular momentum (AM) loss as perturbations to the general two-body problem. Assuming conservative MT, we compare our results to previous models and validate them against numerical integrations. Our model predicts eccentric post-MT systems in wider orbits than classical results. Compared to other eccentric MT frameworks we find a broader parameter space for orbital widening and eccentricity pumping. Accounting for extended bodies yields stronger semimajor axis and eccentricity growth at a given mass ratio, and further broadens the parameter space for orbital widening and eccentricity pumping. Whether extended bodies are considered or not, eccentric MT naturally predicts higher eccentricities at longer orbital periods, a correlation observed in numerous post-MT systems, providing a robust mechanism for their formation. Our model can be integrated into binary evolution and population synthesis codes to consistently treat conservative and non-conservative MT in arbitrarily eccentric orbits with applications ranging from MT on the main sequence to gravitational-wave progenitors.}


   \maketitle
%
%-------------------------------------------------------------------
\section{Introduction}\label{sec:one}

Many binary and multiple-star systems, experience at least one phase of MT during their evolution \citep{2012Sci...337..444S,2017ApJS..230...15M}. Among the mechanisms of mass exchange, such as stellar winds, Roche-lobe overflow (RLOF) stands out for its association with a plethora of observational phenomena. These include X-ray binaries, nova outbursts, cataclysmic variables, Type Ia supernovae, symbiotic stars, and the spin-up of neutron stars. Furthermore, stable MT via RLOF is thought to play a critical role in the formation and evolution of certain populations, such as subdwarf B (sdB) stars \citep{2002MNRAS.336..449H,2003MNRAS.341..669H,2008ASPC..392...15P,2009ARA&A..47..211H,2015A&A...579A..49V,2017A&A...605A.109V,2020A&A...641A.163V,2022A&A...658A.122M}, blue stragglers \citep{2011Natur.478..356G,2009Natur.462.1032M}, Barium stars \citep{2019A&A...626A.127J}, CH and CEMP-s stars \citep{2016A&A...586A.158J,2016A&A...588A...3H,2016ApJ...826...85S}, gravitational wave (GW) sources \citep{2017MNRAS.471.4256V,2021A&A...650A.107M,2025AAS...24541904S,2024A&A...681A..31P} and compact double white dwarfs \citep{2012ApJ...744...12W,2019ApJ...871..148L,2020ApJ...893....2L}.

Current binary evolution models continue to face difficulties in reproducing the orbital properties of many post-MT systems, particularly those with wide and eccentric orbits \citep{2003ASPC..303..290P,2008A&A...480..797B, 2013A&A...551A..50D,2015A&A...579A..49V,2018A&A...620A..85O,2020A&A...642A.234O,2022A&A...658A.122M}. Numerical models of binary evolution during RLOF have traditionally neglected orbital eccentricity, assuming that tidal forces universally circularize orbits before the onset of MT \citep{1996A&A...309..179P,2002MNRAS.329..897H,2003ASPC..303..290P,2008ApJS..174..223B,2012A&A...546A..70T}. This assumption, however, is challenged by observations of interacting binaries with non-zero eccentricities \citep{1999AJ....117..587P,2005A&AT...24..151R} and by inconsistencies in the implementation of tidal prescriptions \citep{2024A&A...681L...1S} or weak tides \citep[e.g.,][]{2009MNRAS.400L..20E,2022ApJ...933...25P}. Moreover, the wide and eccentric nature of many post-MT systems \citep[e.q., see][and references therein]{2016A&A...586A.158J,2018AJ....155..144K,2019A&A...626A.127J,2020A&A...641A.163V,2022A&A...658A.122M,2024MNRAS.52711719Y,2024MNRAS.529.3729S} suggests that RLOF may not only preserve but in some cases even develop eccentricities in these systems.

Analytical expressions for orbital evolution are crucial for studying the secular evolution of mass-transferring systems in large-scale population studies. Working within the framework established by \cite{1969Ap&SS...3...31H}, \cite{2007ApJ...667.1170S, 2009ApJ...702.1387S, 2016ApJ...825...70D, 2016ApJ...825...71D} derived equations describing the secular evolution of orbital elements due to MT via RLOF in eccentric binaries. They assumed a delta-function model (hereafter referred to as the `$\delta$-function' model) centered at the periapsis of the orbit, a model physically motivated for systems with extremely high eccentricities, but not valid for orbits with lower eccentricities. \cite{2019ApJ...872..119H} demonstrated that the equations describing the secular evolution of the orbit are invalid in the case of circular orbits and derived a new set of analytical equations assuming a phase-dependent MT rate. Their updated eccentric mass transfer model (hereafter referred to as the `emt' model), eliminates the problematic behavior at the limit of circular orbits, and it is valid for any eccentricity. However, this formalism is limited to conservative MT.

In this paper, we present a new semi-analytical framework to describe the orbital evolution of mass-transferring binaries, building on the physically motivated MT rate derived by \cite{2019ApJ...872..119H}. The General Mass Transfer model (hereafter referred to as the GeMT-model) is designed to address both conservative and non-conservative MT scenarios across the full range of orbital eccentricities, including circular orbits. In Section~\ref{sec:two}, we establish the foundation of our approach by treating the effects of MT as perturbations to the instantaneous orbit of the binary. In Section~\ref{sec:three}, we outline the key components of the model and highlight improvements over previous studies. In Section~\ref{sec:four}, we derive the orbit-averaged equations of the model. In Section~\ref{sec:five}, we explore the model's behavior in various limiting cases, we compare its predictions with earlier frameworks. In Section~\ref{sec:six}, we apply the model to isolated binaries assuming conservative MT. Finally, we discuss the limitations and implications of our work in Sect.~\ref{sec:seven} and conclude in Sect.~\ref{sec:eight}.

%--------------------------------------------------------------------

\section{The perturbed two body problem}\label{sec:two}

Two-body systems, such as binary stars, are susceptible to gravitational perturbations; tidal dissipation, relativistic corrections, gravitational wave radiation, magnetic fields, inertial forces, mass loss/mass transfer processes, and others are examples of such agents. All of these physical processes operate as perturbing forces on the general two-body problem \citep{2016ApJ...825...70D}, hence each star's perturbed (actual) orbit diverges from its osculating orbit, that is, the one it would have if perturbations were absent.

The perturbing force in principle depends on both the relative position, $\vec{r}$, and velocity,  $\vec{\dot{r}}$, of the binary components. Therefore, the relative acceleration of the perturbed two-body problem is written as
\begin{equation}\label{eq:relative_acceleration_general}
    \centering
    \frac{d^2\vec{r}}{dt^2} = - \frac{GM}{r^3} \vec{r} + \vec{f}(\vec{r},\vec{\dot{r}}),
\end{equation}
where $\vec{f}(\vec{r},\vec{\dot{r}})$ is the perturbing force per unit mass. Of course, in the absence of any perturbation $\vec{f}(\vec{r},\vec{\dot{r}}) =0$ and Eq.~\eqref{eq:relative_acceleration_general} describes the general reduced unperturbed two-body problem.

\subsection{Perturbations during mass transfer}\label{sub:mass_exchange_perurbation}

MT between binary components can occur through various mechanisms. During the detached phase, a star may lose mass via stellar winds, with a fraction of the escaping material potentially accreted by the companion. In contrast, if the system transitions into a semi-detached phase, the donor star can predominantly transfer mass to its companion, the accretor, through the inner Lagrangian point $L_1$, via RLOF. During RLOF, material is ejected and accreted at specific points and with characteristic velocities, generating reaction forces on both the donor and the accretor. Hereafter, the subscripts `${\rm don}$' and `${\rm acc}$' denote parameters associated with the donor and accretor stars, respectively.

Consider a semi-detached binary system. Both stars are assumed to be centrally condensed and spherically symmetric, with masses donated as $M_{\rm don}$ and $M_{\rm acc}$, respectively. The stars rotate around each other, defining an eccentric orbit with semimajor axis $a$, eccentricity $e$ and period $P_{\rm orb}$. The system's total mass is $M= M_{\rm don} + M_{\rm acc}$, and the mass ratio is defined as $q=M_{\rm don}/M_{\rm acc}$. We assume the binary components rotate uniformly at spin angular velocities $\vec{\Omega}_{\rm don}$ and $\vec{\Omega}_{\rm acc}$, parallel to each other and to the orbital angular velocity $\vec{\Omega}_{\rm orb}$. Note that the magnitude of the vector $\vec{\Omega}_{\rm orb}$ varies over time for eccentric orbits, but it remains directed along the orbital AM vector $\hat{\vec{h}}$, where $\vec{h} \equiv \vec{r} \times \vec{\dot{r}}$.  Specifically, $\vec{\Omega}_{\rm orb} = n\sqrt{1-e^2}(a/r)^2 \hat{\vec{h}}$ and $n = 2 \pi/P_{\rm orb}$. 

Assume that the donor loses mass at a rate $\dot{M}_{\rm don}$ from the position $\vec{r}_{\rm don}$ (ejection point) relative to its center of mass. Similarly, the accretor gains mass at a rate $\dot{M}_{\rm acc}$ from the position $\vec{r}_{\rm acc}$ (accretion point) relative to its center of mass. The absolute velocities (i.e, with respect to an inertial reference frame) of the ejected and accreted matter are $\vec{W}_{\rm don}$ and $\vec{W}_{\rm acc}$, respectively. The velocities of the ejected and accreted mass relative to the donor and accretor are $\vec{w}_{\rm don}= \vec{W}_{\rm don} -d\vec{R}_{\rm don}/dt$ and $\vec{w}_{\rm acc} = \vec{W}_{\rm acc} -d\vec{R}_{\rm acc}/dt$, respectively, where $d\vec{R}_{\rm don}/dt$ and $d\vec{R}_{\rm acc}/dt$ denote the absolute velocities of the donor's and accretor's centers of mass. Consequently, a perturbing acceleration acts on the system. Following \cite{1969Ap&SS...3...31H,2007ApJ...667.1170S}, the perturbing acceleration is written as
\begin{flalign}\label{eq:perturbation_acceleration_MT}
    \vec{f} &= \frac{\vec{g}_{\rm acc}}{M_{\rm acc}} - \frac{\vec{g}_{\rm don}}{M_{\rm don}} \nonumber\\
    &+ \frac{\dot{M}_{\rm acc}}{M_{\rm acc}}(\vec{w}_{\rm acc} + \vec{\Omega}_{\rm orb} \times \vec{r}_{\rm acc}) - \frac{\dot{M}_{\rm don}}{M_{\rm don}}(\vec{w}_{\rm don} + \vec{\Omega}_{\rm orb} \times \vec{r}_{\rm don}) \\ 
    &+ \frac{\ddot{M}_{\rm acc}}{M_{\rm acc}} \vec{r}_{\rm acc} - \frac{\ddot{M}_{\rm don}}{M_{\rm don}} \vec{r}_{\rm don}, \nonumber
\end{flalign}
where $\vec{g}_{\rm acc}$ and $\vec{g}_{\rm don}$ represent orbital perturbations caused by particles in the MT stream, the terms in the second line represent the change in linear momentum of the accretor and donor caused by mass ejection and accretion, respectively, and the terms in the third line account for the acceleration of the centers of mass of the accretor and donor, resulting from the instantaneous changes in their masses. The derivation is based on the assumption that second- or higher-order terms in $\Delta M_{\rm don}$ and $\Delta M_{\rm acc}$ are ignored.

It is important to emphasize that Eq.~\eqref{eq:perturbation_acceleration_MT} is generic. In the idealized case of isotropic mass ejection from the donor and accretion onto the companion, such as isotropic stellar winds, the points of ejection and accretion are assumed to coincide with the centers of mass of the two stars (i.e., $\vec{r}_{\rm acc} = \vec{r}_{\rm don} = 0$). However, in the more general case of anisotropic mass ejection and accretion, such as RLOF, the points of ejection and accretion are offset from the centers of mass, introducing the perturbing terms proportional to $\vec{r}_{\rm acc}$ and $ \vec{r}_{\rm don}$.

In the case of conservative MT, all transferred mass is accreted by the companion, preserving the system’s total mass. However, in most cases, some transferred mass escapes the system--this is referred to as non-conservative MT. We parameterize the fraction of the transferred mass that is accreted as $\beta$\footnote{Traditionally, $\beta$ is defined as the fraction of mass transferred from the donor star that is ejected from the system in the vicinity of the accretor \citep[e.g.,][]{1997A&A...327..620S}. Here, we follow the notation introduced in Section 7.2, pages 9–12, of \href{https://www.astro.ru.nl/~onnop/education/binaries_utrecht_notes/Binaries_ch6-8.pdf}{lecture notes on binary star evolution} by Onno Pols.}, where $0 \leq \beta \leq 1$, thus
\begin{equation}\label{eq:beta_parametrization}
    \dot{M}_{\rm acc} = - \beta \dot{M}_{\rm don} \; \text{  and  } \; \dot{M} = \dot{M}_{\rm acc} + \dot{M}_{\rm don} = (1-\beta)\dot{M}_{\rm don}.
\end{equation}
Note that $\beta = 0$ and $\beta =1$ correspond to fully non-conservative and conservative MT, respectively. Moreover, the donor star is losing mass, thus $\dot{M}_{\rm don} < 0$.

In the case of conservative MT, if mass ejection and accretion are isotropic ($\vec{r}_{\rm acc} = \vec{r}_{\rm don} = 0$), the orbital AM is conserved. However, in the more general case, MT is non-conservative, meaning that some of the transferred mass escapes the system, carrying away orbital AM. Additionally, mass ejection from the donor and accretion onto the companion can be anisotropic, as in the case of RLOF, further altering the orbital AM.

We denote the change in orbital AM due to mass loss as $\dot{J}_{\rm orb,ml}$, and the change due to anisotropic mass ejection and accretion as $\dot{J}_{\rm orb,rf}$ such that
\begin{equation}\label{eq:change_of_angular_momentum_total_1}
    \frac{\dot{J}_{\rm orb}}{J_{\rm orb}} = \frac{\dot{J}_{\rm orb,ml} + \dot{J}_{\rm orb,rf}}{J_{\rm orb}},
\end{equation}
where $J_{\rm orb} = \mu \sqrt{GMa(1-e^2)}$ and $\mu = M_{\rm don} M_{\rm acc}/M$. The amount of AM that is carried away by the lost mass can be parametrized in different ways. Following \cite {1997A&A...327..620S}, we define it to be $\gamma$ times the specific AM of the binary, such as 
\begin{equation}\label{eq:angular_momentum_parametrization} 
    \frac{\dot{J}_{\rm orb,ml}}{J_{\rm orb}} \equiv \gamma \frac{\dot{M}}{M} = \gamma (1-\beta)\frac{\dot{M}_{\rm don}}{M_{\rm don} + M_{\rm acc}}.
\end{equation}
The $\gamma$ parameter is defined relative to the system's center of mass frame, it can be phase-dependent, and it is related to the specific assumptions made about how the mass is lost from the system \citep{1997A&A...327..620S}. In this setting, $\gamma \geq 0$, since $\dot{M}_{\rm don} < 0$, so that $\dot{J}_{\rm orb,ml} \leq 0$. The $\dot{J}_{\rm orb,rf}$ term arises through the perturbing terms proportional to $\vec{r}_{\rm acc}$ and $ \vec{r}_{\rm don}$, which account for the impact of the reaction forces on both the donor and the accretor. Thus, $\dot{J}_{\rm orb,rf}$ should be zero in the limit of isotropic mass ejection and accretion, or equivalently in the limit of point masses, namely, $\vec{r}_{\rm acc} = \vec{r}_{\rm don}=0$.

Considering all perturbations and using Eqs.~\eqref{eq:perturbation_acceleration_MT} to \eqref{eq:angular_momentum_parametrization}, we write the total perturbing acceleration as
\begin{flalign}\label{eq:total_perturbation_non_conservative_MT}
    \vec{f}_{total} &=  \frac{\dot{M}_{\rm don}}{M_{\rm don}} \biggl(\frac{(1-\beta)(\gamma + \frac{1}{2})q}{1+q} \biggr) \vec{\dot{r}} \nonumber \\
    &- \frac{1}{M_{\rm don}}(\vec{g}_{\rm don} - \vec{g}_{\rm acc}q) \nonumber \\ 
    &- \frac{\dot{M}_{\rm don}}{M_{\rm don}} (\vec{w}_{\rm don} + \beta q \vec{w}_{\rm acc})  \\
    &- \frac{\dot{M}_{\rm don}}{M_{\rm don}} (\vec{\Omega}_{\rm orb} \times \vec{r}_{\rm don} + \vec{\Omega}_{\rm orb} \times \beta q \vec{r}_{\rm acc}) \nonumber \\ 
    &- \frac{\ddot{M}_{\rm don}}{M_{\rm don}} (\vec{r}_{\rm don} + \beta q \vec{r}_{\rm acc}), \nonumber
\end{flalign} 
where the new term in the first line accounts for mass lost from the system and the additional AM it may carry away.

\section{Equations of motion}\label{sec:three}

The relative acceleration of the binary components, as influenced by the total perturbation from the MT, is given by substituting Eq.~\eqref{eq:total_perturbation_non_conservative_MT} into Eq.~\eqref{eq:relative_acceleration_general}. The equations are general; they are valid for any eccentricity and can be manipulated to account for different mass exchange and mass loss scenarios. In the following section, we derive the variation of the orbital elements due to MT. First, we present simplifying assumptions that we adopt in our modeling to construct analytically tractable equations. Second, we present a new prescription for the magnitude of the ejection point, $\vec{r}_{\rm don}$, based on the position of the Lagrangian $L_1$ point as a function of various parameters of the system. Furthermore, we demonstrate that using Eq.~\eqref{eq:total_perturbation_non_conservative_MT}  under the aforementioned basic assumptions, we recover the parameterization given by Eq.~\eqref{eq:change_of_angular_momentum_total_1} and the form of $\dot{J}_{\rm orb,rf}$. Finally, we briefly present the phase-dependent mass-transfer model we utilize in this work.

\subsection{Basic assumptions}\label{subsec:assumptions}

Determining the final accretion points and velocities from the initial conditions requires solving the full two-body problem coupled with the dynamics of the MT stream \citep{1975ApJ...198..383L,2010ApJ...724..546S,2023MNRAS.524.4315H}. In general, the transferred mass can: (1) impact directly on the surface of the accretor (direct impact), (2) intersect with itself and form an accretion disk around the accretor, (3) be re-accreted onto the surface of the donor (self-accretion), or (4) escape from the system entirely. In this work, we make two simplifying assumptions to derive analytically tractable equations. The assumptions are listed below:
\begin{enumerate}
    \item We assume that any gravitational attractions exerted by the particles in the MT stream on the binary components are negligible, thus $\vec{g}_{\rm don} = \vec{g}_{\rm acc} = 0$.
    \item We assume that the donor ejects mass with a relative velocity $\vec{w}_{\rm don} = \vec{\dot{r}}$, the accretor accretes mass at $\vec{w}_{\rm acc} = -\vec{\dot{r}}$ and that $\vec{r}_{\rm don}, \vec{r}_{\rm acc}$ corotate with the orbit. 
\end{enumerate}
The first assumption is valid as long as $M_{\rm stream} \ll M_{\rm don}, M_{\rm acc}$ \citep{2007ApJ...667.1170S, 2009ApJ...702.1387S, 2016ApJ...825...70D, 2016ApJ...825...71D, 2019ApJ...872..119H}. In Section~\ref{subsec:point_masses_circ_orbits}, we show that by adopting the second assumption, the total perturbing acceleration leads to the canonical relation for the change in semimajor axis caused by non-conservative and conservative MT in circular orbits. Simultaneously, it reproduces the canonical expectation that the rate of change of the eccentricity is zero at exactly zero eccentricity.

Applying assumptions 1 and 2, the total perturbation arising due to MT simplifies to
\begin{flalign}\label{eq:total_perturbation_non_conservative_MT_simplified}
    \vec{f}_{RLOF} &=  - \frac{\dot{M}_{\rm don}}{M_{\rm don}} \biggl(1- \beta q -\frac{(1-\beta)(\gamma + \frac{1}{2})q}{1+q} \biggr) \vec{\dot{r}} \nonumber\\
    &- \frac{\dot{M}_{\rm don}}{M_{\rm don}} \vec{\Omega}_{\rm orb} \times (\vec{r}_{\rm don} +  \beta q \vec{r}_{\rm acc}) \\
    &- \frac{\ddot{M}_{\rm don}}{M_{\rm don}} (\vec{r}_{\rm don} + \beta q \vec{r}_{\rm acc}). \nonumber
\end{flalign} 
For the vector $\vec{r}_{\rm acc}$, we investigate two cases where the accretion point is located on the line connecting the two stars, with $\vec{r}_{\rm acc} = \pm r_{\rm acc} \hat{\vec{r}}$. The case of $\vec{r}_{\rm acc} = r_{\rm acc} \hat{\vec{r}}$ applies when the initial velocity of the ejected mass, $\vec{w}_{\rm don}$, is such that the mass stream follows a curved trajectory and lands on the side of the accretor that does not face the donor.

\subsection{Ejection point: a new model for the $L_1$ Lagrangian point}\label{subsec:L_1}

Traditionally, the position of the Lagrangian point $L_1$ is defined for circular orbits with synchronously rotating component stars. In eccentric orbits, though, the stars cannot remain synchronous with the orbit at all times due to the time-varying orbital angular velocity. The donor's asynchronous rotation causes the companion to exert time-dependent tidal forces, resulting in a time-dependent potential. However, \cite{2007ApJ...660.1624S} demonstrated that this potential can be approximated as quasi-static. In this approximation, the donor's shape conforms instantaneously to the quasi-static potential, provided the donor's dynamical timescale is much shorter than the timescales associated with the orbital angular velocity and the donor's rotation, a condition often referred to as the first approximation \citep{1963ApJ...138.1112L}.

Following \cite{2007ApJ...660.1624S}, the position, $X$, of the Lagrangian $L_1$ point--relative to the donor's center of mass--can be determined by solving the equation
\begin{equation}\label{eq:Lagrangian_points}
    \frac{q}{X_{\rm L}^2} - \frac{1}{(1-X_{\rm L}^2)} - X_{\rm L}(1+q)\mathcal{A}(f_{\rm don},e,\mathcal{E}) +1 = 0,
\end{equation}
where $X_{\rm L}$ is expressed in units of the instantaneous orbital separation (i.e., $X_{\rm L} = X/r$), $f_{\rm don}$ represents the donor's spin angular velocity normalized to the orbital angular velocity at periapsis, so that
\begin{equation}
    f_{\rm don}  \equiv  \frac{\Omega_{\rm don}}{n} \frac{(1-e)^{3/2}}{(1+e)^{1/2}},
\end{equation}
 and 
\begin{equation}\label{eq:A_function}
    \mathcal{A}(f_{\rm don},e,\mathcal{E})= f_{\rm don}^2 \frac{1+e}{(1-e)^3} \biggl(\frac{r}{a} \biggr)^3,
\end{equation}
quantifies the deviation of the donor's spin velocity from the orbital angular velocity at periapsis as a function of the eccentricity $e$ and eccentric anomaly $\mathcal{E}$. $\mathcal{A}(f_{\rm don},e,\mathcal{E})$ reaches its maximum for $e$ approaching 1 at $\mathcal{E}=\pi$, while it is minimal for $e=0$, \citep[see][Fig. 3]{2007ApJ...660.1624S}. Note that we choose the eccentric anomaly to express $\mathcal{A}(f_{\rm don},e,\mathcal{E})$, however other angle parameters, such as true anomaly, are equally valid parameters.

Similar to \cite{2007ApJ...667.1170S}, we solve for the periapsis of the orbit, such as $\mathcal{A}(f_{\rm don},e,\mathcal{E} =0)$, and thus Eq.~\eqref{eq:Lagrangian_points} is written as
\begin{equation}\label{eq:Lagrangian_points_2}
    \frac{q}{X_{\rm L}^2} - \frac{1}{(1-X_{\rm L}^2)} - X_{\rm L}(1+q)f_{\rm don}^2(1+e) +1 = 0.
\end{equation}
We fit a prescription, $X_{\rm L1}(f_{\rm don},q,e)$, to the numerical solutions of Eq.~\eqref{eq:Lagrangian_points_2}, such as $X_{\rm L} = X_{\rm L1}(f_{\rm don},q,e)$ and therefore, the phase-dependent position of $L_1$ in natural units is 
\begin{equation}\label{eq:prescription_L1}
    \vec{r}_{\rm don} = X_{\rm L1}(f_{\rm don},q,e) \vec{r}.
\end{equation}
The $X_{\rm L1}(f_{\rm don},q,e)$ function gives the position of the $L_1$ point at the periapsis of the binary orbit in units of the instantaneous distance between the two stars, and it is given explicitly in Appendix \ref{app:analytical_prescription_L1}. Hereafter, we refer to our prescription as the `Global-$L_1$' fit.  We note that an alternative, overall less accurate, fit to the numerical solution of Eq.~\eqref{eq:Lagrangian_points_2} is provided by \cite{2007ApJ...667.1170S} (Eq. A15 in their Appendix A). A comparison of the two is given in Fig.~\ref{fig:comparison_sep} in our Appendix \ref{app:analytical_prescription_L1}.

Solving Eq.~\eqref{eq:Lagrangian_points} is non-trivial. \cite{2019ApJ...872..119H} proposed two analytical solutions based on distinct limiting assumptions: negligible donor spin velocity, that is, $f_{\rm don} \approx 0$, and large mass ratio, namely, $q \gg 1$. These solutions are hereafter referred to as the `Low $f_{\rm don}$' and `High $q$' models, respectively. In this work, we use the Global-$L_1$ prescription for the position of $L_1$, which is applicable across the entire range of mass ratios while retaining sensitivity to the donor's spin velocity.

\begin{figure}[!htbp]
    \includegraphics[width=\linewidth]{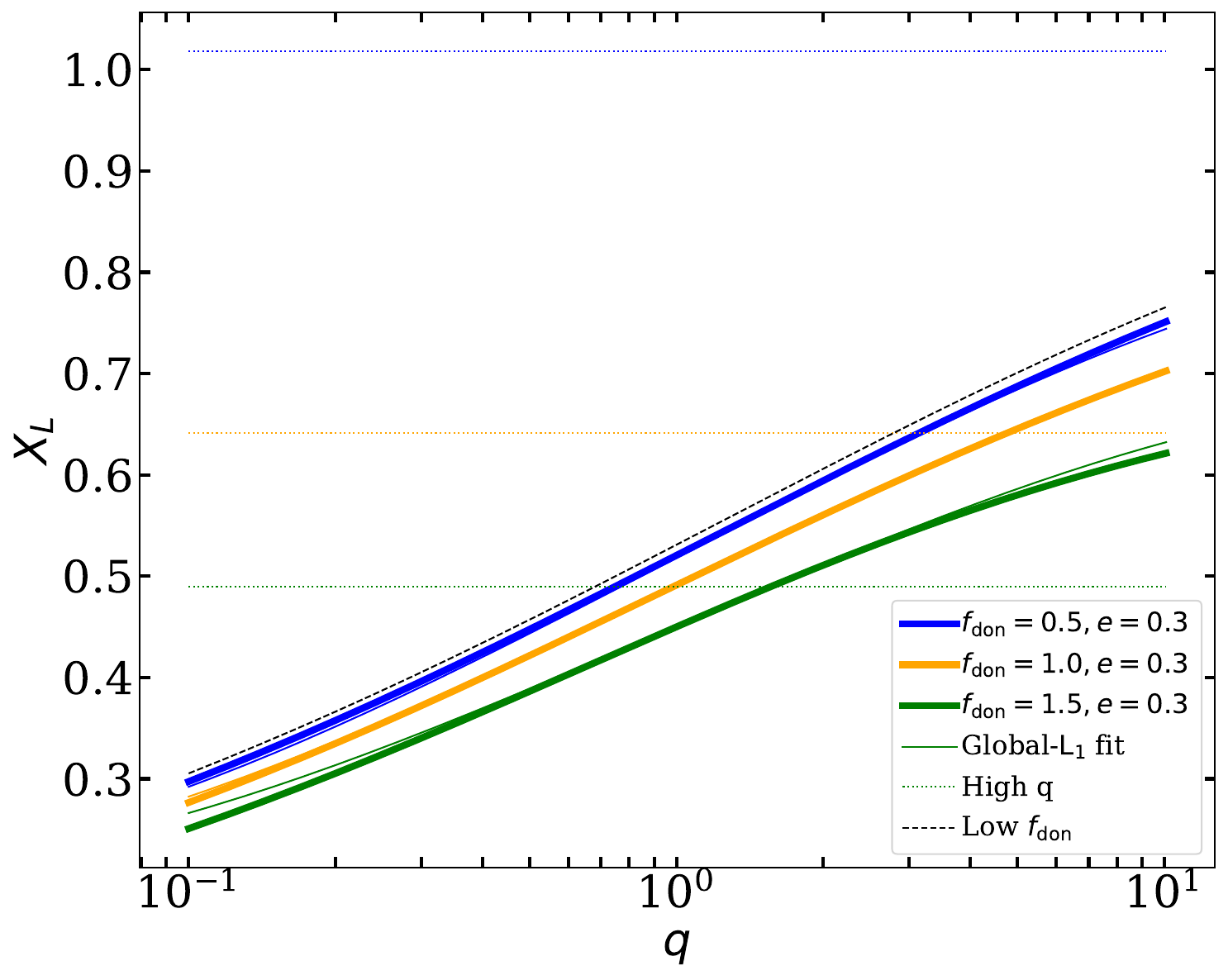}
    \caption{Position of the $L_1$ point, relative to the donor's center of mass, at the periapsis of the binary orbit in units of the instantaneous binary separation. The thick lines are the numerical solutions of Eq.~\eqref{eq:Lagrangian_points_2}. The dotted, dashed, and thin lines illustrate the High-$q$, Low-$f_{\rm don}$, and Global-$L_1$ models, respectively. Blue, orange, and green colors correspond to $f_{\rm don}=0.5,1.0,1.5$, respectively, while $e=0.3$ for all models. Note that the Low-$f_{\rm don}$ prescription is independent of $f_{\rm don}$.}
    \label{fig:fit_varying_f_d}
\end{figure}

In Figure~\ref{fig:fit_varying_f_d}, we compare the Global-$L_1$ fit predictions to the numerical solutions of Eq.~\eqref{eq:Lagrangian_points_2} for $e=0.3$ and varying donor spins $f_{\rm don}$. As $q$ increases, the position of the $L_1$ point shifts further from the donor's center of mass. Conversely, higher donor spins bring the $L_1$ point closer to the donor. This behavior aligns with expectations, as the centrifugal acceleration (captured by $\mathcal{A}(f_{\rm don},e,\mathcal{E})$) increases with $f_{\rm don}$, making it easier for surface material to escape. The High-$q$ model is independent of $q$ and breaks down for subsynchronous donors, incorrectly placing $L_1$ behind the accretor. The Low-$f_{\rm don}$ is independent of $f_{\rm don}$ and becomes increasingly inaccurate as $f_{\rm don}$ increases.  This model effectively represents the limiting case of the Global-$L_1$ model at $f_{\rm don} = 0$.

\begin{figure}[!htbp]
    \centering
    \includegraphics[width=\linewidth]{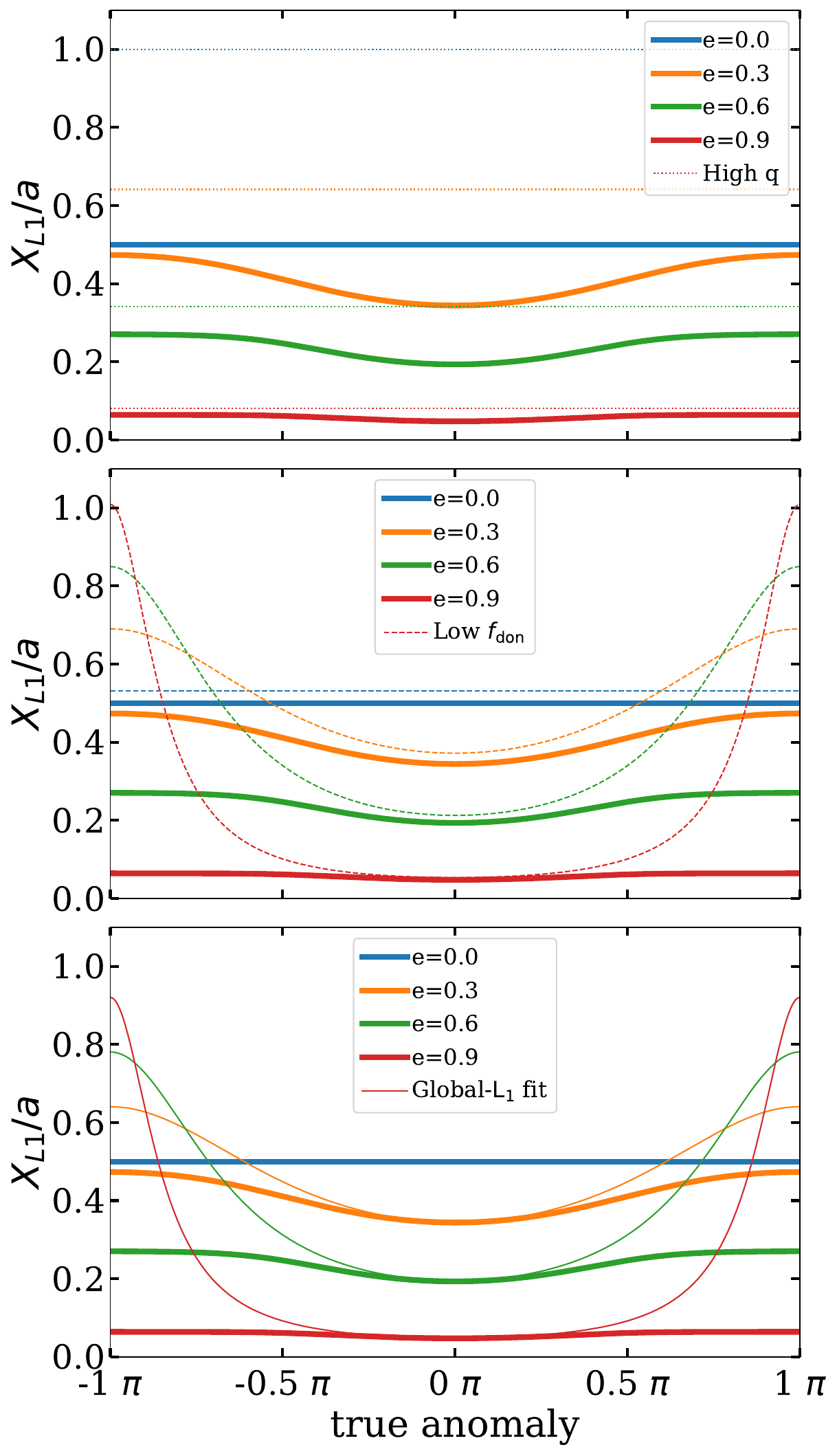}
    \caption{Position of the $L_1$ point, relative to the donor's center of mass, in units of the semimajor axis $a$, for $q=1$ and $f_{\rm don}=1$ as a function of true anomaly. The thick lines are the numerical solutions of Eq.~\eqref{eq:Lagrangian_points}. From top to bottom, the dotted, dashed and thin lines illustrate the High-$q$, Low-$f_{\rm don}$, and Global-$L_1$ models, respectively. Blue, orange, green and red colors correspond to $e=0.0,0.3,0.6,0.9$, respectively.}
    \label{fig:position_L1_during_one_orbit}
\end{figure}

In Figure~\ref{fig:position_L1_during_one_orbit}, we compare our Global-$L_1$ model to the High-$q$ and Low-$f_{\rm don}$ models by tracking the predicted position of the $L_1$ point over one orbital cycle for different eccentricities. In the High-$q$ model, the location of $L_1$ is independent of the orbital phase, and the approximation breaks down for circular orbits with synchronous donors, incorrectly placing $L_1$ in the center of the accretor, and for $f_{\rm don} < 1$, even behind it. The Low-$f_{\rm don}$ model is more accurate. In this case, the location of $L_1$ itself varies during one orbit, and it is always between the binary components, but no information about the donor's rotation is included. For circular orbits, the Global-$L_1$ model provides an accurate position of the $L_1$ point (note that the dashed blue line completely overlaps with the thick blue line), while both Low-$f_{\rm don}$ and High-$q$ models show visible offsets. Additionally, for any $e>0$ our prescription is less accurate during orbital phases when the MT rate is low, while it becomes more accurate close to the periapsis of the orbit where the MT rate is expected to be maximum. A detailed discussion is presented in Appendix~\ref{app:analytical_prescription_L1}.

\subsection{Orbital-element variations during mass transfer}\label{subsec:orbital_evolution_equations}

A perturbation induced on a binary system can give rise to changes in the orbit's Keplerian elements. We highlight that Eq.~\eqref{eq:total_perturbation_non_conservative_MT_simplified} does not contain any components that are out of the orbital plane. A perturbing acceleration of this form does not change the inclination of the orbit $i$, nor the longitude of the ascending node $\Omega$. However, the semimajor axis, $a$, the eccentricity, $e$, and the argument of periapsis, $\omega$, will evolve.

Following \cite{1963Icar....2..440H,2016ApJ...825...70D}, we calculate the evolution of the orbital elements as
\begin{flalign}
    \frac{\dot{a}}{a} &= \frac{2}{n^2a^2} \vec{\dot{r}} \cdot \vec{f}, \label{eq:semimajor_axis_derivative} \\
    \vec{\dot{e}} &= \frac{1}{n^2a^3} \Biggl(2 \vec{r}(\vec{\dot{r}} \cdot \vec{f}) -\vec{f}(\vec{r} \cdot \vec{\dot{r}})-\vec{\dot{r}}(\vec{r}  \cdot \vec{f})\Biggr)\label{eq:eccentricity_derivative}, \\
    \vec{\dot{\omega}} &= \frac{\hat{\vec{p}} \cdot \vec{\dot{e}}}{e},\label{eq:argument_periapsis_derivative}
\end{flalign}
where $\vec{e} = n^{-2} a^{-3} [\vec{r} (\vec{\dot{r}} \cdot \vec{\dot{r}}) - \vec{\dot{r}}(\vec{r} \cdot \vec{\dot{r}})] - \hat{\vec{r}}$ and $\hat{\vec{p}} \equiv \hat{\vec{h}} \times \hat{\vec{e}}$.

In principle, the derivation of Eqs.~\eqref{eq:semimajor_axis_derivative} to \eqref{eq:argument_periapsis_derivative} assumes that the total mass of the binary is constant. However, the effects of both mass loss and AM loss are explicitly incorporated into the perturbing acceleration $f_{\rm total}$. Specifically, consider a binary system with a constant total mass and assume that an external force per unit mass, as described by Eq.~\eqref{eq:total_perturbation_non_conservative_MT_simplified}, acts on the system. Under these conditions, the osculating orbit would undergo the same changes as it does when the total mass varies \citep{1963Icar....2..440H,2016ApJ...825...70D}.

In Appendix \ref{app:angular_momentum_loss}, we show that using Eqs.~\eqref{eq:semimajor_axis_derivative}, \eqref{eq:eccentricity_derivative}, and \eqref{eq:total_perturbation_non_conservative_MT_simplified}, we recover
\begin{equation}\label{eq:change_of_angular_momentum_total}
    \frac{\dot{J}_{\rm orb}}{J_{\rm orb}} = \gamma (1-\beta)\frac{\dot{M}_{\rm don}}{M_{\rm don} + M_{\rm acc}} -\biggl( \frac{r_{\rm don}}{r} \pm \beta q \frac{r_{\rm acc}}{r} \biggr)\frac{\dot{M}_{\rm don}}{M_{\rm don}}.
\end{equation}

From Eqs.~\eqref{eq:change_of_angular_momentum_total_1} and \eqref{eq:change_of_angular_momentum_total}, we see that
\begin{equation}
    \frac{ \dot{J}_{\rm orb,rf}}{J_{\rm orb}} = -\biggl( \frac{r_{\rm don}}{r} \pm \beta q \frac{r_{\rm acc}}{r} \biggr)\frac{\dot{M}_{\rm don}}{M_{\rm don}},
\end{equation}
and in the limit of point masses (i.e., $\vec{r}_{\rm acc} = \vec{r}_{\rm don}=0$), $\dot{J}_{\rm orb,rf}=0$, thus Eq.~\eqref{eq:change_of_angular_momentum_total} reduces to Eq.\eqref{eq:angular_momentum_parametrization} as expected.

We emphasize that for isotropic ejection and accretion (i.e., $\vec{r}_{\rm acc} = \vec{r}_{\rm don}=0$), $J_{\rm orb}$ is an integral of motion both in fully conservative MT ($\beta =1$) and in the non-conservative case if the ejected mass carries no net AM ($\gamma = 0$). However, the orbital evolution differs because, when $\beta < 1$, the total system mass varies, even though the orbital AM does not. Essentially, $\gamma = 0$ represents an idealized scenario in which the total mass of the system varies in such a way that it does not carry away net AM; such an example would be an isotropic wind originating from the system's center of mass. In this set-up, the orbital AM can evolve only if mass is lost from the system\footnote{Perturbations arising from other physical processes, such as gravitational waves, can remove orbital AM without mass loss from the system. Such perturbations are beyond the scope of this paper.}. However, mass can be lost from the system without removing AM ($\gamma = 0$).  Finally, the derivation of Eq.~\eqref{eq:change_of_angular_momentum_total} is independent of the prescription of the mass-loss rate.

\subsection{Phase-dependent mass transfer rate}\label{subsec:mass_transfer_model}

In a circular binary with synchronously rotating stars, the Roche lobe radius is phase-independent, and it is given in good approximation by the fit of \cite{1983ApJ...268..368E},
\begin{equation}\label{eq:roche_lobe_circ}
    \frac{R_{\rm L}^c}{a} = \frac{0.49 q^{2/3}}{0.6 q ^{2/3} + ln(1+q ^{1/3})}.
\end{equation}
If the physical radius of the donor overflows the Roche lobe radius (i.e., $\Delta R = R_{\rm don} - R_{\rm L}^c \geq 0$), then MT occurs. If $\Delta R < 0$, no mass is transferred. The MT rate is extremely sensitive to the extent to which the physical radius overflows the Roche lobe radius.

In this work, we adopt the phase-dependent MT model developed by \cite{2019ApJ...872..119H}. In this model, the MT rate is well-defined at high and low eccentricities, including circular orbits, where MT occurs continuously throughout the orbit. Below, we summarize the key aspects of the model relevant to interpreting our results; a more detailed description can be found in \cite{2019ApJ...872..119H}.

The model is based on the magnitude of $\Delta R$. By assuming a polytropic equation of state for the donor (with a polytrope index $n_p = 1.5$) and applying Bernoulli’s equation \citep{1972AcA....22...73P,1987MNRAS.229..383E}, the MT rate is given by
\begin{equation}\label{eq:instant_mass_transfer_rate}
\dot{M}_{\rm don} =\dot{M}_{\rm don,0} \Biggl(\frac{R_{\rm don} - R_{\rm L}(t)}{R_{\rm don}}\Biggr)^3,
\end{equation}
where $\dot{M}_{\rm don,0}$ represents a phase-independent MT rate and 
\begin{equation}
R_{\rm L}(t) = \frac{R_{\rm L}^c}{a} r(t)
\end{equation}
is an approximation for the instantaneous Roche lobe radius.

Using $r(\mathcal{E}) = a (1-e \cos{\mathcal{E}})$, the MT rate is written as
\begin{equation}\label{eq:mass_transfer_rate}
\dot{M}_{\rm don} = \dot{M}_{\rm don,0}[1-x(1-e\cos{\mathcal{E}})]^3, \; \text{where} \; x\equiv \frac{R_{\rm L}^c}{{R_{\rm don}}}.
\end{equation}
As a result, the MT rate is expected to be maximum at periapsis and minimum at apoapsis, where the instantaneous Roche lobe radius is minimum and maximum, respectively.

\begin{figure}[!htbp]
    \centering
    \includegraphics[width=\linewidth]{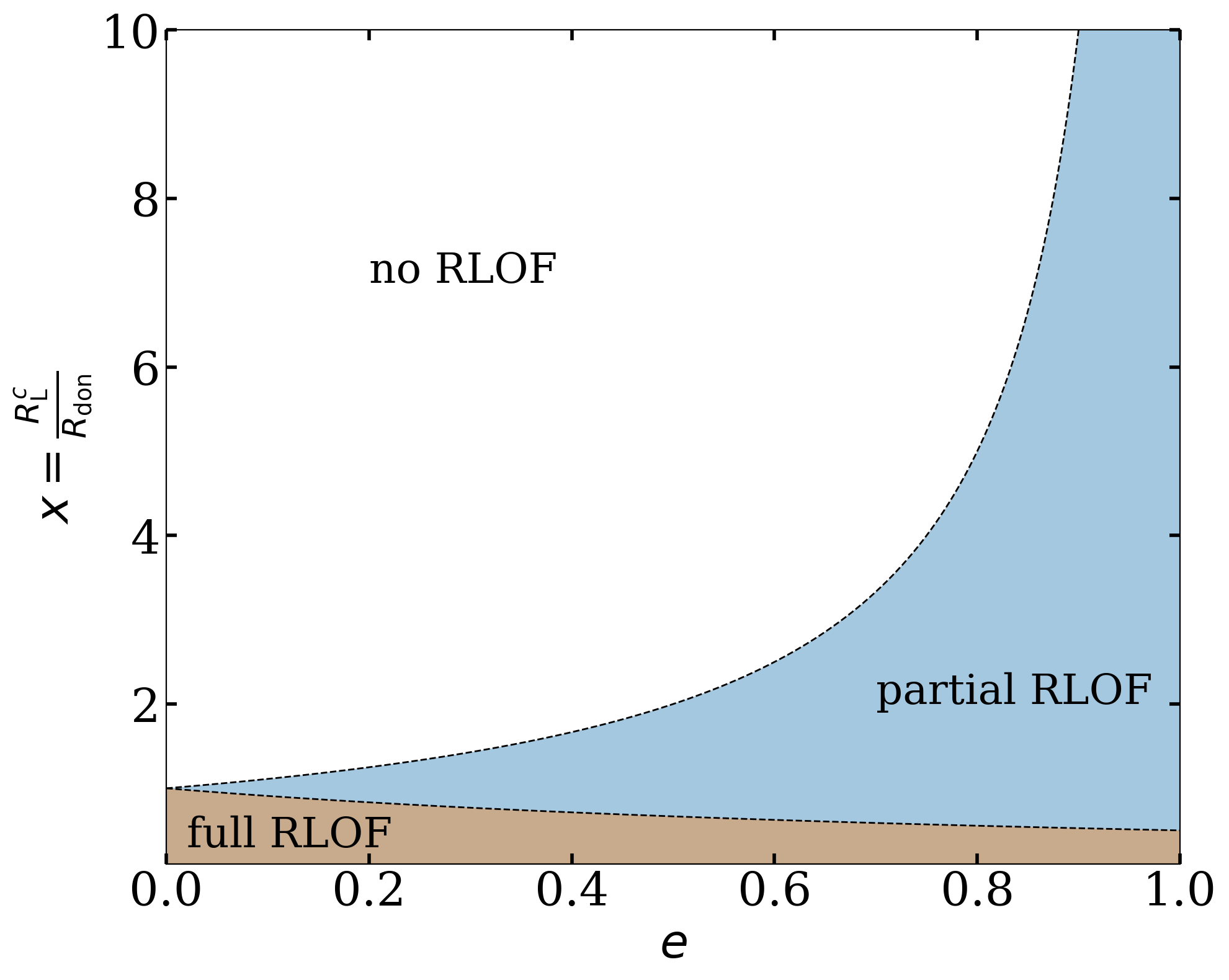}
    \caption{Graphical representation of the mass transfer regimes based on the mass transfer rate formulation by \cite{2019ApJ...872..119H}. The white region indicates no mass transfer, i.e. no RLOF. The light blue region corresponds to partial RLOF, where mass transfer occurs during part of the orbit. The light brown region represents full RLOF, with continuous mass transfer throughout the entire orbit.}
    \label{fig:RLOF_plane}
\end{figure}

During one orbit, for a given $e$ and $x$, a system can either not transfer mass at all (`no RLOF'), transfer during the whole orbit (`full RLOF') or transfer during part of the orbit (`partial RLOF'). Figure~\ref{fig:RLOF_plane} visually depicts the three scenarios on the $e-x$ plane. Additionally, for circular orbits, there is only full RLOF or no RLOF, which correspond to $x<1$ or $x \geq 1$, respectively. This is to be expected, since--for a circular orbit--$x$ is an alternative way to measure how much or if the physical radius $R_{\rm don}$ overflows the Roche lobe radius $R_{\rm L}^c$. 

The part of the orbit in which MT takes place is given by $-\mathcal{E}_0< \mathcal{E}< \mathcal{E}_0$. For MT occurring over the entire orbit (full RLOF), $\mathcal{E}_0 = \pi$. For MT limited to a portion of the orbit (partial RLOF), $\mathcal{E}_0$ is given by
\begin{equation}\label{eq:calc_E0}
\cos\mathcal{E}_0 = \frac{1}{e} (1- \frac{1}{x}).
\end{equation}
We note here that $\mathcal{E}_0$ essentially defines the limits of the integration when we calculate the orbit-averaged MT rate (see below Sect.~\ref{sec:four}), because the MT rate is assumed to be zero for $\mathcal{E} < -\mathcal{E}_0$ and $\mathcal{E} > \mathcal{E}_0$. 

Assuming $\dot{M}_{\rm don,0}, e$ and $x$ are constant over one orbit, $\ddot{M}_{\rm don}$ becomes
\begin{equation}\label{eq:second derivative_Mdot}
\ddot{M}_{\rm don} = -3nxe\dot{M}_{\rm don,0}[1-x(1-e \cos{\mathcal{E}})]^2 \frac{\sin\mathcal{E}}{1-e\cos\mathcal{E}}.
\end{equation}

\section{Orbit-averaged equations}\label{sec:four}

Our aim is to retrieve the secular evolution of the orbital elements, thus we remove periodic terms by averaging over one orbit. We define orbit-averaged quantities in the following way,
\begin{equation}\label{eq:orbit_average}
    \langle (...) \rangle = \frac{1}{2 \pi}\int_{-\pi}^{\pi} (...) (\frac{r}{a}) d \mathcal{E},
\end{equation}
where $(...)$ denotes the quantity to be averaged.

The perturbing acceleration of Eq.~\eqref{eq:total_perturbation_non_conservative_MT_simplified}, which is responsible for the variation of the orbital elements, is inversely proportional to the MT timescale $M_{\rm don}/\dot{M}_{\rm don} = \tau_{\dot{M}_{\rm don}}$. When $P_{\rm orb} \ll \tau_{\dot{M}}$, we can assume that systematic parameters, such as $M_{\rm acc},M_{\rm don},a,e,x$ etc., are approximately constant over one orbit (i.e., adiabatic approximation).
We substitute Eq.~\eqref{eq:mass_transfer_rate} into Eq.~\eqref{eq:total_perturbation_non_conservative_MT_simplified} and from Eqs.~\eqref{eq:orbit_average}, \eqref{eq:semimajor_axis_derivative} to \eqref{eq:argument_periapsis_derivative} we derive the orbit-averaged equations of motion in the adiabatic regime as
\begin{flalign}
 \frac{\langle \dot{a} \rangle}{a} &= -\frac{2  \dot{M}_{\rm don,0}}{M_{\rm don}} \Biggl[ \Biggl(1-\beta q-(1-\beta)\frac{(\gamma+\frac{1}{2})q}{1+q}\Biggr)f_{a}(e,x) \nonumber\\
 &+ X_{\rm L1}(f_{\rm don},q,e) g_{a}(e,x) \pm \beta q \frac{r_{\rm acc}}{a} h_{a}(e,x) \Biggr],\label{eq:orbit_averaged_semimajor_axis_0}\\
 \langle \dot{e} \rangle &= -\frac{2 \dot{M}_{\rm don,0}}{M_{\rm don}} \Biggl[ \Biggl(1-\beta q-(1-\beta)\frac{(\gamma+\frac{1}{2})q}{1+q}\Biggr)f_{e}(e,x) \nonumber\\
 &+ X_{\rm L1}(f_{\rm don},q,e) g_{e}(e,x) \pm \beta q \frac{r_{\rm acc}}{a} h_{e}(e,x) \Biggr],\label{eq:orbit_averaged_eccentricity_0}\\
 \langle \dot{\omega} \rangle &= 0, \label{eq:orbit_averaged_argument_of_periapsis_0}
\end{flalign}
where $f_{\dot{M}_{\rm don}}(e,x), \; f_{a}(e,x), \; f_{e}(e,x), \; g_{a}(e,x), \; g_{e}(e,x), \; h_{a}(e,x),$ and $ \; h_{e}(e,x)$ are dimensionless functions given explicitly in Appendix~\ref{app:dimensionless_functions}. The negative sign in front of the term associated with the accretion point corresponds to $\vec{r}_{\rm acc} = - r_{\rm acc} \hat{\vec{r}}$, while the positive sign to $\vec{r}_{\rm acc} = r_{\rm acc} \hat{\vec{r}}$. The dimensionless functions are equivalent to those derived by \cite{2019ApJ...872..119H}. It is important to note that there is a typo in \citep[][eq. 52 in their Appendix B]{2019ApJ...872..119H}. As a result, we recommend the verification of the $h_{e}(e,x)$ integral before using the emt-model. Furthermore, \cite{2021MNRAS.502.4479H} present an ad hoc extension of the emt-model to non-conservative MT. However, the effects of mass loss and AM loss are not taken into account when computing the evolution of the orbital elements \citep[see][]{2025arXiv251107190P}.

As described in Sect.~\ref{subsec:mass_transfer_model}, the MT rate contains a periodic term and a phase-independent MT rate (Eq.~\ref{eq:mass_transfer_rate}). The periodic term is a consequence of the distance between the stars varying over one orbit (see Eq.~\ref{eq:mass_transfer_rate}). When applying Eqs. \eqref{eq:orbit_averaged_semimajor_axis_0} to  \eqref{eq:orbit_averaged_argument_of_periapsis_0} within detailed or rapid stellar evolution codes, one may not have direct access to the normalization parameter $\dot{M}_{\rm don,0}$. For this reason, we re-express the equations in terms of the orbit-averaged MT rate $\langle \dot{M}_{\rm don} \rangle$ (in the adiabatic regime). The orbit-averaged MT rate is defined via  Eqs.~\eqref{eq:mass_transfer_rate} and \eqref{eq:orbit_average}, as  
\begin{flalign}
    \langle \dot{M}_{\rm don} \rangle &\equiv \frac{1}{2 \pi}\int_{-\pi}^{\pi} \dot{M}_{\rm don,0} [1-x(1-e\cos{\mathcal{E}})]^3 (\frac{r}{a}) d \mathcal{E} \nonumber\\
    &= \dot{M}_{\rm don,0} f_{\dot{M}_{\rm don}}(e,x),\label{eq:mass_loss_rate_normalized} 
\end{flalign}
where $f_{\dot{M}_{\rm don}}(e,x)$ is a dimensionless function acting as a normalization factor. We note that the limits of integration are effectively determined by $\mathcal{E}_0$ via Eq.~\eqref{eq:calc_E0}, since the MT rate is assumed to vanish outside the range $-\mathcal{E}_0< \mathcal{E}< \mathcal{E}_0$. Consequently, the secular evolution equations of motion are given as

\begin{flalign}
 \frac{\langle \dot{a} \rangle}{a} &= -\frac{2  \langle \dot{M}_{\rm don} \rangle}{M_{\rm don}} \frac{1}{f_{\dot{M}_{\rm don}}(e,x)} \Biggl[ \Biggl(1-\beta q-(1-\beta)\frac{(\gamma+\frac{1}{2})q}{1+q}\Biggr)f_{a}(e,x) \nonumber\\
 &+ X_{\rm L1}(f_{\rm don},q,e) g_{a}(e,x) \pm \beta q \frac{r_{\rm acc}}{a} h_{a}(e,x) \Biggr],\label{eq:orbit_averaged_semimajor_axis}\\
 \langle \dot{e} \rangle &= -\frac{2 \langle \dot{M}_{\rm don} \rangle}{M_{\rm don}} \frac{1}{f_{\dot{M}_{\rm don}}(e,x)} \Biggl[ \Biggl(1-\beta q-(1-\beta)\frac{(\gamma+\frac{1}{2})q}{1+q}\Biggr)f_{e}(e,x) \nonumber\\
 &+ X_{\rm L1}(f_{\rm don},q,e) g_{e}(e,x) \pm \beta q \frac{r_{\rm acc}}{a} h_{e}(e,x) \Biggr],\label{eq:orbit_averaged_eccentricity}\\
 \langle \dot{\omega} \rangle &= 0. \label{eq:orbit_averaged_argument_of_periapsis}
\end{flalign}
Furthermore, $\langle \dot{M}_{\rm don} \rangle$ is assumed to be known and constant throughout a single orbit, serving as a free parameter within the model. It is important to note that $\langle \dot{M}_{\rm don} \rangle$ can change over long timescales due to the donor's response to MT or as a result of stellar evolution. Therefore, $\langle \dot{M}_{\rm don} \rangle$ should ideally be calculated self-consistently \citep[e.g.,][]{2025ApJ...983...39R}. Nevertheless, the qualitative behavior of Eqs.~\eqref{eq:orbit_averaged_semimajor_axis} to \eqref{eq:orbit_averaged_argument_of_periapsis} is independent of the exact choice of $\langle \dot{M}_{\rm don} \rangle$.
 
We implement the secular equations of motion (Eqs.~\ref{eq:orbit_averaged_semimajor_axis} to \ref{eq:orbit_averaged_argument_of_periapsis}) into a code named General Mass Transfer ({\sc GeMT}). To ensure that our equations are only applied in the parts of the parameter space in which MT occurs, we introduce three stopping conditions. Specifically, we do not evolve the orbital elements if (1) a system detaches (i.e., is located in the no RLOF regime in Fig.~\ref{fig:RLOF_plane}; mathematically ${R_{\rm L}^c}(1-e) > R_{\rm don}$), (2) the radius of the donor is equal to or larger than the periastron distance (i.e., the system merges; mathematically $R_{\rm don} \geq  a (1-e$)), and (3) the orbit becomes parabolic (mathematically $e > 0.99$). In the following section, we examine the behavior of the GeMT-model in various limiting cases and compare its predictions with those of earlier models.

\section{Properties of the orbit-averaged equations}\label{sec:five}

In this section, we examine the secular rates of change for the semimajor axis $a$ (Eq.~\ref{eq:orbit_averaged_semimajor_axis}) and eccentricity $e$ (Eq.~\ref{eq:orbit_averaged_eccentricity}) under various limiting conditions. Section~\ref{subsec:point_masses_circ_orbits} demonstrates that the GeMT-model reproduces the classical RLOF results in the limit of circular orbits and point masses. Sections~\ref{subsec:extended_bodies_circ_orbits} and \ref{subsec:extended_bodies_ecc_orbits} present the predicted orbital evolution for circular and eccentric orbits, respectively. Moreover, we compare the predictions of the GeMT-model with the results derived in the limit of point masses, as well as with the $\delta$-function and emt models. Hereafter, when referring to the case of `extended bodies'(i.e., $r_{\rm don}, r_{\rm acc} \neq 0$), we implicitly include the impact of anisotropic mass ejection from the donor and accretion onto the companion on the secular evolution. Considering the emt-model, we adopt the default Low-$f_{\rm don}$ prescription throughout this work. Moreover, we adopt $\cos(\phi_{P}) = -1$ for the $\delta$-function throughout this work. In Table~\ref{tab:colormaps_parameters} we summarize the initial conditions at the onset of RLOF for the different examples presented.

\begingroup
\begin{table*}[!htbp]
\caption{ Initial conditions at the onset of RLOF.}
  \centering
  \begin{tabular}{ccccccccc}
  \hline
  Fig. & $\beta$ & $\langle \dot{M}_{\rm don} \rangle$ & $M_{\rm don}$ & $a$ & $e$ & x & $f_{\rm don}$ & $r_{\rm acc}$\\
  &  & (M$_{\odot} \; \rm yr^{-1}$) & (M$_{\odot}$) & (au) & & & & ($R_{\odot}$) \\
  \hline \hline
  \ref{fig:colormap_semimajor_axis_circ} & 1.0 & $ 10^{-8}$ & 1.1 &  1.0 & 0.0 &–& $[0.0,2.0]$ &  0.62  \\
  \ref{fig:colormap_semimajor_axis_ecc} & 1.0 & $ 10^{-8}$ & 1.1 &  1.0 & $[0.0,0.99]$ & 0.95 & 1.0 &  0.62  \\
  \ref{fig:colormap_eccentricity_ecc} & 1.0 & $ 10^{-8}$ & 1.1 &  1.0 & $[0.01,0.99]$ &  0.95 & 1.0 &  0.62  \\
  \hline
  \end{tabular}\label{tab:colormaps_parameters}
  \tablefoot{The $\langle \dot{M}_{\rm don} \rangle $ term is not applicable to the $\delta$-function model, instead an instantenaneous MT rate $\dot{M}_{0} = 10^{-8}$ M$_{\odot} \; \rm yr^{-1}$ is used. The spin of the ejection point $f_{\rm don}$ is only applicable to the $\delta$-function and GeMT models in the general case of extended bodies. The level at which the donor overflows the Roche-lobe $x$ is only applicable to the emt and GeMT models.}
\end{table*}

\endgroup

\subsection{Point masses and circular orbits}\label{subsec:point_masses_circ_orbits}

In the classical picture of RLOF, the orbit is circular, and the binary components are modeled by point masses \citep[see][]{2014LRR....17....3P}. Following \cite {1997A&A...327..620S} and using the orbital AM, $J_{\rm orb} = \mu \sqrt{GMa(1-e^2)}$, the change of the semimajor axis of the orbit is given by
\begin{equation}\label{eq:semimajor_axis_non_conservative_points}
    \frac{\dot{a}}{a} = -2 \frac{\dot{M}_{\rm don}}{M_{\rm don}} \Biggl(1-\beta \frac{M_{\rm don}}{M_{\rm acc}} - (1-\beta)(\gamma + \frac{1}{2}) \frac{M_{\rm don}}{M_{\rm acc}+M_{\rm don}}\Biggr).
\end{equation}

In the limit of point masses (i.e., $\vec{r}_{\rm don} = \vec{r}_{\rm acc} =0$), the orbit-averaged equations of motion (Eqs.~\ref{eq:orbit_averaged_semimajor_axis} and \ref{eq:orbit_averaged_eccentricity}) reduce to
\begin{flalign}
     \frac{\langle \dot{a} \rangle}{a} &= -\frac{2 \langle \dot{M}_{\rm don} \rangle}{M_{\rm don}} \frac{f_{a}(e,x)}{f_{\dot{M}_{\rm don}}(e,x)} \Biggl[ 1-\beta q-(1-\beta)\frac{(\gamma+\frac{1}{2})q}{1+q} \Biggr], \label{eq:orbit_averaged_semimajor_axis_MT_points}\\
     \langle \dot{e} \rangle &= -\frac{2 \langle \dot{M}_{\rm don} \rangle}{M_{\rm don}} \frac{f_{e}(e,x)}{f_{\dot{M}_{\rm don}}(e,x)} \Biggl[ 1-\beta q-(1-\beta)\frac{(\gamma+\frac{1}{2})q}{1+q} \Biggr].\label{eq:orbit_averaged_eccentricity_MT_points}
\end{flalign}
Furthermore, if the orbit is circular (see Appendix~\ref{app:limits}) the resulting secular rates of change simplify to
\begin{flalign}
     \frac{\langle \dot{a} \rangle}{a} &= -\frac{2 \langle \dot{M}_{\rm don} \rangle }{M_{\rm don}} \Biggl[ 1-\beta q-(1-\beta)\frac{(\gamma+\frac{1}{2})q}{1+q} \Biggr], \label{eq:orbit_averaged_semimajor_axis_points_circ}\\
     \langle \dot{e} \rangle &= 0.\label{eq:orbit_averaged_eccentricity_points_circ}
\end{flalign}
Consequently, in the limit of point masses and circular orbits, the GeMT-model reproduces the canonical relation for the rate of change of the semimajor axis, given by Eq.~\eqref{eq:semimajor_axis_non_conservative_points}. The latter is widely used in studies of non-conservative MT in circular orbits \citep[e.g.][]{1997A&A...327..620S,2014LRR....17....3P}.  Notably, in the limit of $\mathcal{E}_{0} \rightarrow \pi$, $\langle \dot{e} \rangle \propto e$, indicating that an initially circular orbit, that is mathematically $e=0$, will remain circular. On the other hand, if the system has some seed eccentricity (i.e., $e\neq 0$), we expect $\langle \dot{e} \rangle \neq 0$. 

For $\beta =1$, Eq.~\eqref{eq:orbit_averaged_semimajor_axis_points_circ} reduces to,
\begin{equation}\label{eq:semimajor_axis_conservative_points}
    \frac{\langle \dot{a} \rangle}{a} = -2 \frac{\langle \dot{M}_{\rm don} \rangle}{M_{\rm don}} \Biggl(1- \frac{M_{\rm don}}{M_{\rm acc}}\Biggr),
\end{equation}
which is the canonical relation used in studies of conservative MT \citep[e.g.,][]{pringle1985interacting,2002MNRAS.329..897H,2018MNRAS.480.3195K}. According to Eq.~\eqref{eq:semimajor_axis_conservative_points}, the orbit expands when the donor is less massive than the accretor, and contracts otherwise. Consequently, the evolution of the semimajor axis is consistent with the conservation of $M_{\rm don}^{2} M_{\rm acc}^{2} a$.

We note that the $\delta$-function model for eccentric RLOF used in numerous studies \citep{2007ApJ...667.1170S,2009ApJ...702.1387S,2016ApJ...825...70D,2016ApJ...825...71D,2025ApJ...983...39R} is invalid in this regime. Specifically, at $e=0.0$, the eccentricity derivative is non-zero, becoming negative when  $q>1$ and positive when $q<1$, as shown by \cite{2019ApJ...872..119H}.

\subsection{Circular orbits and the effect of the donor's spin velocity }\label{subsec:extended_bodies_circ_orbits}

For extended bodies $\vec{r}_{\rm don} = X_{\rm L1}(f_{\rm don},q,e) \vec{r}$ and we assume that $\vec{r}_{\rm acc} =\pm r_{\rm acc} \hat{\vec{r}}$. In the limit of circular orbits (see Appendix \ref{app:limits}), the GeMT-model simplifies to
\begin{flalign}
     \frac{\langle \dot{a} \rangle}{a} &= -\frac{2 \langle \dot{M}_{\rm don} \rangle }{M_{\rm don}} \Biggl[ \Biggl(1-\beta q-(1-\beta)\frac{(\gamma+\frac{1}{2})q}{1+q}\Biggr) \nonumber \\
     &+ X_{\rm L1}(f_{\rm don},q,e \rightarrow 0) \pm \beta q \frac{r_{\rm acc}}{a}  \Biggr],\label{eq:limit_orbit_averaged_semimajor_axis}\\
     \langle \dot{e} \rangle &= 0, \label{eq:limit_orbit_averaged_eccentricity} \\ \nonumber
\end{flalign}
where $X_{\rm L1}(f_{\rm don},q,e)$ is given explicitly in Appendix \ref{app:analytical_prescription_L1}. From Eq.~\eqref{eq:limit_orbit_averaged_semimajor_axis}, we observe that the rate of change of the semimajor axis is independent of $x$ (see also Fig.~\ref{fig:RLOF_plane}). Notably, this rate retains information about the spin angular velocity of the donor ($f_{\rm don}$) star. 

\begin{figure}[!htbp]
    \centering
    \includegraphics[width=\linewidth]{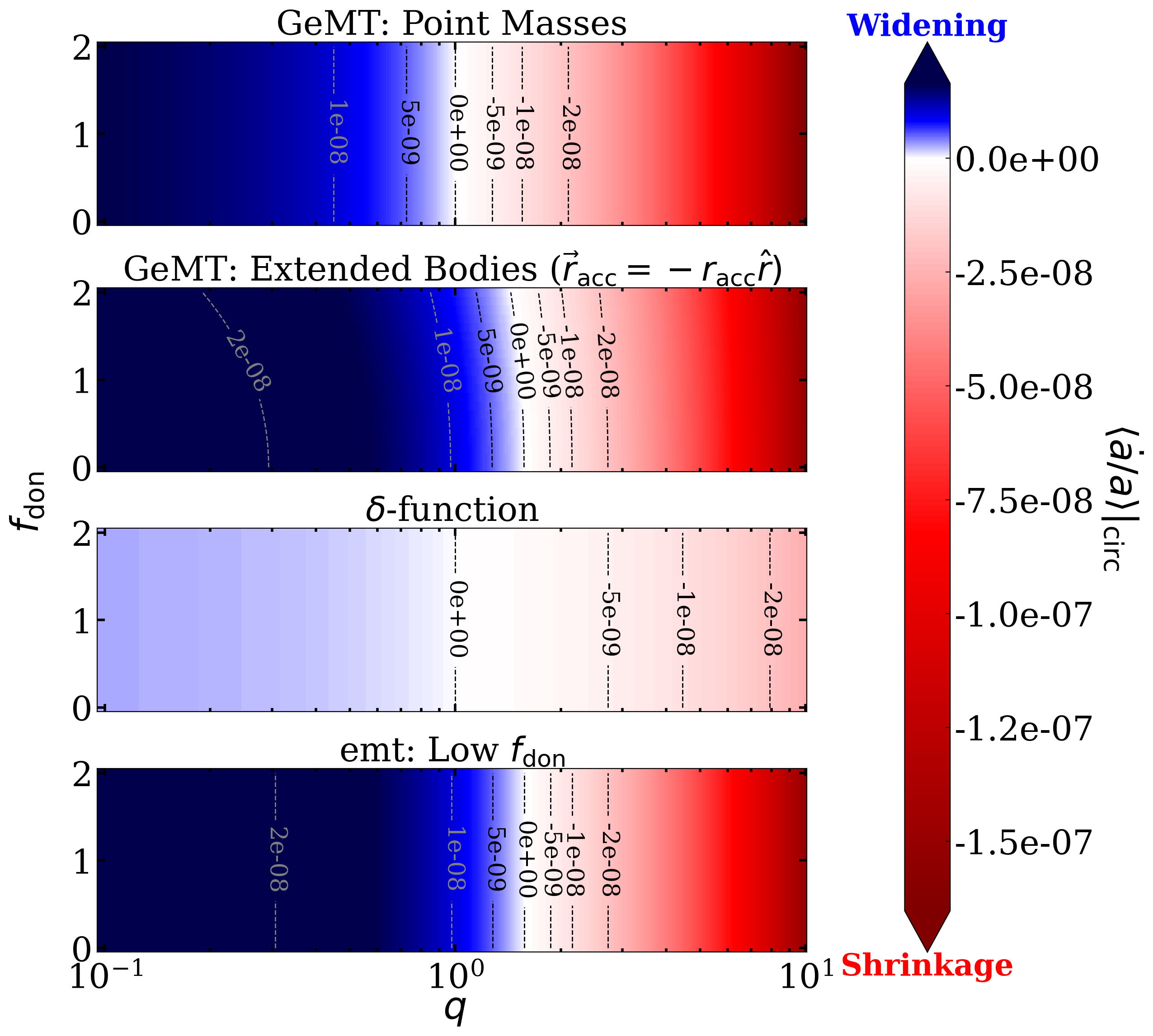}
    \caption{Secular rate of change of the semimajor axis as a function of mass ratio $q$, and the donor's level of synchronism $f_{\rm don}$, in the limit of circular orbits.  From top to bottom: the GeMT-model in the limit of point masses, extended bodies for $\vec{r}_{\rm acc} = - r_{\rm acc} \hat{\vec{r}}$, the $\delta$-function, and emt models. The values of the relevant parameters are provided in Table~\ref{tab:colormaps_parameters}.}
    \label{fig:colormap_semimajor_axis_circ}
\end{figure}

In Figure~\ref{fig:colormap_semimajor_axis_circ}, we show the rate of change of the semimajor axis (Eq.~\ref{eq:limit_orbit_averaged_semimajor_axis}) as function of mass ratio $q$ varying $f_{\rm don}$. Additionally, we compare these results with the rates predicted by the GeMT-model in the limit of point masses, as well as the $\delta$-function and the emt-model. Red regions indicate orbital shrinkage, while blue regions represent orbital widening, with the color intensity reflecting the magnitude of the rate: deeper red and blue correspond to stronger shrinkage and widening, respectively. 

In the point-mass and $\delta$-function models, the transitional mass ratio $q_{\rm trans,a}$ separating orbital widening and shrinkage occurs at equal masses ($q_{\rm trans,a}=1$) and is independent of donor's spin velocity. This result is expected since, in the $\delta$-function model, the terms associated with $\vec{r}_{\rm don}, \vec{r}_{\rm acc}$ are proportional to the eccentricity $e$ and vanish as $e\rightarrow 0$, whether extended bodies are considered or not. Consequently, the model simplifies to the canonical relation for the semimajor axis evolution in this regime, given by Eq.~\eqref{eq:semimajor_axis_conservative_points}. However, the color intensity indicates that the $\delta$-function model predicts weaker evolution of the semimajor axis across all mass ratios. The GeMT model for the point-mass case and $\delta$-function model would be exactly equivalent only if $\langle \dot{M}_{\rm don} \rangle$ in Eq.~\eqref{eq:semimajor_axis_conservative_points} is interpreted as $\langle \dot{M}_{\rm don} \rangle =  \dot{M}_{0}/(2 \pi)$ \citep[see also Sect. 5 in][]{2007ApJ...667.1170S}, which applies only to circular orbits (see Appendix \ref{app:delta_function}).

The GeMT-model takes into account the position of the $L_1$ (Global-$L_1$ model) point, and the orbital evolution deviates from the classical RLOF picture. Specifically, the parameter space for orbital widening as well as the intensity of the blue region increase. $q_{\rm trans,a}$ shifts to higher values, e.g. for synchronous donors ($f_{\rm don} = 1.0$), our model predicts $q_{\rm trans,a} \approx 1.53$. Furthermore, increasingly subsynchronous donors result in higher $q_{\rm trans,a}$ and vice versa; with $q_{\rm trans,a} \approx 1.57$ for $f_{\rm don}=0.0$ and $q_{\rm trans,a} \approx 1.44$ for $f_{\rm don}=2.0$, respectively. Finally, the bluer color shows a stronger orbital widening compared to the point-mass and $\delta$-function models. 

The emt-model also predicts $q_{\rm trans,a} > 1$, but unlike the GeMT-model, the rate of change of the semimajor axis is independent of the donor’s spin velocity. This is because the Low-$f_{\rm don}$ prescription used for the position of the $L_1$ point assumes $f_{\rm don} \approx 0$. In summary, for circular orbits, the semimajor axis in the emt-model is entirely independent of $f_{\rm don}$. Essentially, the emt-model is a subset of the GeMT-model under the specific condition of non-rotating donors ($f_{\rm don} = 0$).

\subsection{Eccentric orbits}\label{subsec:extended_bodies_ecc_orbits}

\begin{figure}[!htbp]
    \centering
    \includegraphics[width=\linewidth]{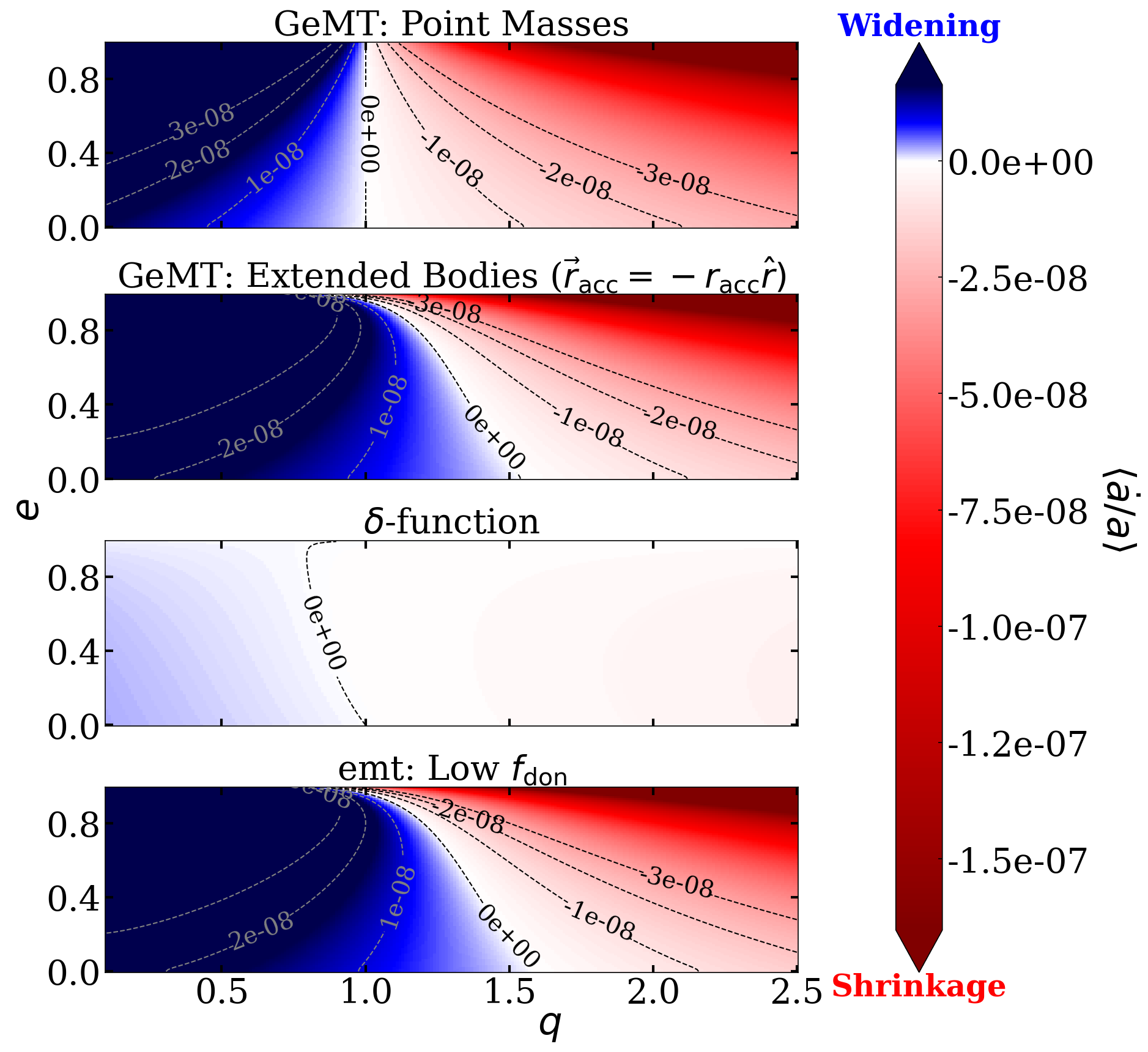}
    \caption{Secular rate of change of the semimajor axis in the limit of conservative mass transfer as a function of mass ratio $q$, and eccentricity, $e$. From top to bottom: the GeMT-model in the limit of point masses, extended bodies for $\vec{r}_{\rm acc} = - r_{\rm acc} \hat{\vec{r}}$, the $\delta$-function and emt models. The values of the relevant parameters are provided in Table~\ref{tab:colormaps_parameters}. }
    \label{fig:colormap_semimajor_axis_ecc}
\end{figure}

\begin{figure}[!htbp]
    \centering
    \includegraphics[width=\linewidth]{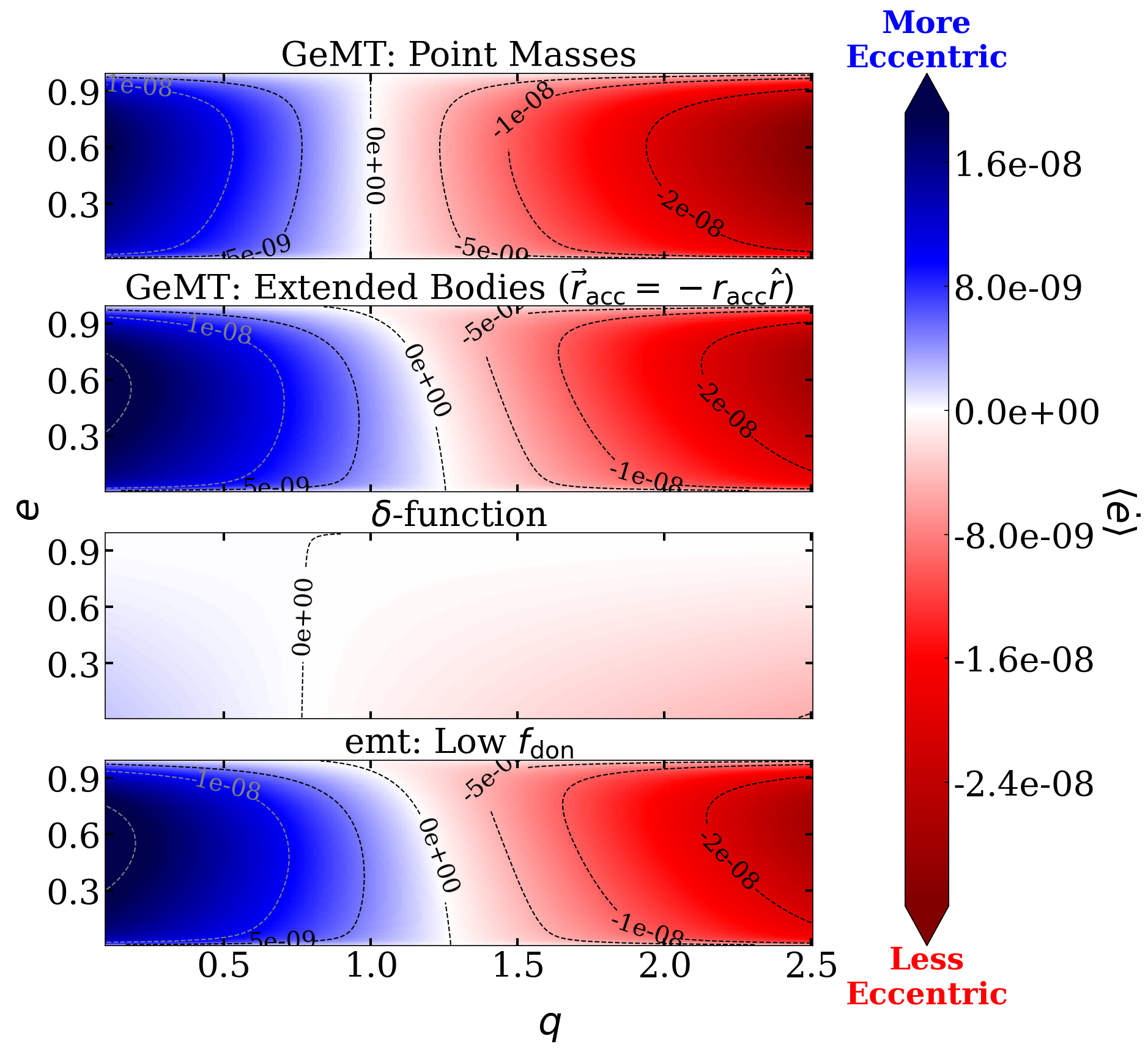}
    \caption{ Similar to Fig.~\ref{fig:colormap_semimajor_axis_ecc}, but now the color gradient illustrates the secular rate of change of the eccentricity.}
    \label{fig:colormap_eccentricity_ecc}
\end{figure}

For non-zero eccentricities, the secular rates of change of the semimajor axis $a$ and eccentricity $e$ are determined by Eqs.~\eqref{eq:orbit_averaged_semimajor_axis} and \eqref{eq:orbit_averaged_eccentricity}. In Figures~\ref{fig:colormap_semimajor_axis_ecc} and \ref{fig:colormap_eccentricity_ecc}, we show the aforementioned rates in the limit of conservative MT as functions of the mass ratio $q$ and the eccentricity $e$, respectively. Additionally, we compare these results with the rates predicted by the GeMT-model in the limit of point masses, as well as with those from the $\delta$-function and the emt models.  
For non-zero eccentricities, the terms associated with ejection and accretion points ($\vec{r}_{\rm don}, \vec{r}_{\rm acc}$) in the $\delta$-function model are non-zero. Consequently, we employ the prescription of the $L_1$ point $X_{\rm L1,sep}(f_{\rm don},q,e)$ as derived by \cite{2007ApJ...667.1170S} (Eq. A15 in their Appendix A). Specifically, $|\vec{r}_{A_{1},P}| = X_{\rm L1,sep}(f_{\rm don},q,e)a(1-e)$.

In the limit of point masses, the GeMT-model predicts that the orbit expands when $q < 1$ and shrinks when $q > 1$, independent of $e$ (Fig.~\ref{fig:colormap_semimajor_axis_ecc}). However, when accounting for extended bodies, the parameter space for orbital widening and shrinkage changes. Specifically, the GeMT-model predicts that for lower eccentricities, the transitional mass ratio, $q_{\rm trans,a}$, shifts to higher values and vice versa for higher eccentricities. The $\delta$-function model also shows a decrease in $q_{\rm trans,a}$ with increasing eccentricity for non-zero eccentricities. However, the predicted $q_{\rm trans,a}$ values are significantly lower. Additionally, the color gradient in the $\delta$-function model suggests a weaker evolution of the semimajor axis compared to the GeMT-model. Notably, in the limit of conservative MT ($\beta = 1$) and non-rotating donors ($f_{\rm don}=0.0$) the GeMT-model and the emt-model are equivalent.

The GeMT model predicts an eccentricity-evolution parameter space that closely resembles that of the semimajor axis (Figs.~\ref{fig:colormap_eccentricity_ecc} and \ref{fig:colormap_semimajor_axis_ecc}). In the limit of point masses, $\dot{e}$ is positive for $q<1$ and negative for $q>1$, independent of $e$. When extended bodies are considered, the parameter space for eccentricity pumping broadens. Specifically, the transitional mass ratio, $q_{\rm trans,e}$, shifts to higher values, a trend more prominent for lower eccentricities. In contrast, the $\delta$-function model shows that $q_{\rm trans,e}$ remains largely independent of $e$, with the eccentricity-pumping regime confined to $q \leq 0.74$. Additionally, the color gradient suggests a weaker evolution of the eccentricity compared to the GeMT-model across all mass ratios and eccentricities. Finally, for $\beta = 1$ and $f_{\rm don}=0.0$ the GeMT and the emt models are equivalent.

\section{Applications: conservative mass transfer}\label{sec:six}

In this section, we examine the orbital evolution of isolated binary systems undergoing conservative RLOF ($\beta = 1$). Using the GeMT code, we numerically integrate Eqs.~\eqref{eq:orbit_averaged_semimajor_axis} and \eqref{eq:orbit_averaged_eccentricity}. Here, we neglect additional physical processes such as tides or stellar evolution to isolate the effects of MT via RLOF. For simplicity, we treat the stars as rigid spheres and assume that both the donor's radius $R_{\rm don}$ and the accretor's radius (or the outer edge of the accreting disc) $r_{\rm acc}$ remain constant throughout the integration. 

\subsection{Comparison to different mass transfer frameworks}\label{subsec:eccentric_orbits}

\begin{figure}[!htbp]
    \centering
    \includegraphics[width=\linewidth]{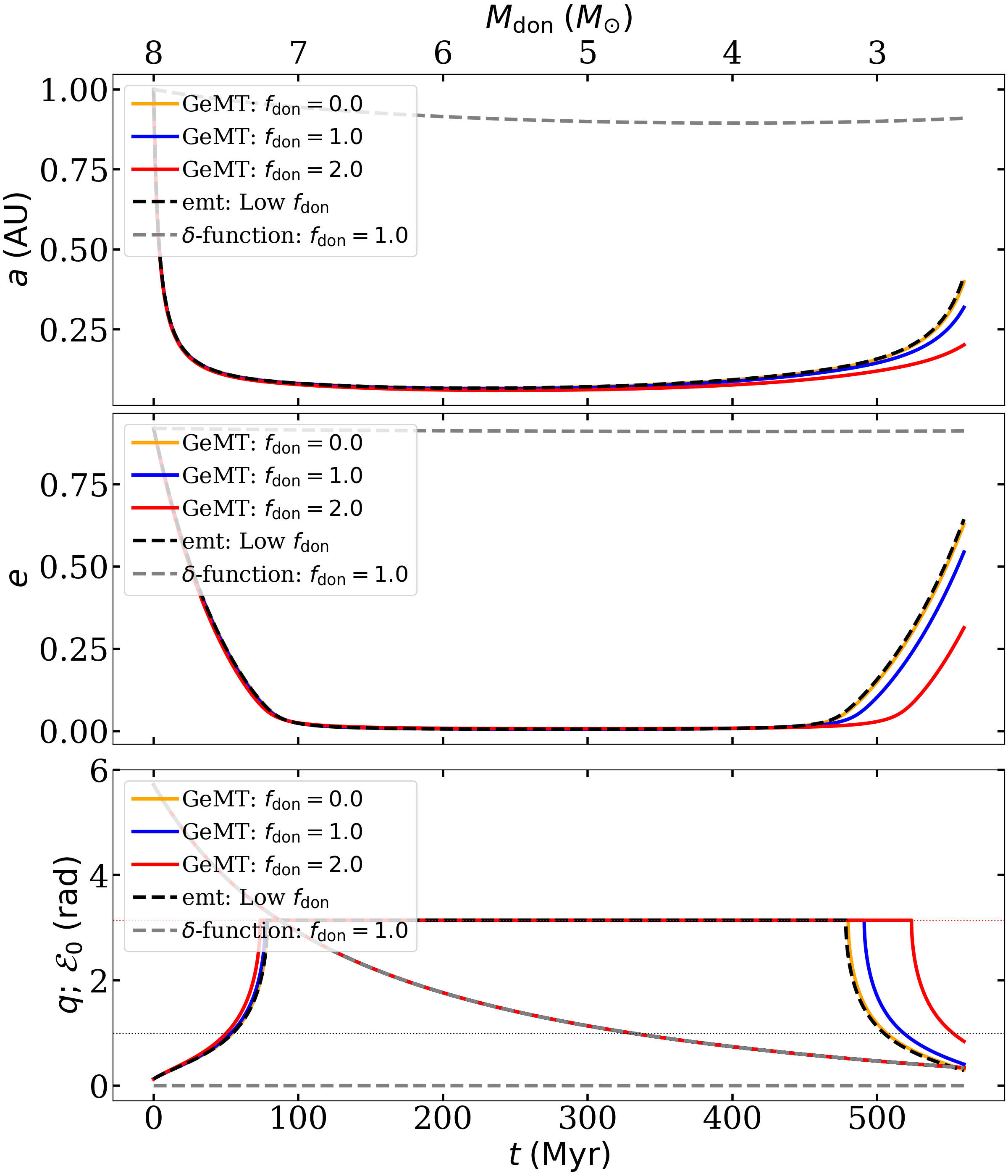}
    \caption{ Evolution of the semimajor axis (top), eccentricity (middle) and mass ratio $q$ and angle $\mathcal{E}$ (bottom) as a function of time during mass transfer. The dashed black and gray lines correspond to the emt and $\delta$-function ($f_{\rm don}=1.0$) models, respectively. The orange, blue and red lines correspond to the GeMT-model, for subsynchronous ($f_{\rm don}=0.0$), synchronous ($f_{\rm don}=1.0$) and supersynchronous donors ($f_{\rm don}=2.0$), respectively. In the bottom subfigure, the two horizontal dotted lines indicate $\mathcal{E}_0 = \pi$ and $q = 1$.}
    \label{fig:comparison_hamers_evolution}
\end{figure}

We consider a system with initial parameters: $M_{\rm don} = 8$ M$_\odot$, $M_{\rm acc} = 1.4$ M$_\odot$, $R_{\rm don}=10$ R$_\odot$, $r_{\rm acc}=0.01 $ R$_\odot$, $a = 1$ au and $e=0.92$, equal to the example system of \cite{2019ApJ...872..119H}. In this configuration, the accretor represents a neutron star. The donor initiates RLOF near periapsis at $x \approx 11.4$. We assume $\langle \dot{M}_{\rm don} \rangle = 10^{-8}$ M$_\odot$ yr$^{-1}$ for the emt and GeMT models, and $\dot{M}_{0} = 10^{-8}$ M$_\odot$ yr$^{-1}$ for the $\delta$-function model. For a direct comparison with previous models, we set $\vec{r}_{\rm acc}= - r_{\rm acc} \hat{\vec{r}}$ for the GeMT-model. The evolution of the system is presented in Fig.~\ref{fig:comparison_hamers_evolution}.

\begin{figure}[!htbp]
    \centering
    \includegraphics[width=\linewidth]{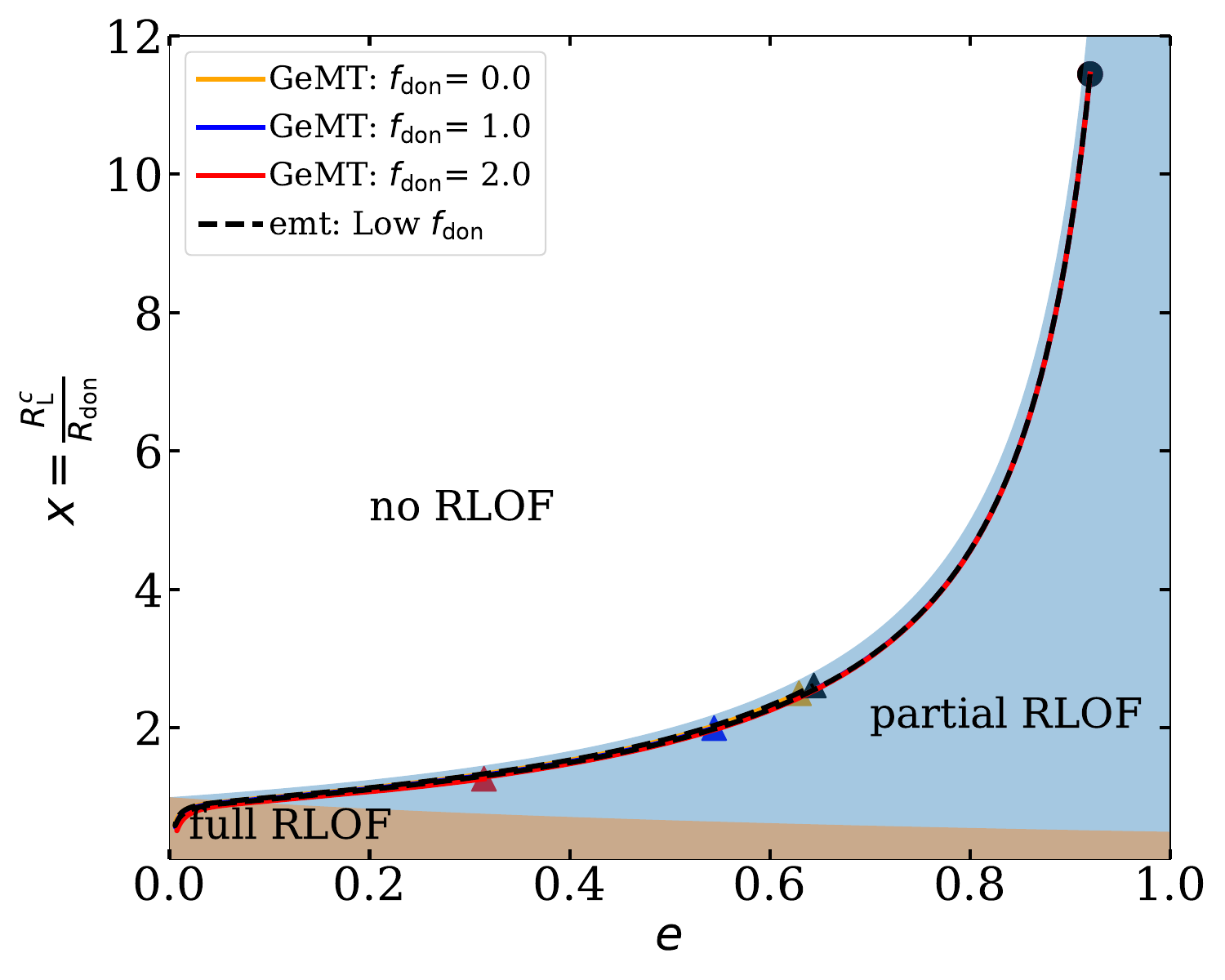}
    \caption{Evolution of the systems presented in Fig.\ref{fig:comparison_hamers_evolution} on the $e-x$ plane for subsynchronous ($f_{\rm don}=0.0$), synchronous ($f_{\rm don}=1.0$) and supersynchronous donors ($f_{\rm don}=2.0$). The circles and triangles indicate the initial and final positions of the systems, respectively.}
    \label{fig:comparison_classical_RLOF_ecc_plane}
\end{figure}

Initially, the system undergoes partial RLOF, during which both the semimajor axis and the eccentricity decrease across all models. In the emt and GeMT models, the orbit becomes almost circular and the system transitions to full RLOF by $t \approx 100$ Myr. Meanwhile, the orbit keeps shrinking until $q \approx 1.5$ at $t \in [200,250]$ Myr. Beyond this point, the orbit begins to expand, and the system transitions back to partial RLOF at $t \approx 500$ Myr. This transition marks the point that the eccentricity starts increasing again. Figure~\ref{fig:comparison_classical_RLOF_ecc_plane} illustrates how the system transitions between the partial and full RLOF regimes during its evolution.

Slower donor rotation leads to greater orbital widening and eccentricity growth because lower $f_{\rm don}$ places the $L_1$ point farther from the donor (see Fig.~\ref{fig:fit_varying_f_d}), yielding to larger initial $\dot{a}/a$ and $\dot{e}$ (see Fig.~\ref{fig:colormap_semimajor_axis_ecc}). Consequently, for rotating donors with lower spin (i.e., lower $f_{\rm don}$), the system evolves into wider and more eccentric orbits. Finally, the GeMT-model reproduces the results of the emt-model for $f_{\rm don} = 0$ (limit of a non-rotating donor).   

The evolution predicted by the $\delta$-function model is significantly different from other models (see also Sect.~\ref{sec:evolution_AM} and Appendix \ref{app:delta_function}). Initially, both the semimajor axis and the eccentricity decrease, with mass reversal occurring at $t \approx 330$ Myr. Following mass reversal, the orbit slowly widens and eccentricity slowly increases. However, the rates of change for both parameters are notably weaker than in the GeMT-model. Unlike other predictions, the orbit remains highly eccentric, with $e_{\rm min} \gtrapprox 0.9$.

\subsection{Formation of wide post-MT binaries in eccentric orbits}\label{subsec:circular_orbits}

\begin{figure}[!htbp]
    \centering
    \includegraphics[width=\linewidth]{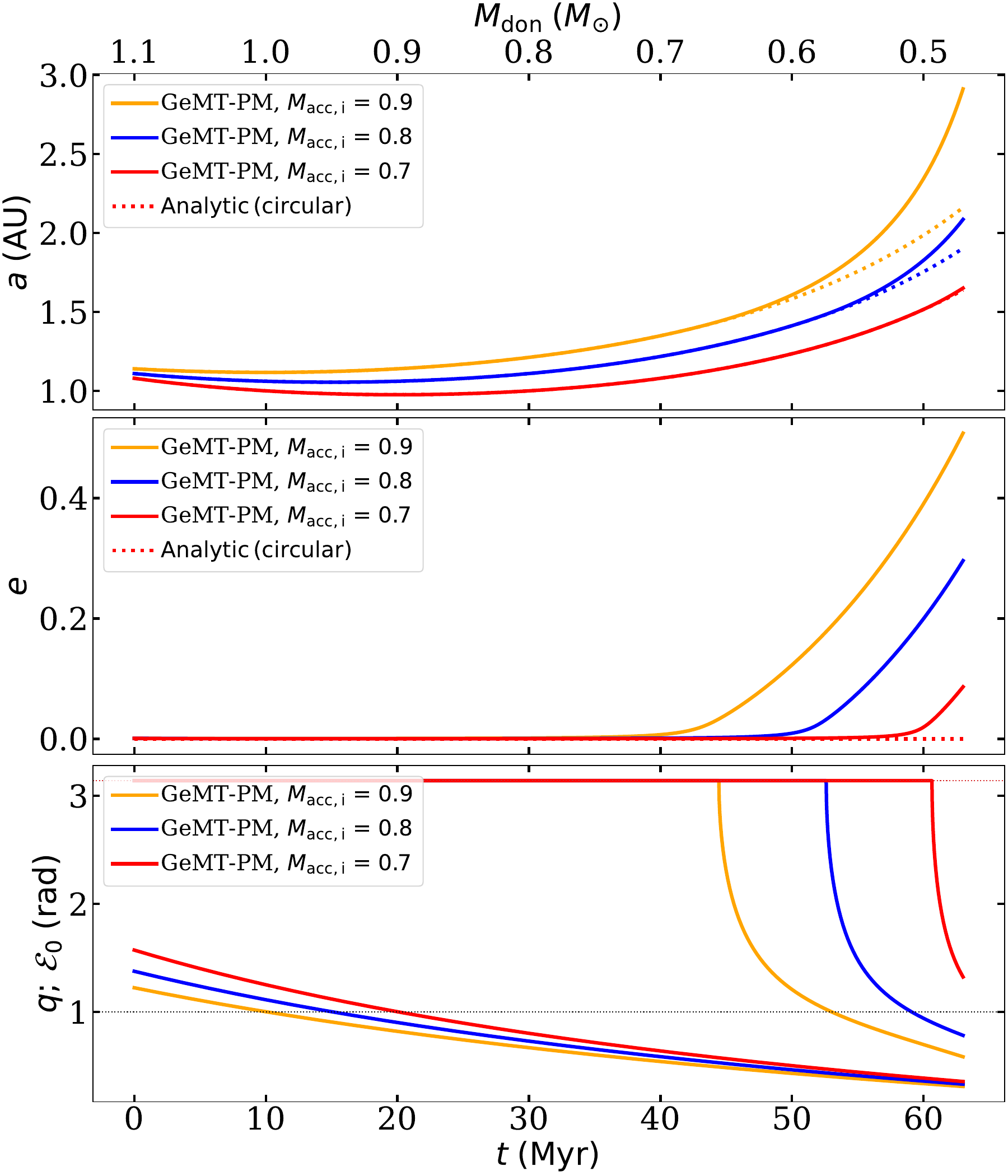}
    \caption{Similar to Fig.~\ref{fig:comparison_hamers_evolution}, but now for an initially nearly circular binary ($e=0.001$). The orange, blue and red lines correspond to the GeMT-model, for $a = 1.08, 1.11, 1.14$ au and $M_{\rm acc} = 0.7,0.8,0.9$ M$_\odot$, respectively. In the top subfigure, the dotted lines illustrate the classical analytic expectation $M_{\rm don}^{2} M_{\rm acc}^{2} a$ is constant for circular orbits, assuming instantaneous circularization; for initial semimajor axis $a_{\rm circ} = 1-e$ au.}
    \label{fig:circular_evolution}
\end{figure}

We consider a donor $M_{\rm don} = 1.1$ M$_\odot$ in a nearly circular orbit of $e=0.001$. The initial semimajor axis and accretor mass are 
$a_{\rm i} = 1.08, 1.11, 1.14$ au and $M_{\rm acc,i} = 0.7,0.8,0.9$ M$_\odot$, respectively. In all three configurations, the donor fills its Roche lobe around the tip of the red giant branch (RGB), when $R_{\rm don} \approx 102$ R$_\odot$ (initially, $x \approx 0.95)$, and initiates MT; late Case B examples. \cite{2023A&A...669A..45T} showed that for such mass ratios, the MT proceeds in a stable manner (they assumed a point mass accretor). During the ascent of the RGB, the donor develops a degenerate helium core that grows in mass until the occurrence of the helium flash when the core mass $M_{\rm c} \approx 0.47$ M$_{\odot}$ \citep{2002MNRAS.336..449H}, while the accretor is still on the main sequence. In contrast to Sect.~\ref{subsec:eccentric_orbits}, we track the orbital evolution assuming point masses under the assumption that the AM stored in the stars in negligible compared to the orbital AM. Under the assumption of conservative MT and point masses, our model is equivalent to the emt-model. We assume $\langle \dot{M}_{\rm don} \rangle = 10^{-8}$ M$_\odot$ yr$^{-1}$. In addition, we follow the orbital evolution under the classical RLOF framework (dotted lines in Fig.~\ref{fig:circular_evolution}), assuming circularization at the onset of MT. The initial semimajor axis is set equal to the original orbital separation at periapsis\footnote{In a similar approach the initial semimajor axis is determined by conserving orbital AM, yielding $a_{\rm circ} = a(1-e^2)$, which also ensures circular post-MT orbits.}, that is, $a_{\rm circ} = a(1-e)$. The evolution of the system is presented in Fig.~\ref{fig:circular_evolution}.

\begin{figure}[!htbp]
    \centering
    \includegraphics[width=\linewidth]{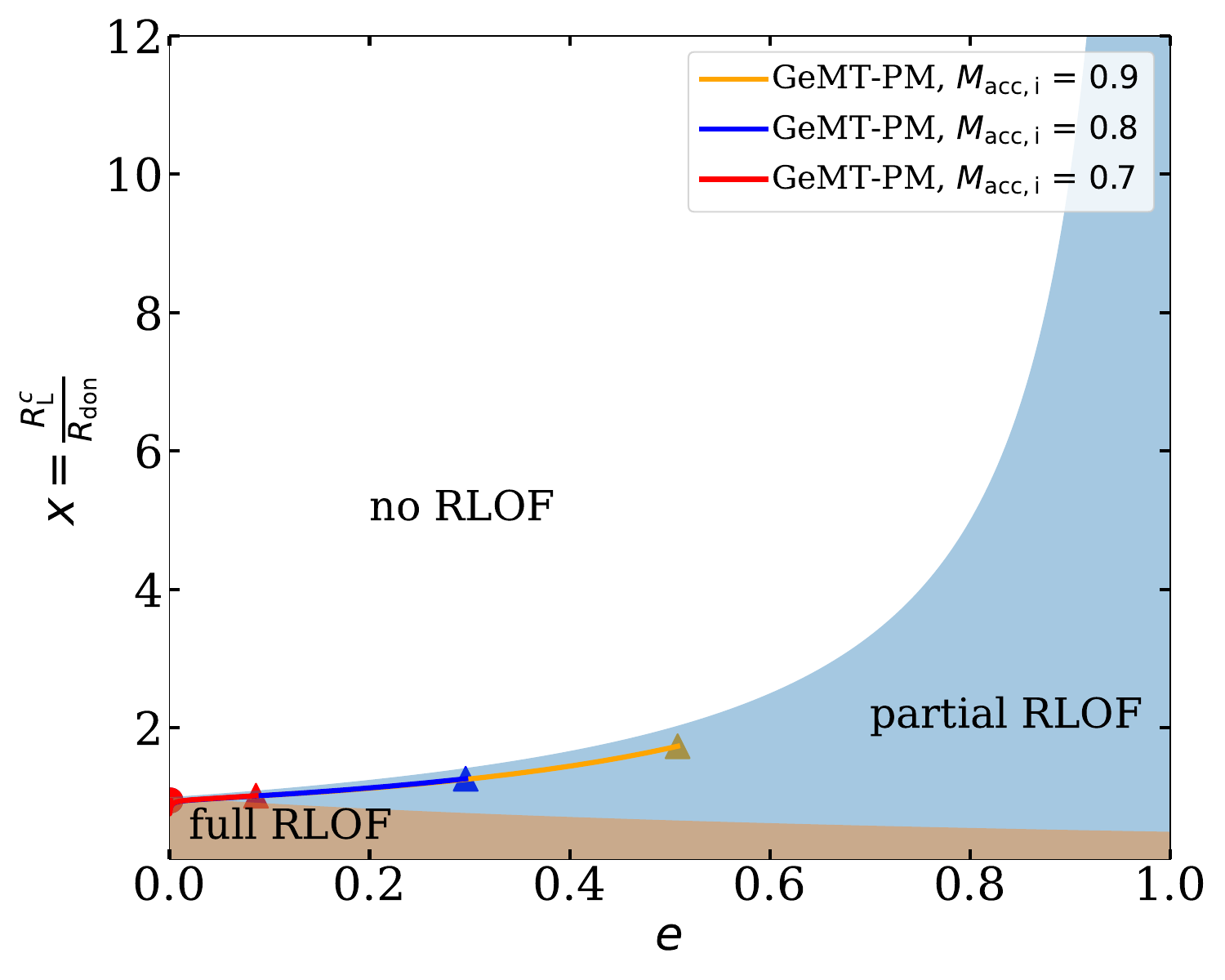}
    \caption{Similar to Fig.~\ref{fig:comparison_classical_RLOF_ecc_plane}, but now for the systems presented in Fig.~\ref{fig:circular_evolution}.}\label{fig:comparison_classical_RLOF}
\end{figure}

In the classical RLOF model, the orbit initially shrinks across all three models (dotted lines) until mass reversal at $t \in [10-20]$ Myr, after which it widens. By the end of the evolution, all three post-MT systems are circular with longer orbital periods than before MT. For the GeMT-model, all three systems initially undergo full RLOF, during which the semimajor axis decreases, similar to the classical model. However, the orbital evolution diverges once the models transition from full to partial RLOF: the semimajor axis changes more rapidly than in the classical case, and the eccentricity increases. Figure~\ref{fig:comparison_classical_RLOF} shows how the systems alternate between partial and full RLOF during its evolution.

Systems with larger initial accretor mass $M_{\rm acc}$  (i.e., smaller initial mass ratios $q$) evolve to wider, more eccentric final orbits (Fig.~\ref{fig:circular_evolution}). These systems initiate RLOF with $q$ closer to the critical $q_{\rm trans,a}=q_{\rm trans,e}=1$ (bottom subfigure in Fig.~\ref{fig:circular_evolution}), while the total transferred mass is the same across all models. Consequently, they spend a larger fraction of their evolution in the orbital widening and eccentricity pumping regions than systems with larger initial $q$ (see also Figs.~\ref{fig:colormap_semimajor_axis_ecc} and \ref{fig:colormap_eccentricity_ecc}).

In summary, $\delta$-function framework yields the weakest secular evolution among the three models; in Sect.~\ref{sec:evolution_AM} and Appendix \ref{app:delta_function} we show why this is the case. Moreover, the emt-model is a subset of the GeMT in the limit of conservative MT ($\beta = 1$) and non-rotating donors ($f_{\rm don} = 0$). In the GeMT framework, for point masses under conservative MT, the secular changes in semimajor axis and eccentricity are correlated: orbits either widen while becoming more eccentric, or shrink while circularizing. 

\section{Discussion}\label{sec:seven}

\subsection{Model limitations}\label{subsec:limitations}

Working in the framework established by \cite{1969Ap&SS...3...31H} and by adopting the assumptions outlined in Sect.~\ref{subsec:assumptions}, we present a novel,  semi-analytical framework for describing the secular orbital evolution of semi-detached systems undergoing RLOF. As \cite{2019ApJ...872..119H} previously emphasized, the validity of assumption 2--imposed ejection ($\vec{w}_{\rm don} = \vec{\dot{r}}$) and accretion velocities ($\vec{w}_{\rm acc} = - \vec{\dot{r}}$)--requires careful evaluation \citep[see][]{2008ARep...52..680L}. Notably, by adopting this assumption, we recover the canonical relation for changes in the semimajor axis due to both non-conservative and conservative MT in the limit of circular orbits. While our assumptions are physically motivated, future studies are essential to thoroughly assess their validity in other regimes.

Hydrodynamical simulations of mass transferring binaries via RLOF \citep[e.g.,][]{2005MNRAS.358..544R,2009MNRAS.395.1127C,2011ApJ...726...67L,2016MNRAS.455..462V,2017MNRAS.467.3556B} would be important in estimating the validity of both assumptions 1 and 2. Furthermore, N-body simulations optimized for MT \citep[e.g.,][]{2010ApJ...724..546S,2014A&A...570A..25D,2017ApJ...844...12D,2023MNRAS.524.4315H} might be employed to assess the secular evolution by computing the MT stream trajectories and their impact on the orbit. Finally, our examples currently assume that the MT rate peaks at periapsis. However, hydrodynamical simulations suggest it may instead peak just after periapsis \citep[e.g.][]{2009MNRAS.395.1127C,2011ApJ...726...67L,2016MNRAS.455..462V}.

\subsection{What physical mechanism drives the evolution of the eccentricity?}\label{subsec:physical_mechanism}

In Section~\ref{subsec:point_masses_circ_orbits}, we demonstrate that orbital eccentricity evolves as long as $e\neq 0$ at the onset of RLOF, whether the stars are treated as extended bodies or not. To illustrate this, let us consider the simpler case of a system undergoing conservative MT ($\beta = 1$) in an eccentric orbit, where both stars are approximated as point masses. Under these assumptions, the total AM must be conserved, and Eqs.~\eqref{eq:orbit_averaged_semimajor_axis_MT_points} and \eqref{eq:orbit_averaged_eccentricity_MT_points} reduce to
\begin{flalign}
     \frac{\langle \dot{a} \rangle}{a} &= -\frac{2 \langle \dot{M}_{\rm don} \rangle}{M_{\rm don}} (1- q) \frac{f_{a}(e,x)}{f_{\dot{M}_{\rm don}}(e,x)} , \label{eq:orbit_averaged_semimajor_axis_MT_points_conserv}\\
     \langle \dot{e} \rangle &= -\frac{2 \langle \dot{M}_{\rm don} \rangle}{M_{\rm don}} (1- q) \frac{f_{e}(e,x)}{f_{\dot{M}_{\rm don}}(e,x)} .\label{eq:orbit_averaged_eccentricity_MT_points_conserv}
\end{flalign}
The terms in these equations are always positive, except for the factor $(1-q)$, which is negative as long as the donor is more massive than the accretor. Over the course of MT, $q$ decreases, hence the orbit shrinks and circularizes up to the point when $q<1$. From that point on, the orbit expands and becomes more eccentric. Essentially, the sign change in $(1-q)$ follows the conservation of orbital AM.

To understand the physical mechanism driving eccentricity evolution, we focus on the terms $f_{a}(e,x), f_{e}(e,x)$ and $f_{\dot{M}_{\rm don}}(e,x)$. The first two arise from assumption 2 in Section~\ref{subsec:assumptions}, where we assume that the velocities of the ejected and accreted material, relative to the donor and accretor, are proportional to the binary's relative velocity. In circular orbits, this velocity is constant and phase-independent. However, in eccentric orbits, it varies with orbital phase, peaking at periapsis and reaching a minimum at apoapsis. Consequently, in eccentric orbits, the velocity of the ejected and accreted mass is assumed higher at periapsis and lower at other orbital phases, introducing an asymmetry absent in circular orbits. In addition, the normalization term $f_{\dot{M}_{\rm don}}(e,x)$ reflects how the mass transfer rate varies with orbital phase for a given $e$ and $x$ (see Fig.~\ref{fig:mass_transfer_rate}). In a circular orbit, the mass transfer rate remains constant and phase-independent. However, in an eccentric orbit, it fluctuates, peaking at periapsis and reaching a minimum at apoapsis. As a result, in eccentric systems, the donor/accretor experiences a higher mass loss/accretion rate at periapsis and a lower rate at other orbital phases, further reinforcing the asymmetry. These constructive asymmetries persist as long as there is a non-zero eccentricity at the onset of MT, yielding phase-dependent RLOF, and their combined effect act as a `kick' around periastron driving changes in eccentricity. 

In the case of extended bodies, the additional terms $g_{a}(e,x), \; g_{e}(e,x), \; h_{a}(e,x),$ and $ \; h_{e}(e,x)$ are associated with the reaction forces exerted on the binary components due to the anisotropic mass ejection from the donor and accretion onto the companion. In Figures~\ref{fig:colormap_semimajor_axis_ecc} and \ref{fig:colormap_eccentricity_ecc}, we illustrate the qualitative impact of these additional perturbations on the rate of change of the semimajor axis and eccentricity, while a discussion about their quantitative effect is presented in Sect.~\ref{sec:evolution_AM}. Importantly, whether assuming extended bodies or not, the underlying mechanism behind the evolution of the eccentricity remains unchanged. In summary, the physical mechanism responsible for the non-zero rate of change of eccentricity arises from the combined effect of the aforementioned asymmetries, which emerge only in the presence of non-zero eccentricity at the onset of MT--whether assuming point masses or extended bodies. Furthermore, conservation of orbital AM dictates the signs of $\dot{a}/a$ and $\dot{e}$.

\subsection{Conservation of angular momentum}\label{sec:evolution_AM}

\begin{figure}[!htbp]
    \includegraphics[width=\linewidth]{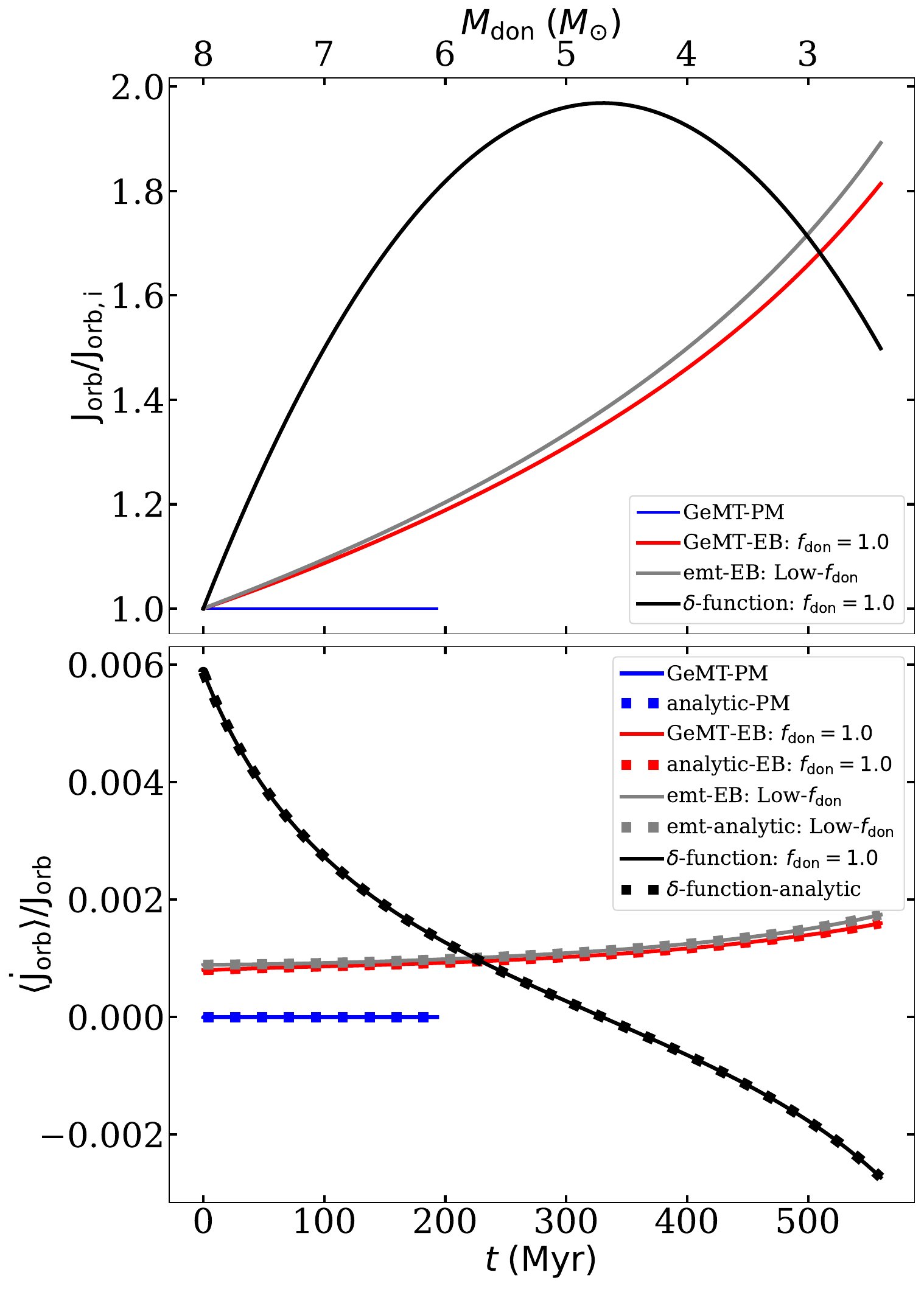}
    \caption{Evolution of the orbital angular momentum (top) and its rate of change (bottom) for the examples presented in Fig.~\ref{fig:comparison_hamers_evolution}. The blue and red colors correspond to the GeMT-model for point masses and extended bodies, respectively. The black and gray colors correspond to the $\delta$-function and emt models, respectively. The solid lines illustrate the numerical solutions, and the square markers the analytical expectations. }
    \label{fig:angular_momentum_evolution}
\end{figure}

Here, we investigate the evolution of the orbital AM momentum predicted by the different MT frameworks and perform a test of the consistency between the numerical and analytical expectations. For this comparison, we use the different models presented in Fig.~\ref{fig:comparison_hamers_evolution}. In the top panel of the Figure.~\ref{fig:angular_momentum_evolution} we present the orbital AM evolution predicted by the GeMT (red and blue lines), the $\delta$-function (black line) and emt (gray line) models. Specifically, the red line corresponds to the synchronous donor case ($f_{\rm don}=1$; blue line in Fig.~\ref{fig:comparison_hamers_evolution}), and the blue line to the evolution of the system assuming point masses (labeled `PM'), which neglects reaction-force contributions to the orbital evolution. In the bottom panel of Figure.~\ref{fig:angular_momentum_evolution}, we compare the rate of change of the orbital AM during conservative MT with the analytical expectations. 

We start our discussion of the orbital AM evolution with the $\delta$-function model. Contrary to the other models, the $\delta$-function model has been derived mathematically under the assumption of conservation of the orbital AM \citep{2007ApJ...667.1170S}. Yet, in the limit of conservative MT, orbital AM is not conserved, as shown by the black line in the top panel of Fig.~\ref{fig:angular_momentum_evolution}. The evolution of the orbital AM predicted by the $\delta$-function model can be described analytically. Whether assuming point masses or not, using Eqs. (29), (39), and (40) from \cite{2007ApJ...667.1170S} it follows that
\begin{equation}\label{eq:sep_analytic}
    \frac{\langle \dot{J}_{\rm orb} \rangle}{J_{\rm orb}} = \frac{\dot{M}_{\rm 0}}{M_{1}} (1-q) \Biggl[ 1+ \frac{e}{\pi} \sqrt{\frac{1-e}{1+e}} - \frac{\sqrt{1-e^2}}{2 \pi} \Biggr],
\end{equation}
where $\dot{M}_{\rm 0}$ is the instantaneous MT rate and $M_{1}$ is the mass of the donor (i.e., equivalent to $M_{\rm don}$ in our notation). In the above derivation, the term $\langle \dot{M}_{1} \rangle$ has been replaced directly by $\dot{M}_{\rm 0}$. Equation \eqref{eq:sep_analytic} is shown in the bottom panel of Fig.~\ref{fig:angular_momentum_evolution} (black squares) verifying our numerical calculations.

We conclude that in the current implementation of the $\delta$-function model, AM is not conserved in the limit of conservative MT--either assuming point masses or not--as can be seen in Fig.~\ref{fig:angular_momentum_evolution}. Consequently, the results of studies using the $\delta$-function should be interpreted with caution when it comes to treatment of orbital AM. In Appendix \ref{app:delta_function} we provide a detailed discussion about the $\delta$-function model and how to apply it such that orbital AM is conserved in the limit of conservative MT.   

The GeMT and emt models can be used either assuming point masses or extended bodies.  In the limit of point masses and conservative MT ($\beta$=1) the models are equivalent, thus the blue line in Fig.~\ref{fig:angular_momentum_evolution} corresponds to both models, and the orbital AM is conserved, as expected (blue line in the top panel of Fig.~\ref{fig:angular_momentum_evolution}). In addition, the numerical results are in very good agreement with the analytical predictions ($\langle \dot{J}_{\rm orb} \rangle /J_{\rm orb}$; Eq.~\ref{eq:change_of_angular_momentum_total} overlaid in square markers in the bottom panel of Fig.~\ref{fig:angular_momentum_evolution}) verifying the robustness of the models. 

When accounting for extended bodies (e.g., red and gray lines in Fig.~\ref{fig:angular_momentum_evolution}), reaction forces drive the secular evolution of the orbital AM; in this specific setup, it increases, and the predicted evolution differs significantly. Notably, in the point-mass approximation, the system merges at $t \approx 200$ Myr, illustrating that neglecting reaction forces can significantly alter the inferred orbital evolution. Moreover, the numerical results are once again in agreement with the analytical predictions both for the GeMT and emt models. Note that when $\beta=1$ and $r_{\rm don}$ is given by Eq.~40 in the Appendix of \cite{2019ApJ...872..119H} (i.e., Low-$f_{\rm don}$ model), the orbit-averaged Eq.~\eqref{eq:change_of_angular_momentum_total} provides also the analytical expectation for the emt model.

In general, we can thus interpret the AM evolution in the following way:\\
Point masses: on the one hand, the classical point-mass approximation represents a scenario where mass ejection and accretion are isotropic, making Eq.~\eqref{eq:angular_momentum_parametrization} well suited for studying orbital evolution due to isotropic winds but less accurate for RLOF. On the other hand, if the AM stored in the rotation of the two stars is negligible compared to the orbital AM, then point masses are valid assumptions. Fig.~\ref{fig:angular_momentum_evolution} demonstrates that the predictions of the GeMT-model are robust in the point-mass limit, confirming that the examples in Fig.~\ref{fig:circular_evolution} capture both the qualitative and quantitative impact of RLOF on the orbital evolution.

Extended bodies: in Section~\ref{sec:three}, we demonstrate that under fully conservative MT ($\beta =1$) the total mass is conserved, yet the orbital AM can still evolve when ejection from the donor and accretion onto the companion are anisotropic (Eq.~\ref{eq:change_of_angular_momentum_total}). During RLOF, the donor star loses mass anisotropically via the $L_1$ point. The accretor also gains mass anisotropically in the case of direct accretion, while the situation becomes more complex when an accretion disk is involved. Nevertheless, if MT is conservative, the total AM must also be conserved, so any orbital gain or loss must be balanced by corresponding changes in the spin AM of the stars. In our setup (and for the emt-model), the reaction force on the donor always increases the orbital AM (Eq.~\ref{eq:change_of_angular_momentum_total}; note that $\dot{M}_{\rm don} < 0$), whereas the reaction force on the accretor can either increase ($\vec{r}_{\rm acc} = r_{\rm acc} \hat{\vec{r}}$) or decrease it ($\vec{r}_{\rm acc} = -r{_{\rm acc}} \hat{\vec{r}}$). 

The extent of orbital AM that can be gained is limited by the donor’s spin AM reservoir, while the amount that can be absorbed is constrained by the accretor's critical rotational velocity \cite[e.g.,][]{1981A&A...102...17P} and its response to accretion \cite[e.g.,][]{2024ApJ...966L...7L}. Fig.~\ref{fig:angular_momentum_evolution} shows that in this simplified example, the orbital AM has increased by $\sim80\%$ (red line), which is not physically expected. Therefore, the evolution presented in Fig.~\ref{fig:comparison_hamers_evolution} should be interpreted qualitatively, as a self-consistent treatment of the total AM requires modeling the spin evolution of both stars. Nevertheless, Fig.~\ref{fig:comparison_hamers_evolution} highlights qualitatively the impact of the donor spin $f_{\rm don}$ on the orbital evolution. Observations of systems shortly after MT with slowly spinning donors and rapidly rotating companions \citep[e.g.,][]{2022A&A...659A...9P,2025A&A...701A...9M} support the relevance of this mechanism.

In summary, Eqs.~\eqref{eq:orbit_averaged_semimajor_axis_MT_points_conserv} and \eqref{eq:orbit_averaged_eccentricity_MT_points_conserv} (point-mass limit) are appropriate for studying the secular evolution of mass-transferring systems (in the limit of conservative MT, $\beta = 1$) assuming that the AM stored in the rotation of the two stars is negligible. However, if this assumption does not hold, accounting for anisotropic mass ejection and accretion is necessary and Eqs.~\eqref{eq:orbit_averaged_semimajor_axis} and \eqref{eq:orbit_averaged_eccentricity} must be adopted. A fully self-consistent treatment requires modeling the spin evolution of both stars, and enforcing conservation of the total AM, and coupling the GeMT-model to detailed stellar evolution, which lies beyond the scope of this work. Nevertheless, the GeMT framework presented here captures the qualitative impact of the reaction forces on orbital AM, making its integration with binary evolution codes an important future direction.

\subsection{Stability of mass transfer}\label{subsec:stability}

RLOF in a binary system leads to either stable or unstable MT, shaping its future evolution and the properties of the final remnants. Unstable MT followed by a common-envelope (CE) phase typically results in a close binary or a merged object \citep[e.g.][]{1976IAUS...73...75P}, while stable MT tends to produce wider binaries \citep[e.g.][]{1997A&A...327..620S}. The stability of the MT process and its outcomes depend mainly on two factors\footnote{A third factor is the accretor response to MT \citep{2024ApJ...966L...7L}, but it is considered of secondary importance.}: (1) how the donor’s radius responds to mass loss and (2) how the orbit—and consequently the Roche lobe—responds to MT. Several observed systems contradict the standard understanding of MT stability. This includes systems that appear to have experienced MT from donors on the RGB (Case B) or asymptotic giant branch (AGB; Case C), yet have relatively wide orbits \citep{1989SSRv...50..165E}, despite classical results predicting otherwise.

Numerous studies have investigated the stability of MT, suggesting that it is often severely underestimated \citep{2012ApJ...744...12W,2012ApJ...760...90P,2015MNRAS.449.4415P,2010ApJ...717..724G,2015ApJ...812...40G,2020ApJ...899..132G,2021A&A...645A..54K,2023A&A...669A..45T}. In these studies, there has been great effort to model more accurately the response of the donor to mass loss, however the response of the orbit is still modeled under classical assumptions. Traditionally, the orbital response is modeled within the classical RLOF framework, assuming circular orbits and point masses. The GeMT framework improves upon these limitations in two key ways: (1) relaxing the point-mass approximation, allowing for anisotropic mass ejection and accretion, thereby accounting for the offset location of the $L_1$ point, (2) it does not impose instantaneous circularization, enabling self-consistent modeling of orbital evolution in eccentric systems.

The deviation from the classical RLOF picture has significant implications for MT stability criteria. For instance, by relaxing the point-mass approximation, the GeMT-model predicts values up to $q_{\rm trans,a} \approx 1.53$ for circular orbits with synchronously rotating binary components. Consequently, the critical mass ratio separating stable from unstable MT needs to increase. In summary, implementing the GeMT-model in studies of mass transferring systems using detailed evolution codes \cite[e.g.][]{2014A&A...570A..25D,2025arXiv250921521P} can provide a direct comparison to observations of interacting and post-interaction binaries and help constrain binary physics.

\subsection{Observed post-interaction wide binaries}

Recently, it has become clear that eccentric orbits are common in wide post-interaction binary systems \citep[see e.g.,][]{2024MNRAS.529.3729S}. Observations reveal a notable trend: the range of eccentricities increases with orbital period \citep[see][Fig. 8]{2024MNRAS.529.3729S}, with the maximum observed eccentricities also rising as the orbital period grows. This pattern is not confined to a specific population of binaries, but is evident across diverse post-interaction systems, including long-period sdB binaries \citep[][and references therein]{2015A&A...579A..49V,2017A&A...605A.109V,2020A&A...641A.163V,2022A&A...658A.122M}, Barium stars \citep{1998A&A...332..877J,2010A&A...523A..10I,2019A&A...626A.127J,2018A&A...620A..85O,2019A&A...626A.128E}, blue stragglers \citep{2011Natur.478..356G,2009Natur.462.1032M}, CH and CEMP stars \citep{2016A&A...586A.158J,2018A&A...620A..85O,2019A&A...623A.128H}. Despite their prevalence, the formation of these systems remains a challenge for existing models, as neither their long periods nor their high eccentricities can be reproduced by theoretical models. While several eccentricity-pumping mechanisms have been proposed, synthetic models still struggle to reproduce the general orbital properties of post-interaction binaries. Specifically:
\begin{enumerate}
    \item The tidally enhanced wind mass-loss mechanism \citep{2000A&A...357..557S,2008A&A...480..797B} can generate eccentric He-WD binaries but not sdB systems, as extreme mass loss prevents helium ignition \citep{2015A&A...579A..49V}.
    \item Circumbinary disk (CD) interactions tend to produce high eccentricities at shorter periods, contradicting observations \citep{2013A&A...551A..50D,2015A&A...579A..49V,2018MNRAS.474..433D}. Additionally, \cite{2020A&A...642A.234O} demonstrated that binary interactions with a CB disk cannot account for the observed eccentric orbits in post-AGB binaries.
    \item White dwarf kicks \citep[e.g.,][]{2010A&A...523A..10I,2025arXiv250821805C} could increase eccentricity; however, the kick mechanism remains unclear, and it is not relevant for sdB binaries. 
    \item Mergers in triples \citep{2009ApJ...697.1048P} and dynamical interactions with a tertiary companion \citep{2020A&A...640A..16T} can lead to the formation of eccentric binaries. In the first scenario, the surviving binary originates from the former outer orbit, with the merger product and the original tertiary companion as its components. In the second scenario, the Lidov-Kozai mechanism drives eccentricity growth in the inner binary. However, it is unlikely that triple interactions alone can account for all the eccentric orbits observed across the entire population of post-interaction binaries.
\end{enumerate}

The long orbital periods observed ($P_{\rm orb} \gtrsim 10^2$ days) rule out a CE phase for these systems, as it would have resulted in much tighter orbits. Instead, stable MT appears to be the more plausible interaction mechanism. However, classical circularization theory does not predict eccentric post-RGB and post-AGB binaries from stable RLOF (see Fig.~\ref{fig:circular_evolution}). In summary, no proposed mechanism to date can fully explain the observed correlation between longer orbital periods and higher eccentricities, a trend that appears to hold across the entire post-interaction binary population.
 
\begin{figure}[!htbp]
    \centering
    \includegraphics[width=\linewidth]{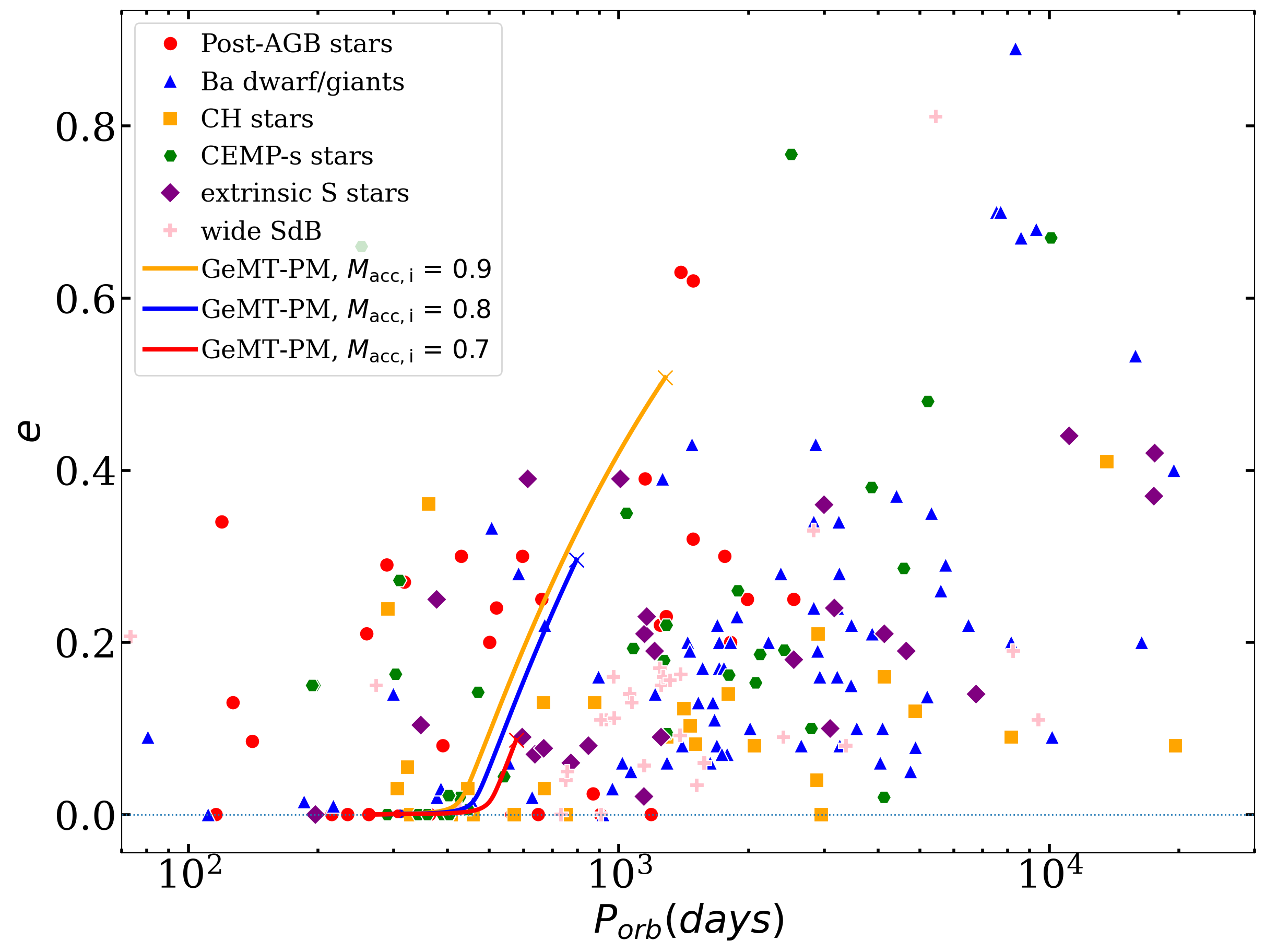}
    \caption{Evolution of the system presented in Fig.~\ref{fig:comparison_hamers_evolution} on the $P_{\rm orb}-e$ plane. The orange, blue and red lines correspond to the GeMT-model, for subsynchronous ($f_{\rm don}=0.0$), synchronous ($f_{\rm don}=1.0$) and supersynchronous donors ($f_{\rm don}=2.0$), respectively. The dashed black line corresponds to the emt-model. The small circles and x symbols, indicate the initial and final positions of the systems, respectively. Red circles represent the post-AGB stars \citep{2018A&A...620A..85O}. Blue triangles represent the barium (Ba) dwarfs and giants \citep{2019A&A...626A.127J,2019A&A...626A.128E}, orange squares are CH subgiants \citep{2019A&A...626A.128E}, green pentagons are CEMP-s stars \citep{2016A&A...588A...3H,2016A&A...586A.158J,2016ApJ...826...85S}, purple diamonds are extrinsic S stars \citep{2000AJ....119.1375F,2019A&A...626A.127J}, pink plusses are wide sdB binaries \citep{2017A&A...605A.109V}.}
    \label{fig:circular_obs}
\end{figure}

At the onset of RLOF, some systems may have sufficiently low eccentricities to be classified observationally as circular binaries \citep{1992RSPTA.341...39P,2024MNRAS.534..455C}. Isolating the effects of MT via RLOF from other physical processes, the GeMT framework, which accounts for MT in eccentric orbits, demonstrates that partial RLOF can amplify small undetectable eccentricities to measurable levels while simultaneously widening the orbit, see Fig.~\ref{fig:circular_evolution}. For instance, in Figure~\ref{fig:circular_obs}, we illustrate how the late Case B MT example presented in Fig.~\ref{fig:circular_evolution} evolves on the $P_{\rm orb}-e$ plane. This evolutionary path is consistent with observed systems. We note that other physical processes, such as tides, stellar evolution or the response of the donor and accretor to mass loss and accretion, will affect our results. Nevertheless, the GeMT-model predicts a type of evolution that is relevant to all post-interaction observed systems, since it naturally predicts higher eccentricities at longer orbital periods, aligning well with numerous observed systems \citep[see][and references therein]{2017A&A...605A.109V,2019A&A...626A.127J,2018AJ....155..144K,2022A&A...658A.122M,2023A&A...671A..97E,2024MNRAS.52711719Y}.

We note, that while partial-RLOF can act as an eccentricity-pumping mechanism, the formation of wide and eccentric binaries is challenging using the $\delta$-function model of \cite{2007ApJ...667.1170S}, except in cases where the systems start wide and eccentric at the onset of RLOF. For systems with initially low eccentricities or these that might circularize during RLOF ($e \rightarrow 0$), the $\delta$-function model is invalid, and the transition to the classical point-mass RLOF model cannot reproduce the formation of such systems.

Lastly, a discrepancy exists between the observed period distribution of double white dwarf (DWD) binaries and predictions from synthetic models of systems formed via stable MT. Specifically, theoretical models fail to reproduce the longest orbital periods observed in DWD binaries \citep[see][Fig. 11]{2022MNRAS.515.1228K}. The GeMT-model’s natural tendency to predict stronger orbital widening (see Fig.~\ref{fig:circular_evolution}) suggests that it may help resolve this discrepancy.

\section{Conclusions}\label{sec:eight}

We presented the General Mass Transfer ({\sc GeMT}) model, a comprehensive semi-analytic framework for the orbital evolution of mass-transferring binaries. For the first time, an eccentric mass transfer model applies to both conservative and non-conservative mass transfer across the full range of eccentricities, while also accounting for the spin of the donor. Our main conclusions are given below.
\begin{enumerate} 
    \item We demonstrated that in the case of anisotropic mass ejection and accretion, such as mass transfer via RLOF, reaction forces on the donor and accretor affect the orbital evolution even in the limit of conservative mass transfer. Therefore, in detailed stellar evolution studies where the component spins and angular momentum can be modeled, these additional perturbations must be included when investigating orbital evolution driven by mass transfer.
    \item For the position of the ejection point (the $L_1$ Lagrangian point) relative to the donor, the previous prescriptions introduced by \cite{2019ApJ...872..119H} were limited to cases of either massive donors ($q \gg 1$) or static donors ($f_{\rm don} = 0.0$). We introduced an accurate prescription (most accurate to date) for the position of the Lagrangian $L_1$ point (Global-$L_1$ model), applicable for any $e \in [0.0, 0.99]$, $q \in [0.1, 10.1]$ and $f_{\rm don} \in [0.0, 2.0]$. Lastly, we introduced a novel mass accretion scenario in which the ejected mass follows a curved trajectory due to its initial velocity and lands on the side of the accretor that faces away from the donor.
    \item  The classical RLOF framework typically assumes circularization before the onset of RLOF. In contrast, the GeMT-model does not impose this assumption and instead predicts that due to phase-dependent RLOF, the orbital eccentricity can either increase or decrease. Mass-transferring binaries with up to $q = M_{\rm don}/M_{\rm acc} \leq 1.0$ naturally evolve toward wider and more eccentric orbits, in contrast to classical expectations (Fig.~\ref{fig:circular_evolution}).
    \item  When accounting for extended bodies, the parameter space for orbital widening and eccentricity pumping increases (Figs.~\ref{fig:colormap_semimajor_axis_circ}, \ref{fig:colormap_semimajor_axis_ecc}, \ref{fig:colormap_eccentricity_ecc}). For circular orbits, the transitional mass ratio, $q_{\rm trans,a}$, which separates orbital widening from shrinkage, can increase from $q_{\rm trans,a}=1$ up to $q_{\rm trans,a} \sim 1.5$. This implies that the critical mass ratio distinguishing stable from unstable mass transfer is systematically underestimated, regardless of the donor’s evolutionary phase. Detailed stellar evolution is necessary to test quantitatively the contribution of stellar angular momentums to the orbital angular momentum evolution.
    \item The GeMT-model predicts qualitatively and quantitatively distinct evolutionary pathways for circular and eccentric orbits. Nonetheless, we showed that in the limit of non-rotating donors ($f_{\rm don} = 0$) and conservative mass transfer, GeMT reproduces the predictions of the emt-model \citep{2019ApJ...872..119H}. Compared to the $\delta$-function formalism \citep{2007ApJ...667.1170S}, GeMT--whether assuming point masses or extended bodies--yields a broader parameter space for eccentricity pumping and stronger evolution of both semimajor axis and eccentricity for both circular and eccentric orbits. Since the $\delta$-function model \citep{2007ApJ...667.1170S}, as applied in the literature, does not conserve orbital angular momentum in the limit of conservative mass transfer, its predicted orbital evolution should be interpreted with caution (Sect.~\ref{sec:evolution_AM}). Here, we show that in the limit of MT occurring at periastron, the functional shape of the mass transfer rate predicted by the GeMT-model is a $\delta$-function and orbital angular momentum is conserved (Appendix \ref{app:delta_function}). Moreover, in this limit, and adopting point masses, the GeMT-model is equivalent to the nomralized $\delta$-function model (i.e., Eqs.~\ref{eq:sep_semimajor_axis_MT_points_conserv}, and \ref{eq:sep_eccentricity_MT_points_conserv}) verifying that the latter is also a subset of the GeMT-model.
    \item We demonstrated that phase-dependent RLOF can act as an eccentricity-pumping mechanism. Isolating the effects of mass transfer via RLOF from other physical processes, we showed that stable mass transfer can produce post-RLOF systems with wide and eccentric orbits. Importantly, we proved that the observed correlation of higher eccentricities and longer orbital periods naturally arises from our model. The predicted orbital evolution closely aligns with the observed properties of wide, eccentric systems containing blue stragglers \citep{2011Natur.478..356G, 2009Natur.462.1032M}, sdB stars \citep{2015A&A...579A..49V,2017A&A...605A.109V,2020A&A...641A.163V,2022A&A...658A.122M}, Barium stars \citep{1998A&A...332..877J,2010A&A...523A..10I,2019A&A...626A.127J,2018A&A...620A..85O,2019A&A...626A.128E}, CH and CEMP stars \citep{2016A&A...586A.158J,2018A&A...620A..85O,2019A&A...623A.128H} and WDs \citep{2018AJ....155..144K,2024MNRAS.529.3729S,2024MNRAS.52711719Y} (Fig.~\ref{fig:circular_obs}). Our model supports the interpretation that stable mass transfer in eccentric orbits is also relevant for other post-interaction systems with similar orbital characteristics.
\end{enumerate}
Our model can be integrated into detailed and rapid stellar evolution codes to consistently treat both conservative and non-conservative mass transfer in circular and eccentric orbits.

\section{Data availability}

The data necessary to reproduce Figs. 7-11 is available  on \href{https://doi.org/10.5281/zenodo.15198323}{Zenodo}. The GeMT code will be shared upon reasonable request to the authors.

\begin{acknowledgements}
The authors would like to thank Caspar Bruenech, Floris Kummer, Martin Heemskerk, Ed van den Heuvel, Onno Pols, George Voyatzis, and Thomas Tauris for the useful discussions. The authors would like to thank Glenn-Michael Oomen for providing the observational data for Figure 12. AP \& ST acknowledges support from the Netherlands Research Council NWO (VIDI 203.061 grant). FD acknowledges support from the UK’s Science and Technology Facilities Council Grant No. ST/V005618/1. EL acknowledges support through a start-up grant from the Internal Funds KU Leuven (STG/24/073) and through a Veni grant (VI.Veni.232.205) from the Netherlands Organization for Scientific Research (NWO). This work used the following software packages: Matplotlib \citep{Hunter:2007}, Seaborn \citep{Waskom2021}, NumPy \citep{harris2020array}, SciPy \citep{2020SciPy-NMeth} and SymPy \citep{10.7717/peerj-cs.103}.
\end{acknowledgements}

\bibliographystyle{aa}
\bibliography{references}

\begin{appendix}

\section{Parametrizing angular momentum loss}\label{app:angular_momentum_loss}

A perturbation induced on a binary system can give rise to changes in the orbit's Keplerian elements. Using the true anomaly, $\nu$, the binary separation is 
\begin{equation}
    \vec{r} = \frac{a(1-e^2)}{1+e \cos \nu} \vec{\hat{r}}
\end{equation}
and the relative velocity is given by
\begin{equation}\label{eq:relative_velocity}
    \dot{\vec{r}} = \dot{r} \vec{\hat{r}} + r \Omega_{\rm orb} (\hat{\vec{h}} \times \hat{\vec{r}}).
\end{equation}
In addition, the perturbation given by Eq.~\eqref{eq:total_perturbation_non_conservative_MT_simplified} has the form $\vec{f} = c_{1} \vec{\dot{r}} + c_{2} (\hat{\vec{h}} \times \hat{\vec{r}}) + c_{3} \hat{\vec{r}}$, where $c_1 = - \frac{\dot{M}_{\rm don}}{M_{\rm don}} \biggl(1- \beta q -\frac{(1-\beta)(\gamma + \frac{1}{2})q}{1+q} \biggr)$, $c_2 = - \frac{\dot{M}_{\rm don}}{M_{\rm don}} \Omega_{\rm orb} (r_{\rm don} \pm  \beta q r_{\rm acc})$ and $c_3= - \frac{\ddot{M}_{\rm don}}{M_{\rm don}} (r_{\rm don} \pm \beta q r_{\rm acc})$. Using Eq.~\eqref{eq:relative_velocity} we re-write the perturbation as
\begin{equation}\label{eq:perturbation_compact_form}
    \vec{f}_{\rm total} = C \vec{\dot{r}} + C' \vec{\hat{r}},
\end{equation}
where $C = c_1 +\frac{c_{2}}{r\Omega_{\rm orb}}$ and $C' = c_3 - \frac{c_{2} \dot{r}}{r\Omega_{\rm orb}}$.

The semimajor axis and the eccentricity of the orbit will change based on Eqs.~\eqref{eq:semimajor_axis_derivative} and \eqref{eq:eccentricity_derivative}, respectively. Under the influence of $\vec{f}_{total} = C \vec{\dot{r}} + C' \vec{\hat{r}}$, one derives
\begin{equation}\label{eq:semimajor_axis_C}
    \frac{\dot{a}}{a} = 2C \Biggl(1+ \frac{2e (e+ \cos \nu)}{1-e^2} \Biggr) + \frac{2C'e\sqrt{1-e^2}sin(\nu)}{na(1-e^2)},
\end{equation}
\begin{equation}\label{eq:eccentricity_C}
    \dot{e} = 2C (e+ \cos \nu) + \frac{C'\sqrt{1-e^2}sin(\nu)}{n a}.
\end{equation}

Substituting Eq.~\eqref{eq:eccentricity_C} into Eq.~\eqref{eq:semimajor_axis_C} and multiplying both sides by, $M a (1-e^2)$ we have
\begin{flalign}\label{eq:integral_of_the_system}
    M(1-e^2)\dot{a} = 2C a M(1-e^2) + 2 a eM\dot{e}& \nonumber \\
   a(1-e^2)\dot{M}  + M(1-e^2)\dot{a} - 2 a eM\dot{e} = \nonumber\\
    2C a M(1-e^2)+ a(1-e^2)\dot{M} \nonumber\\
    \frac{d}{dt} [M a (1-e^2)]  = [M a (1-e^2)] \Biggl(2C +\frac{\dot{M}}{M}\Biggr)& \nonumber\\
    \frac{\frac{d}{dt} [GM a (1-e^2)]}{[GM a (1-e^2)]}  =  \Biggl(2C +\frac{\dot{M}}{M}\Biggr)& \nonumber\\
    \frac{1}{2}\frac{\frac{d}{dt} [GM a (1-e^2)]}{[GM a (1-e^2)]}  =  \Biggl(C +\frac{1}{2} \frac{\dot{M}}{M}\Biggr)& \nonumber\\
    \frac{\dot{l}_{\rm orb}}{l_{\rm orb}}  =  \Biggl(C +\frac{1}{2} \frac{\dot{M}}{M}\Biggr)& \nonumber ,\\
\end{flalign}
where $l=J/\mu$ the orbital AM per reduced mass. Additionally, by differentiating the orbital AM, we derive
\begin{equation}\label{eq:derivative_orbital_angular_momentum}
    \frac{\dot{J}_{\rm orb}}{J_{\rm orb}} + \frac{\dot{M}}{M} - \frac{\dot{M}_{\rm acc}}{M_{\rm acc}} - \frac{\dot{M}_{\rm don}}{M_{\rm don}} = \frac{1}{2} \frac{\frac{d}{dt} [GM a (1-e^2)]}{[GM a (1-e^2)]},
\end{equation}
and by substituting Eq.~\eqref{eq:integral_of_the_system} into Eq.~\eqref{eq:derivative_orbital_angular_momentum} and using Eq.~\eqref{eq:beta_parametrization}, we find
\begin{equation}\label{eq:derivative_orbital_angular_momentum_2}
    \frac{\dot{J}_{\rm orb}}{J_{\rm orb}} = C + (1-\beta q)\frac{\dot{M}_{\rm don}}{M_{\rm don}} - \frac{1}{2}\frac{(1-\beta)q}{1+q} \frac{\dot{M}_{\rm don}}{M_{\rm don}} .
\end{equation}

At this point, $C$ needs to be specified. As shown by Eqs.~\eqref{eq:perturbation_compact_form} and \eqref{eq:total_perturbation_non_conservative_MT_simplified}, we define 
\begin{flalign}
   C = &- \biggl(1- \beta q -\frac{(1-\beta)(\gamma + \frac{1}{2})q}{1+q} \biggr)\frac{\dot{M}_{\rm don}}{M_{\rm don}} \nonumber \\
   &- \biggl( \frac{r_{\rm don}}{r} \pm \beta q \frac{r_{\rm acc}}{r} \biggr)\frac{\dot{M}_{\rm don}}{M_{\rm don}},  
\end{flalign}
where the negative sign in front of the term associated with the accretion point corresponds to $\vec{r}_{\rm acc} = - r_{\rm acc} \hat{\vec{r}}$, while the positive sign to $\vec{r}_{\rm acc} = r_{\rm acc} \hat{\vec{r}}$. Hence, after substituting $C$ into Eq.~\eqref{eq:derivative_orbital_angular_momentum_2}, we find
 \begin{equation}\label{eq:gradient_J_orb}
    \frac{\dot{J}_{\rm orb}}{J_{\rm orb}} = \gamma (1-\beta)\frac{\dot{M}_{\rm don}}{M_{\rm don} + M_{\rm acc}} -\biggl( \frac{r_{\rm don}}{r} \pm \beta q \frac{r_{\rm acc}}{r} \biggr)\frac{\dot{M}_{\rm don}}{M_{\rm don}}.
\end{equation}

\section{The inner Lagrangian $L_1$ point in asynchronous and eccentric binaries}\label{app:analytical_prescription_L1}

An accurate description of the location of the inner Lagrangian $L_1$ point is essential for understanding the evolution of mass-transferring binaries via RLOF. The Global-$L_1$, $X_{\rm L1}(f_{\rm don},q,e)$, determines the position of the $L_1$ point, relative to the donor's center of mass, at periapsis in units of the instantaneous distance between the two stars, given explicitly by 
\begin{flalign}\label{eq:analytic_formula_L1}
X_{\rm L1}(f_{\rm don},q,e) &= 0.526 + 0.255\log10(q) - 0.024\log(q)^3 \nonumber \\
&+ f_{\rm don}^2\Biggl( -(0.027+0.216 e)(1+ 0.626\log(q)) \\
&+ 0.007 e^2 (1-1.267 \log(q)^2 ) \Biggr) \nonumber .  
\end{flalign}
Equation~\eqref{eq:analytic_formula_L1} achieves an accuracy better than $\sim 9\%$ for $0.0 \leq f_{\rm don} \leq 2.0$, $0.1 \leq q \leq 10.1$ and $0.0 \leq e \leq 0.99$.

\begin{figure}[!htbp]
    \includegraphics[width=\linewidth]{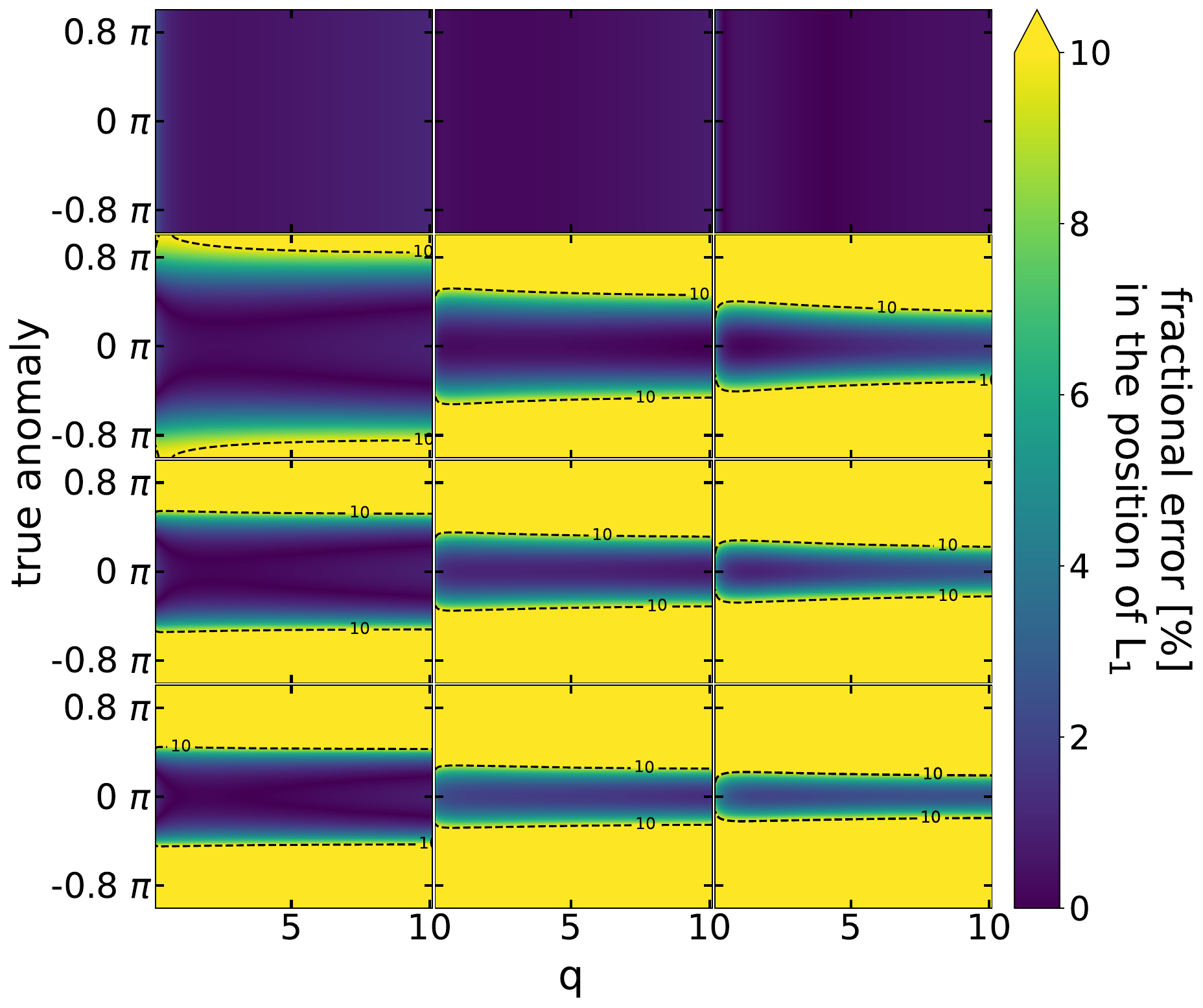}
    \caption{Fractional error of the position of the $L_1$ point over one orbit for the Global-$L_1$ model. From top to bottom, the subfigures correspond to $e=0.0,0.3,0.6,0.9$, respectively. From left to right, the subfigures correspond to $f_{\rm don} =0.5,1.0,1.5$, respectively. The dashed line corresponds to a fractional error of 10\%.}
    \label{fig:gemt_model_L1_error}
\end{figure}

We approximate the position of $L_1$ over a single orbit in natural units using the product $X_{\rm L1}(f_{\rm don},q,e) \vec{r}$. To assess the accuracy of our prescription, we calculate the fractional error at the predicted positions, as
\begin{equation}
    \Delta X_{\rm L1} = \frac{X_{\rm L1}(f_{\rm don},q,e,\mathcal{E}) - X_{\rm L1}(f_{\rm don},q,e)\vec{r}}{X_{\rm L1}(f_{\rm don},q,e,\mathcal{E})} \times 100\%,
\end{equation}
where $X_{\rm L1}(f_{\rm don},q,e,\mathcal{E})$ is the numerical solution  of Eq.~\eqref{eq:Lagrangian_points} for $0.0 \leq f_{\rm don} \leq 2.0$, $0.1 \leq q \leq 10.1$, $0.0 \leq e \leq 0.99$ and $0.0 \leq \mathcal{E} \leq 2\pi$. We have arbitrary defined the region of "good accuracy" as the part of the parameter space where the fractional error is $\leq 10\%$. Figure~\ref{fig:gemt_model_L1_error} illustrates the fractional error at the position of the $L_1$ point over one orbit.

\begin{figure}[!htbp]
    \includegraphics[width=\linewidth]{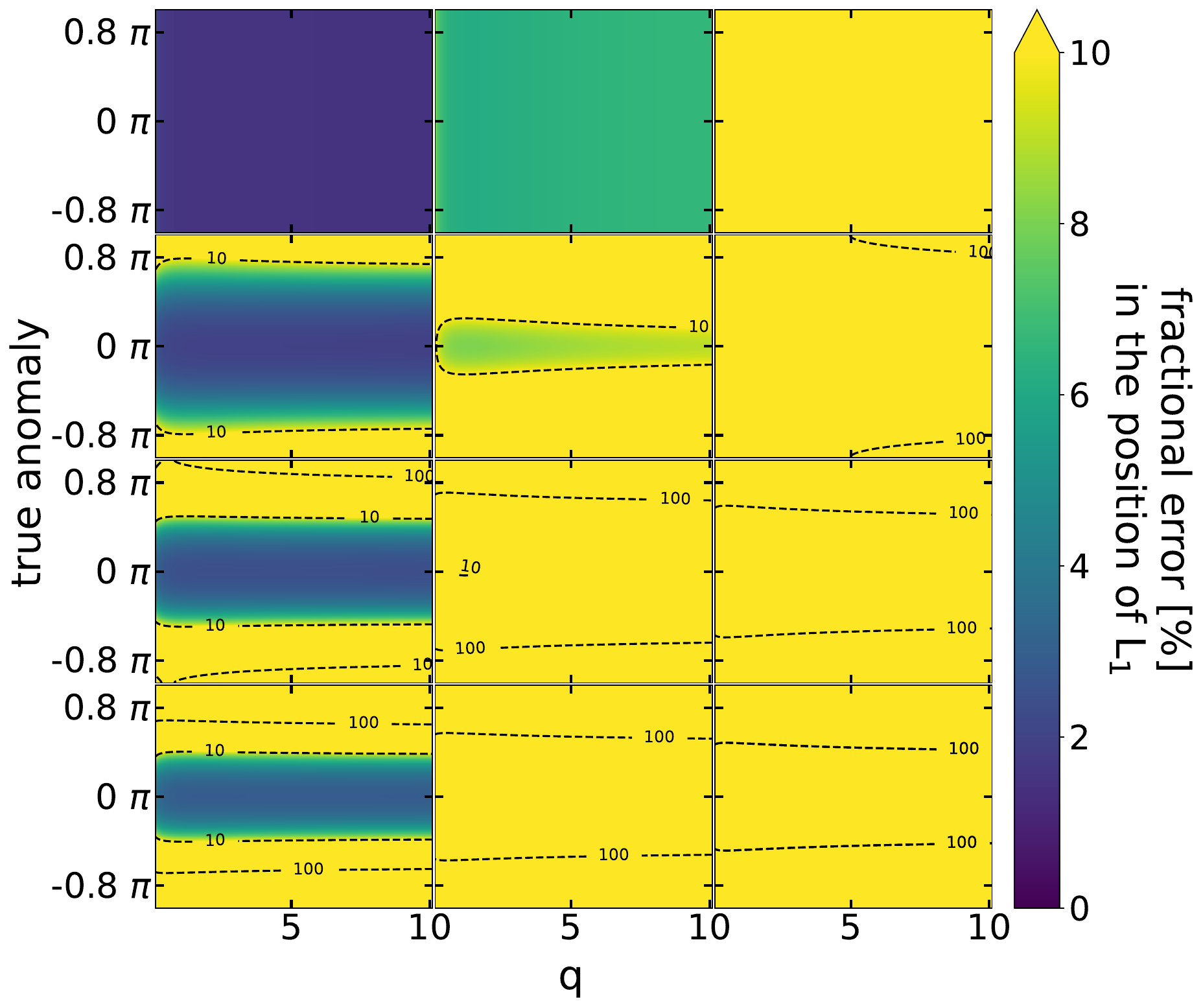}
    \caption{Fractional error of the position of the $L_1$ point over one orbit for the Low-$f_{\rm don}$ model. From top to bottom, the subfigures correspond to $e=0.0,0.3,0.6,0.9$, respectively. From left to right, the subfigures correspond to $f_{\rm don} =0.5,1.0,1.5$, respectively. The dashed lines correspond to fractional errors of 10\% and 100\%.}
    \label{fig:low_f_L1_error}
\end{figure}

In Figure~\ref{fig:gemt_model_L1_error}, we observe that for circular orbits, the fractional error remains within $\leq 3\%$ for $0.0 \leq f_{\rm don} \leq 2.0$ and $0.1 \leq q \leq 10.1$. For eccentric orbits, the region of good accuracy becomes increasingly confined to the vicinity of periapsis as eccentricity increases. The model's accuracy is largely independent of the mass ratio $q$, but exhibits a weak dependence on the degree of asynchronism $f_{\rm don}$. Specifically, the region of good accuracy shrinks around periapsis for donors with progressively higher spin rates (supersynchronous rotation). We highlight that the region of "good accuracy" essentially mirrors the mass loss rate given by Eq.~\eqref{eq:mass_loss_rate_normalized}. The position of the $L_1$ point is less accurate away from periapsis, where the mass loss rate is very low, but remains highly accurate near periapsis, where mass loss peaks. For comparison purposes, we show the same plot in Fig.~\ref{fig:low_f_L1_error}, for the Low-$f_{\rm don}$ model.

\begin{figure}[!htbp]
    \centering
    \includegraphics[width=\linewidth]{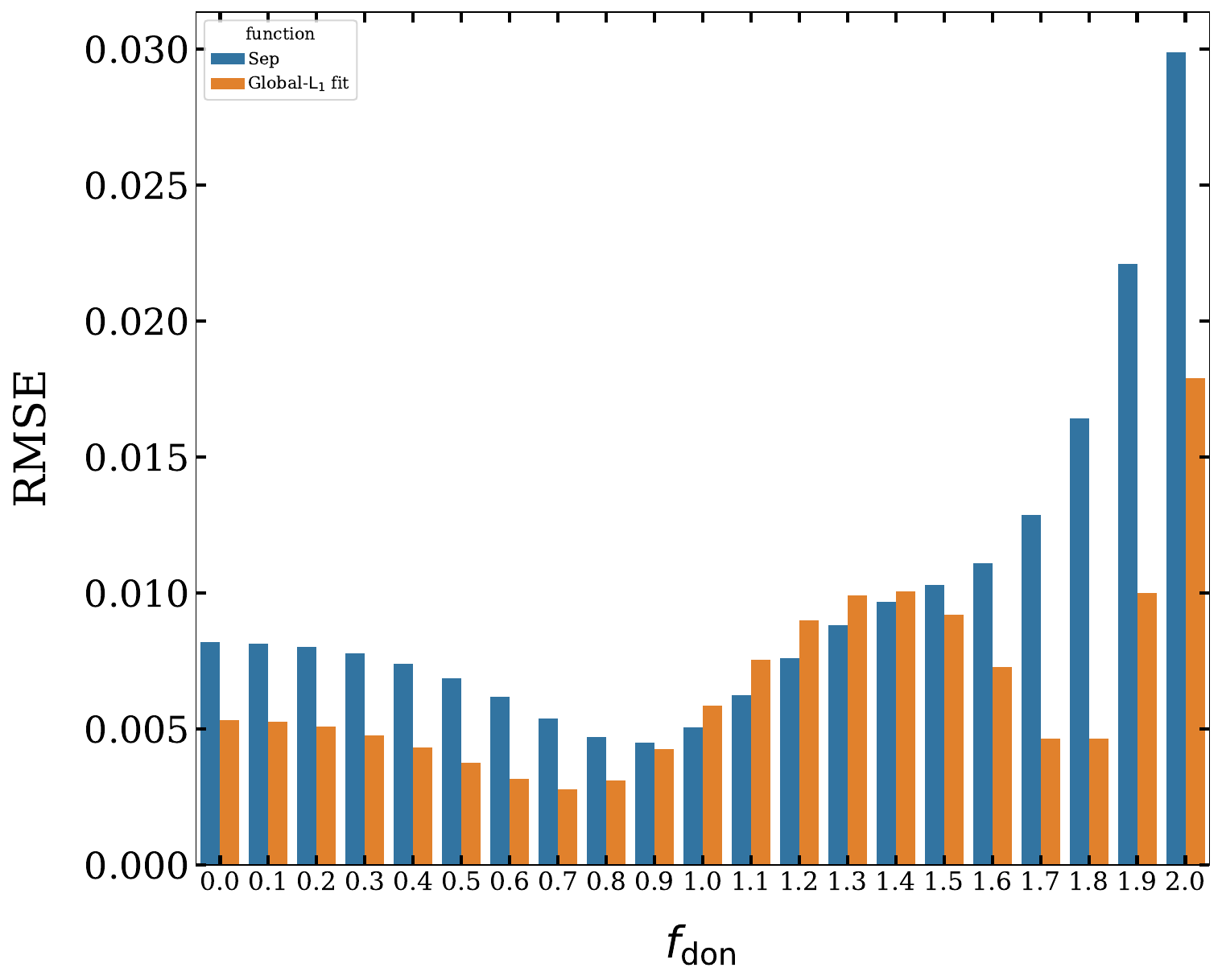}
    \caption{Comparison of the accuracy of the two fits in predicting the position of the $L_1$ point at the periapsis of the binary orbit for varying $f_{\rm don},q,e$. The blue and orange colors correspond to the equation A.15 in Appendix A of \cite{2007ApJ...667.1170S} and the Global-$L_1$ model (Eq.~\ref{eq:analytic_formula_L1}), respectively.}
    \label{fig:comparison_sep}
\end{figure}

The Global-$L_1$ model, $X_{\rm L1}(f_{\rm don},q,e)$, is a fit to the numerical solutions of Eq.~\eqref{eq:Lagrangian_points_2} over the parameter ranges $0.0 \leq f_{\rm don} \leq 2.0$, $0.1 \leq q \leq 10.1$ and $0.0 \leq e \leq 0.99$. An alternative fit, $X_{\rm L1,sep}(f_{\rm don},q,e)$, is provided by \cite{2007ApJ...667.1170S} (Eq. A15 in their Appendix A). To compare the two fits, we compute the predicted positions using both $X_{\rm L1}(f_{\rm don},q,e)$ and $X_{\rm L1,sep}(f_{\rm don},q,e)$ and evaluate their accuracy using the root mean squared error (RMSE), 
\begin{equation}
    RMSE = \sqrt{\frac{1}{N} \sum_{i=1}^{N} (y_i - \hat{y_i})^2},
\end{equation}
where $y_i$ represents the numerical solutions and $\hat{y_i}$ the predicted values from each fit. Figure~\ref{fig:comparison_sep} illustrates the accuracy of the two fits. 
The Global-$L_1$ model illustrates overall a better agreement with the numerical solutions. 

\section{Functions appearing in the orbit-averaged equations of motion in the limit of circular orbits}\label{app:limits}

In the limit of $\mathcal{E}_{0} \rightarrow \pi$, RLOF occurs during the whole orbit. Moreover, as the orbit circularize ($e \rightarrow 0$), the following limits apply to the orbit-averaged equations of motion (Eqs.~\ref{eq:orbit_averaged_semimajor_axis} and \ref{eq:orbit_averaged_eccentricity}):
\begin{flalign}
    &\lim_{\mathcal{E}_{0},e\to\pi,0} \frac{f_{a}(e,x)}{f_{\dot{M}_{\rm don}}(e,x)}=1, 
    \lim_{\mathcal{E}_{0},e\to\pi,0} \frac{f_{e}(e,x)}{f_{\dot{M}_{\rm don}}(e,x)}=0, \nonumber \\
    &\lim_{\mathcal{E}_{0},e\to\pi,0} \frac{g_{a}(e,x)}{f_{\dot{M}_{\rm don}}(e,x)}=1,
    \lim_{\mathcal{E}_{0},e\to\pi,0} \frac{g_{e}(e,x)}{f_{\dot{M}_{\rm don}}(e,x)}=0, \nonumber \\
    &\lim_{\mathcal{E}_{0},e\to\pi,0} \frac{h_{a}(e,x)}{f_{\dot{M}_{\rm don}}(e,x)}=1,
    \lim_{\mathcal{E}_{0},e\to\pi,0} \frac{h_{e}(e,x)}{f_{\dot{M}_{\rm don}}(e,x)}=0, \nonumber 
\end{flalign}
assuming $x\neq 1$.

\section{The limit of $\delta$-function MT rate}\label{app:delta_function}

\begin{figure}[!htbp]
    \includegraphics[width=\linewidth]{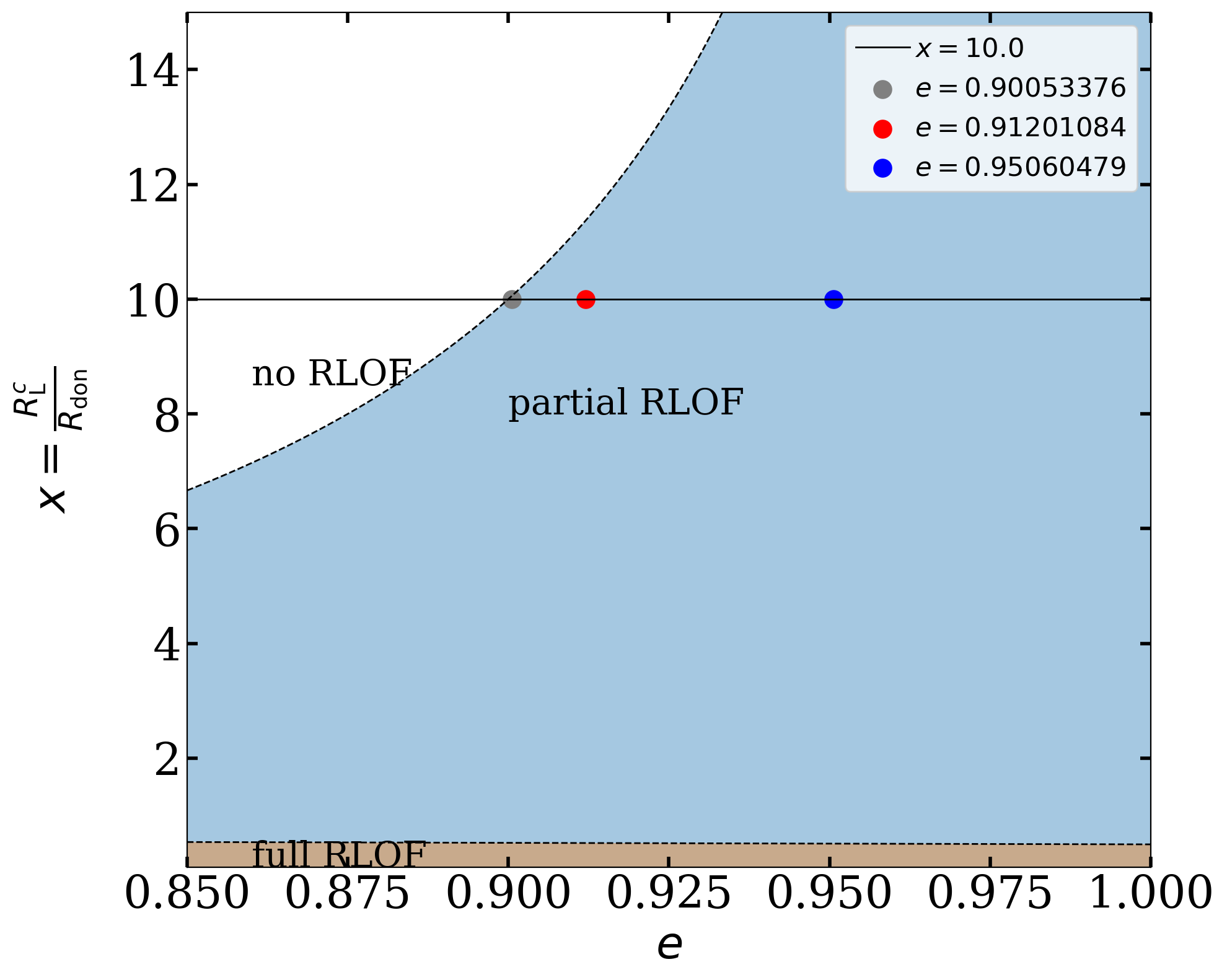}
    \caption{Zoom in region of Fig.~\ref{fig:RLOF_plane} showing the parameter space of applicability of the GeMT-model. The horizontal line corresponds to $x =10.0$.}
    \label{fig:RLOF_plane_zoom}
\end{figure}

The GeMT and the $\delta-$function models are two fundamentally different formalisms. The GeMT model accounts for the degree of RLOF through the parameter $x$ (Eq.~\ref{eq:mass_transfer_rate}), which evolves self-consistently as the Roche lobe equivalent radius changes during the integration. In contrast, this parameter is absent from the secular evolution equations of the $\delta-$function formalism \citep{2007ApJ...667.1170S, 2009ApJ...702.1387S}. This is a fundamental difference between the two models. Nevertheless, the physical motivation behind the $\delta-$function model is reflected along the border separating no from partial RLOF regions (black dashed line in Fig.~\ref{fig:RLOF_plane_zoom}).

\begin{figure}[!htbp]
    \includegraphics[width=\linewidth]{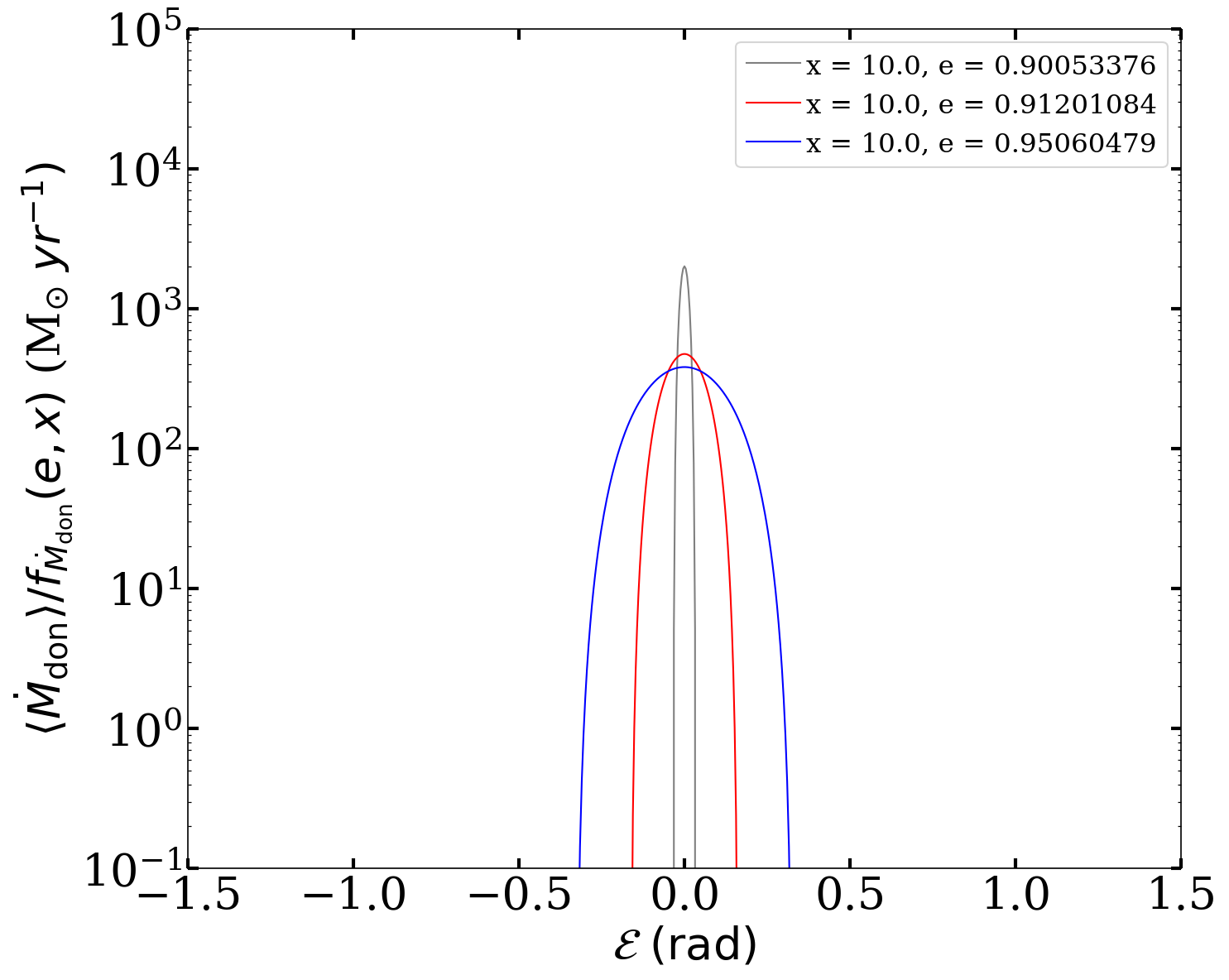}
    \caption{Normalized mass transfer rate as a function of eccentric anomaly. We selected $\langle \dot{M}_{\rm don} \rangle = 1$ M$_{\odot} \; \rm yr^{-1}$ for representation purposes.}
    \label{fig:mass_transfer_rate}
\end{figure}

In Figure~\ref{fig:RLOF_plane_zoom}, we present a small region of the $e-x$ plane and systems for $x = 10.0$ and varying $e$. Moreover, in Fig.~\ref{fig:mass_transfer_rate}, we present the respective normalized MT rate as a function of eccentric anomaly.
The $\delta$-function framework assumes that in eccentric orbits, MT occurs only at periastron, where the stellar separation is at minimum. In the GeMT formalism, this behavior naturally arises at the transition between the no RLOF and partial RLOF regimes (i.e., dashed black line in Fig~\ref{fig:RLOF_plane_zoom}). Near this boundary, the functional shape of the MT rate closely resembles a $\delta$-function (gray point in \ref{fig:RLOF_plane_zoom} and gray line in Fig.~\ref{fig:mass_transfer_rate}) as expected when MT is strongly confined around periastron. However, away from the boundary (i.e., red and blue points in Fig.~\ref{fig:RLOF_plane_zoom}) the functional shape of the MT rate is flatter, as MT occurs during a larger part of the orbit (i.e., red and blue lines in Fig.~\ref{fig:mass_transfer_rate}).

On the boundary (i.e., $\cos(E_0) \rightarrow 1$, see Eq.~\ref{eq:calc_E0}), the functional shape of the MT rate approaches a $\delta$-function. In this limit, the ratios $f_{\alpha}(e,x)/f_{\dot{M}_{\rm d}}(e,x)$ and $f_{e}(e,x)/f_{\dot{M}_{\rm d}}(e,x)$ in Eqs.~\eqref{eq:orbit_averaged_semimajor_axis_MT_points_conserv} and \eqref{eq:orbit_averaged_eccentricity_MT_points_conserv} have a regular singular point\footnote{The regular singular point reflects a physical assumption of the model; when approaching the boundary, the instantaneous MT rate approaches zero because the stellar radius $R_{\rm don}$ tends to become equal to the Roche lobe equivalent radius $R_{L}$ (see Eq.~\ref{eq:instant_mass_transfer_rate}).}, thus the true behavior of the equations is determined as
\begin{flalign} 
    &\lim_{\mathcal{E}_{0}\to0^{+}} \frac{f'_{\alpha}(e,x)}{f'_{\dot{M}_{\rm d}}(e,x)}=  \frac{1 + e}{1-e}, \lim_{\mathcal{E}_{0}\to0^{+}} \frac{f'_{e}(e,x)}{f'_{\dot{M}_{\rm d}}(e,x)} = 1 + e \nonumber
\end{flalign}
and the secular equations of motion (Eqs.~\ref{eq:orbit_averaged_semimajor_axis_MT_points_conserv} and \ref{eq:orbit_averaged_eccentricity_MT_points_conserv}) become
\begin{flalign} 
     \frac{\langle \dot{a} \rangle}{a} &= -\frac{2 \langle \dot{M}_{\rm don} \rangle}{M_{\rm don}} (1- q) \frac{1 + e}{1-e}, \label{eq:delta_func_semimajor_axis_MT_points_conserv}\\
     \langle \dot{e} \rangle &= -\frac{2 \langle \dot{M}_{\rm don} \rangle}{M_{\rm don}} (1- q) (1 + e).\label{eq:delta_func_eccentricity_MT_points_conserv}
\end{flalign}
Consequently, the functional shape of the MT rate predicted by the GeMT-model approaches a $\delta$-function when MT occurs only at periastron (i.e., $\cos(E_0) \rightarrow 1$), then the secular evolution is given by Eqs.~\eqref{eq:delta_func_semimajor_axis_MT_points_conserv}, and \eqref{eq:delta_func_eccentricity_MT_points_conserv}.

We note that when using Eqs. (39) and (40) of \cite{2007ApJ...667.1170S} the orbital AM is not conserved (see Fig.~\ref{fig:angular_momentum_evolution}). The reason stems from the treatment of the instantaneous MT rate $\dot{M}_{0}$, which is assumed to be known and constant in the works of \cite{2007ApJ...667.1170S,2009ApJ...702.1387S} and it is calculated numerically in the work of \cite{2025ApJ...983...39R}. In the GeMT-model (and in the emt-model), we re-express the instantaneous MT rate $\dot{M}_{\rm don,0}$ in Eqs.~\eqref{eq:orbit_averaged_semimajor_axis_0} and \eqref{eq:orbit_averaged_eccentricity_0} in terms of the orbit-averaged MT rate $\langle \dot{M}_{\rm don} \rangle$--that ideally would be calculated numerically and self-consistently--as described in Sect.~\ref{sec:four}. This procedure ensures conservation of the secular orbital AM (in the point-mass limit), as shown in Fig.~\ref{fig:angular_momentum_evolution}. This is expected, since secular AM conservation is obtained only when the donor and accretor masses vary in a way that is normalized to the orbit-averaged MT rate, consistent with the secular changes in $a$ and $e$.

In contrast, Eq. (39) and (40) of \cite{2007ApJ...667.1170S} \citep[and Eqs. 18 and 19 of][]{2009ApJ...702.1387S}, have been used in the literature \citep[e.g.,][]{2007ApJ...667.1170S,2009ApJ...702.1387S,2016ApJ...825...70D,2016ApJ...825...71D,2025ApJ...983...39R} without the proper normalization factor required by orbit averaging Eq. (32) of \cite{2007ApJ...667.1170S}. This is equivalent to substituting $\langle \dot{M}_1 \rangle$ directly with $\dot{M}_0$ during the derivation of Eq.~\eqref{eq:sep_analytic}-- which tracks the orbital AM evolution predicted by Eqs. (39) and (40) of \cite{2007ApJ...667.1170S}--as shown in Fig.~\ref{fig:angular_momentum_evolution}. This leads to a secular evolution, which--although formally derived under the assumption of conserved secular orbital AM--does not actually conserve it in practice, as seen in Fig.~\ref{fig:angular_momentum_evolution}. 

However, if the same normalization procedure used in our model is applied to the $\delta$-function case, then the secular orbital AM momentum is conserved. Specifically, the proper normalization can be derived by orbit averaging Eq. (32), yielding
\begin{equation}\label{eq:normalization_delta_func}
    \langle \dot{M}_{1} \rangle =  \frac{\dot{M}_{0}}{2 \pi} \frac{(1-e^2)^{3/2}}{(1+e)^2},
\end{equation}
and finally Eqs. (39) and (40) should be written as 
\begin{flalign} 
     \langle\frac{da}{dt}\rangle_{\rm sec} &= \frac{2 \langle \dot{M}_{1} \rangle}{M_{1}} \frac{a}{(1-e)^{2}} \Biggl[qe \frac{|\vec{r}_{A_2}|}{a}\cos(\phi_P)+e\frac{|\vec{r}_{A_{1},P}|}{a} \nonumber \\ 
     &+(q-1)(1-e^2)\Biggr],\label{eq:sep_semimajor_axis_MT_points_conserv} \\
     \langle\frac{de}{dt}\rangle_{\rm sec} &= \frac{\langle \dot{M}_{1} \rangle}{M_{1}} \frac{1+e}{1-e} \Biggl[q \frac{|\vec{r}_{A_2}|}{a}\cos(\phi_P)+\frac{|\vec{r}_{A_{1},P}|}{a} \nonumber \\
     &+2(1-e)(q-1)\Biggr].\label{eq:sep_eccentricity_MT_points_conserv}
\end{flalign}
Notably, \eqref{eq:sep_semimajor_axis_MT_points_conserv}, and \eqref{eq:sep_eccentricity_MT_points_conserv} have a different dependency on $e$ than those presented by \cite{2007ApJ...667.1170S}. Using Eq. (29), (39), (40) from \cite{2007ApJ...667.1170S}, and Eqs.~\eqref{eq:normalization_delta_func} it follows that
\begin{equation}\label{eq:sep_analytic_norm}
    \frac{\langle \dot{J}_{\rm orb} \rangle}{J_{\rm orb}} = 0,
\end{equation}
as expected in the limit of conservative MT. Importantly, the terms related to $|\vec{r}_{A_{1},P}|$ and $|\vec{r}_{A_2}|$ cancel each other out both in the derivations of Eqs.~\eqref{eq:sep_analytic_norm} and \eqref{eq:sep_analytic}, demonstrating that--in contrast to the emt and GeMT models--the orbital AM evolution predicted by the $\delta$-function model is equivalent for point masses and extended bodies. 

\begin{figure}[!htbp]
    \centering
    \includegraphics[width=\linewidth]{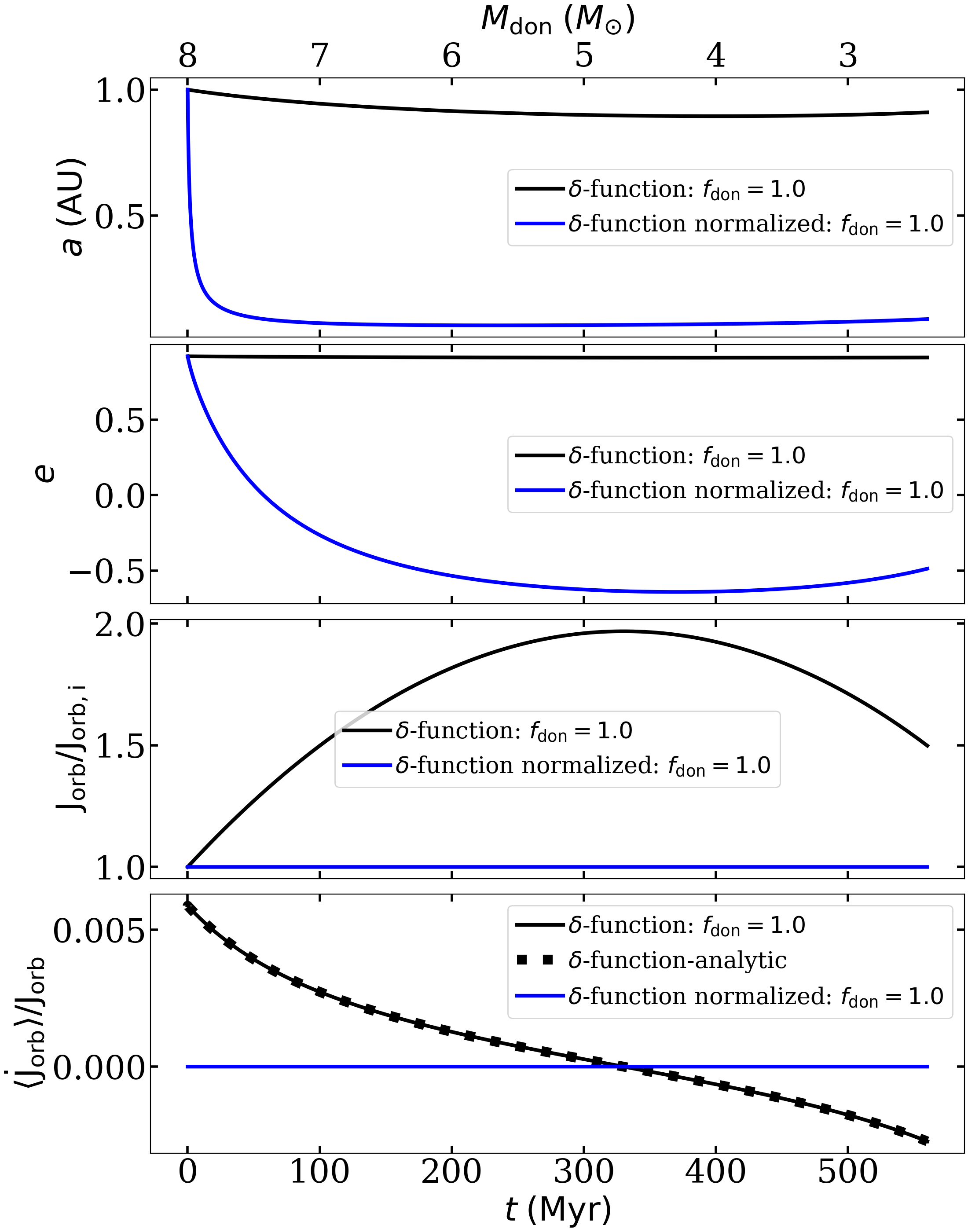}
    \caption{Secular evolution using the $\delta$-function model. The black line shows the evolution predicted by Eqs. (39) and (40) from \cite{2007ApJ...667.1170S}, for $\dot{M}_0 = 10^{-8}$ M$_{\odot} \; \rm yr^{-1}$. The black squares show Eq.~\eqref{eq:sep_analytic}. The blue line shows the evolution predicted by Eqs.~\eqref{eq:sep_semimajor_axis_MT_points_conserv}, and \eqref{eq:sep_eccentricity_MT_points_conserv}, for  $\langle \dot{M}_1 \rangle = 10^{-8}$ M$_{\odot} \; \rm yr^{-1}$. In both cases, $\cos(\phi_P) = -1$.}
    \label{fig:delta_func_corrected}
\end{figure}

In Fig.~\ref{fig:delta_func_corrected} we compare the secular evolution of the system introduced in Sect.~\ref{subsec:eccentric_orbits} using Eqs. (39) and (40) from \cite{2007ApJ...667.1170S} (black line; assuming $\dot{M}_{0} = 10^{-8}$ M$_{\odot} \; \rm yr^{-1}$) with our results obtained from Eqs.~\eqref{eq:sep_semimajor_axis_MT_points_conserv}, and \eqref{eq:sep_eccentricity_MT_points_conserv} (blue line; $\dot{M}_{\rm don} = 10^{-8}$ M$_{\odot} \; \rm yr^{-1}$).
The predicted evolution between the two sets of equations is significantly different, as expected (see Eq.~\ref{eq:normalization_delta_func}). In addition, the evolution shown by the blue line (Eqs.~\ref{eq:sep_semimajor_axis_MT_points_conserv}, and \ref{eq:sep_eccentricity_MT_points_conserv}), is more similar to that predicted by the emt and GeMT models in Fig.~\ref{fig:comparison_hamers_evolution}. Nevertheless, we see that the normalized $\delta$-function model can still lead to negative eccentricities similar to the $\delta$-function model of \cite{2007ApJ...667.1170S}--first noted by \citet{2019ApJ...872..119H}--highlighting that this behavior is inherent to the model. Most importantly, though, we see that when using the normalized $\delta$-function model, the orbital AM is conserved
(blue line) in the limit of conservative MT, contrary to the case when Eqs. (39) and (40) of \cite{2007ApJ...667.1170S} (black line) are used directly. Finally, we note, that Eqs.~\eqref{eq:sep_semimajor_axis_MT_points_conserv}, and \eqref{eq:sep_eccentricity_MT_points_conserv} also represent the $\delta$-function MT rate limit of our model (i.e., $\cos(E_0) \rightarrow 1$); they are equivalent to Eqs.~\eqref{eq:delta_func_semimajor_axis_MT_points_conserv}, and \eqref{eq:delta_func_eccentricity_MT_points_conserv} in the limit of point masses (i.e., $|\vec{r}_{A_{1},P}| = |\vec{r}_{A_2}| = 0$) verifying that the corrected $\delta$-function model is a subset of the GeMT-model.

We conclude by mentioning the $\delta$-function model has been implemented inconsistently in the literature. First, it has been implemented by using directly Eqs. (39) and (40) of \cite{2007ApJ...667.1170S} or \citep[Eqs. 18 and 19 of][]{2009ApJ...702.1387S}--without the proper normalization factor required \citep[e.g.,][]{2007ApJ...667.1170S,2009ApJ...702.1387S,2016ApJ...825...70D,2016ApJ...825...71D,2025ApJ...983...39R,2025arXiv251016201C}--leading to a secular evolution, which--although formally derived under the assumption of conserved secular orbital AM--do not actually conserve it in practice, as seen in Fig.~\ref{fig:delta_func_corrected}. Second, in \citet{2019ApJ...872..119H} the $\delta$-function model was implemented by replacing $\dot{M}_{0}$ by ${2 \pi} \langle\dot{M}_{1} \rangle$, which is not the proper normalization factor, and thus the orbital AM is not conserved--in the limit of conservative MT--in their implementation either. Here, we derive the normalized $\delta$-function model (Eqs.~\ref{eq:sep_semimajor_axis_MT_points_conserv}, and \ref{eq:sep_eccentricity_MT_points_conserv}), which has a different dependency on $e$ and we verify both numerically and analytically that it conserves the orbital AM in the limit of conservative MT. 

In summary, when $\cos(E_0) \rightarrow 1$ (i.e., MT occurs exactly at periastron), the functional shape of the MT rate predicted by the GeMT-model is a $\delta$-function. In this limit, and adopting point masses, the GeMT-model is equivalent to the corrected $\delta$-function model of \cite{2007ApJ...667.1170S} (i.e., \ref{eq:sep_semimajor_axis_MT_points_conserv}, and \ref{eq:sep_eccentricity_MT_points_conserv}).  Nevertheless, even if the initial conditions are set to approximate this limit, the parameter $x$ evolves over time. As a result, the secular evolution predicted by the emt and GeMT models is expected to diverge from the $\delta$-function model as the integration proceeds, highlighting the fundamental difference between these formalisms. 

\section{Explicit expressions for the functions appearing in the orbit-averaged equations of motion}\label{app:dimensionless_functions}

In this section, we explicitly present the dimensionless functions $f_{\dot{M}_{\rm don}}(e,x), \; f_{a}(e,x), \; f_{e}(e,x), \; g_{a}(e,x), \; g_{e}(e,x), \; h_{a}(e,x),$ and $ \; h_{e}(e,x)$, as referenced in Sect.~\ref{sec:four}.

\subsection{Normalization}\label{app:normalization}

The dimensionless normalization function obtained by orbit-averaging the mass loss rate $\dot{M}_{\rm don}$ given by Eq.~\eqref{eq:mass_transfer_rate},
\begin{flalign}
f_{\dot{M}_{\rm don}}&(e,x) = - \frac{1}{96 \pi} \Biggl(36 e^4 x^3 \mathcal{E}_0 + 3e^4  \nonumber x^3\sin{(4\mathcal{E}_0)} \\ \nonumber
&- 32e^3 x^3 \sin{(3\mathcal{E}_0)} + 24 e^3 x^2 \sin{(3\mathcal{E}_0)} \\ \nonumber
&+ 288 e^2 x^3 \mathcal{E}_0 - 432 e^2 x^2 \mathcal{E}_0   \\ \nonumber
&+24 e^2 x(e^2 x^2 +6x^2 -9x +3)\sin(2\mathcal{E}_0) +144 e^2 x \mathcal{E}_0  \\ \nonumber
&-24 e(12e^2x^3-9e^2x^2+16x^3-36x^2+24x-4)\sin{(\mathcal{E}_0)} \\ 
&+96x^3 \mathcal{E}_0 -288x^2\mathcal{E}_0 +288x\mathcal{E}_0 -96 \mathcal{E}_0\Biggr). \\ \nonumber
\end{flalign}

\subsection{Dimensionless functions associated with the relative acceleration of the stars}

The dimensionless functions associated with the terms in $f_{\rm RLOF}$ (Eq.~\ref{eq:total_perturbation_non_conservative_MT_simplified}) proportional to $\vec{\dot{r}}$, which are related to the distribution of the total mass in the system and the effects of mass and angular momentum loss. Furthermore, these terms appear in Eqs.~\eqref{eq:orbit_averaged_semimajor_axis} and \eqref{eq:orbit_averaged_eccentricity} regardless of whether point masses or extended bodies are assumed,
\begin{flalign}
f_{a}&(e,x) =  \frac{1}{96 \pi} \Biggl(36 e^4 x^3 \mathcal{E}_0 + 3e^4 x^3 \nonumber \sin{(4\mathcal{E}_0)} \\ \nonumber
&- 16 e^3 x^3 \sin{(3\mathcal{E}_0)}  + 24 e^3 x^2 \sin{(3\mathcal{E}_0)} - 144 e^2 x^2 \mathcal{E}_0 \\ \nonumber
&+ 24 e^2 x(e^2 x^2 -3x +3) \sin(2\mathcal{E}_0) +144 e^2 x \mathcal{E}_0    \\ \nonumber
&+24 e (-6 e^2 x^3 +9 e^2 x^2 + 8 x^3 -12 x^2 +4)\sin{(\mathcal{E}_0)} \\ 
&-96x^3 \mathcal{E}_0 +288x^2\mathcal{E}_0 -288x\mathcal{E}_0 +96 \mathcal{E}_0\Biggr),  \\ \nonumber
\end{flalign}

\begin{flalign}
f_{e}&(e,x) =  - \frac{e^{2} - 1}{32 \pi} \Biggl(12 e^{3} x^{3} \nonumber \mathcal{E}_0 + e^{3} x^{3} \sin{(4 \mathcal{E}_0 )} \\ \nonumber
&- 8 e^{2} x^{3} \sin{(3 \mathcal{E}_0 )} + 8 e^{2} x^{2} \sin{(3 \mathcal{E}_0 )} + 48 e x^{3} \mathcal{E}_0 \\ \nonumber
&- 96 e x^{2} \mathcal{E}_0 + 8 e x (e^{2} x^{2} + 3 x^{2} - 6 x + 3) \sin{(2 \mathcal{E}_0 )} \\ 
&+ 48 e x \mathcal{E}_0  \\ \nonumber
&- (72 e^{2} x^{3} - 72 e^{2} x^{2} + 32 x^{3} - 96 x^{2} + 96 x - 32) \sin{(\mathcal{E}_0 )}\Biggr). \\ \nonumber
\end{flalign}

\subsection{Dimensionless functions associated with the ejection point}\label{app:funcs_ejection_point}

The dimensionless functions associated with the terms in $f_{\rm RLOF}$ (Eq.~\ref{eq:total_perturbation_non_conservative_MT_simplified}) that are related to the ejection point. These terms appear by modeling the donor as an extended body,

\begin{flalign}
g_{a}&(e,x) =\frac{1}{32 \pi} \Biggl( ex 
\Bigl( - 16(6+e^{2}(x-3) - 4x)x\sin{(\mathcal{E}_0)} \\
&+ e^{2}x [16(1- x) \sin{(3 \mathcal{E}_0 )} + 3 e x \sin{(4 \nonumber \mathcal{E}_0)}] \\ \nonumber 
&+ 8e(((e^{2} + 2)x - 6)x + 3) \sin{(2 \mathcal{E}_0)}\Bigr)  \\ \nonumber
&+ 4 \mathcal{E}_0 x(-8(3+(x-3)x)+e^{2}(12+(e^{2}-8)x^{2})) \\ 
&+ 64 \sqrt{1 - e^{2}} \operatorname{asin}{(\frac{\sqrt{e + 1} \sqrt{\frac{1}{1 - e}} \sin{(\frac{\mathcal{E}_0}{2} )}}{\sqrt{\cos^{2}{(\frac{\mathcal{E}_0}{2} )} + \frac{(e + 1) \sin^{2}{(\frac{\mathcal{E}_0}{2} )}}{1 - e}}} )}\Biggr), \\  \nonumber
\end{flalign}

\begin{flalign}
g_{e}&(e,x) = \frac{(1 - e^{2})}{48 \pi e} \Biggl(e^{2} x \Bigl(e x [3 e x \sin{(4 \nonumber \mathcal{E}_0 )} \\ \nonumber 
&+ (18 - 20 x) \sin{(3 \mathcal{E}_0 )}] \\ \nonumber 
&+ 6([2 (e^{2} + 4)x - 15]x + 6) \sin{(2 \mathcal{E}_0 )}\Bigr) \\ \nonumber 
&- 6 e \Bigl([e^{2} x (14 x - 15) +  8\bigl((x - 3)x + 3\bigr)]x - 4\Bigr) \sin{(\mathcal{E}_0 )} \\ \nonumber 
&+ 12\Bigl(e^{2} x \bigr([(e^{2} + 4)x - 9]x + 6 \bigl) - 2 \Bigr) \mathcal{E}_0 \\  
&+ 48 \sqrt{1 - e^{2}} \operatorname{asin}{(\frac{\sqrt{e + 1} \sqrt{\frac{1}{1 - e}} \sin{(\frac{\mathcal{E}_0}{2} )}}{\sqrt{\cos^{2}{(\frac{\mathcal{E}_0}{2} )} + \frac{(e + 1) \sin^{2}{(\frac{\mathcal{E}_0}{2} )}}{1 - e}}} )}\Biggr). \\ \nonumber
\end{flalign}

\subsection{Dimensionless functions associated with the accretion point}\label{app:funcs_accretion_point}

The dimensionless functions associated with the terms in $f_{\rm RLOF}$ (Eq.~\ref{eq:total_perturbation_non_conservative_MT_simplified}) that are related to the accretion point. These terms appear by modeling the accretor as an extended body,

\begin{flalign}
h_{a}&(e,x) =\frac{1}{8 \pi} \Biggl( -4 \mathcal{E}_0 x \nonumber
\Bigl(6+x(-6+(2+e^{2})x)\Bigr)  \\ \nonumber
&+ 2e \Bigl((-12x-(-4+e^{2})x^{3}-\frac{4}{-1+e\cos{\mathcal{E}_0}} \Bigr)\sin{(\mathcal{E}_0}) \\  \nonumber 
&+ ex^{2}(3(-2+x)\sin{(2\mathcal{E}_0}) -ex\sin{(3\mathcal{E}_0}))\Bigr) \\  
&+\frac{16}{\sqrt{1-e^2}} \operatorname{asin}{(\frac{\sqrt{e + 1} \sqrt{\frac{1}{1 - e}} \sin{(\frac{\mathcal{E}_0}{2} )}}{\sqrt{\cos^{2}{(\frac{\mathcal{E}_0}{2} )} + \frac{(e + 1) \sin^{2}{(\frac{\mathcal{E}_0}{2} )}}{1 - e}}} )}\Biggr), \\  \nonumber  
\end{flalign}

\begin{flalign}
h_{e}&(e,x) =\frac{1}{48 e\pi} \Biggl( \frac{1} {-1+e\cos{(\mathcal{E}_0})}(-1+e^{2})  \nonumber \\  
&\Bigl[12 \mathcal{E}_0 (-2+e^{2}x^{2}(-3+2x)) \nonumber \\ 
&+ 4e \Bigl(3\mathcal{E}_0(2+e^{2}(3-2x)x^{2})\cos{(\mathcal{E}_0}) +6\sin{(\mathcal{E}_0}) \nonumber \\
&+x(-1+e\cos{(\mathcal{E}_0}))\Bigl(4(9+x(-9+(3+2e^{2})x))  \nonumber\\
&+ ex(9(3-2x)\cos{(\mathcal{E}_0)}+4ex\cos{(2\mathcal{E}_0}))\Bigr)\sin{(\mathcal{E}_0})\Bigl)\Bigl] \nonumber \\
&+48\sqrt{1-e^2}\operatorname{asin}{(\frac{\sqrt{e + 1} \sqrt{\frac{1}{1 - e}} \sin{(\frac{\mathcal{E}_0}{2} )}}{\sqrt{\cos^{2}{(\frac{\mathcal{E}_0}{2} )} + \frac{(e + 1) \sin^{2}{(\frac{\mathcal{E}_0}{2} )}}{1 - e}}} )}\Biggr).  \\   \nonumber 
\end{flalign}
\end{appendix}

\end{document}